\documentclass[12pt,twoside]{article}
\usepackage{amsmath,amsfonts,amssymb,amsthm,mathabx}
\usepackage{times}
\usepackage{pdfsync}
\usepackage{cite}
\usepackage{url}
\usepackage{hyperref}
\usepackage{tensor} 
\usepackage{color}
\usepackage{caption}
\usepackage{enumitem}
\usepackage{graphicx}
\usepackage{array}
\usepackage{amsbsy}
\usepackage{appendix}
\usepackage{mathrsfs}
\usepackage{physics}
\usepackage{cancel}
\usepackage{yfonts}
\usepackage{inputenc}
\usepackage{chngcntr}

\advance\textheight\topskip
\numberwithin{equation}{section} \makeatletter
\@addtoreset{equation}{section}

\newcommand{\D}{\text{d}}

\begin{document}

\title{Notes on massless scalar field partition functions, modular inariance
  and Eisenstein series}

\author{Francesco Alessio, Glenn Barnich, Martin Bonte}


\def\mytitle{Notes on massless scalar field partition functions, modular invariance
  and Eisenstein series}

\pagestyle{myheadings} \markboth{\textsc{\small F.~Alessio et al.}}
{\textsc{\small Massless scalar partition functions and Eisenstein series}}

\addtolength{\headsep}{4pt}

\begin{centering}

  \vspace{1cm}

  \textbf{\Large{\mytitle}}

  \vspace{1cm}

  {\large Francesco Alessio${}^{a,b,c}$, Glenn Barnich${}^{b}$, Martin
    Bonte$^{b}$}

  \vspace{1cm}

  \begin{minipage}{.9\textwidth}\small \it  \begin{center}
      ${}^a$ Nordita \\ KTH Royal Institute of Technology and Stockholm
      University \\
      Hannes Alfvéns väg 12, SE-106 91 Stockholm, Sweden\\
      E-mail:
      \href{mailto:francesco.alessio@su.se}{francesco.alessio@su.se}
    \end{center}
  \end{minipage}

  \vspace{.5cm}

  \begin{minipage}{.9\textwidth}\small \it  \begin{center}
      ${}^b$ Physique Th\'eorique et Math\'ematique, Universit\'e libre de
      Bruxelles and \\ International Solvay Institutes, Campus
      Plaine C.P. 231, B-1050 Bruxelles, Belgium\\
      E-mail:
      \href{mailto:gbarnich@ulb.ac.be}{gbarnich@ulb.ac.be},
      \href{mailto:mbonte@ulb.ac.be}{mbonte@ulb.ac.be} 
    \end{center}
  \end{minipage}

  \vspace{.5cm}

  \begin{minipage}{.9\textwidth}\small \it  \begin{center}
      ${}^c$ Dipartimento di Fisica ``E. Pancini'' and \\INFN Universit\`a degli
      studi di Napoli ``Federico II'', I-80125 Napoli, Italy
    \end{center}
  \end{minipage}

\end{centering}

\vspace{1cm}
  
\begin{center}
  \begin{minipage}{.9\textwidth} \textsc{Abstract}. The partition function of a
    massless scalar field on a Euclidean spacetime manifold $\mathbb
    R^{d-1}\times \mathbb{T}^2$ and with momentum operator in the compact
    spatial dimension coupled through a purely imaginary chemical potential is
    computed. It is modular covariant and admits a simple expression in terms of
    a real analytic SL$(2,\mathbb Z)$ Eisenstein series with $s=(d+1)/2$.
    Different techniques for computing the partition function illustrate
    complementary aspects of the Eisenstein series: the functional approach
    gives its series representation, the operator approach yields its Fourier
    series, while the proper time/heat kernel/world-line approach shows that it
    is the Mellin transform of a Riemann theta function. High/low temperature
    duality is generalized to the case of a non-vanishing chemical potential. By
    clarifying the dependence of the partition function on the geometry of the
    torus, we discuss how modular covariance is a consequence of full
    SL$(2,\mathbb Z)$ invariance. When the spacetime manifold is $\mathbb
    R^{p}\times \mathbb{T}^{q+1}$, the partition function is given in terms of a
    SL$(q+1,\mathbb Z)$ Eisenstein series again with $s=(d+1)/2$. In this case,
    we obtain the high/low temperature duality through a suitably adapted dual
    parametrization of the lattice defining the torus. On $\mathbb T^{d+1}$, the
    computation is more subtle. An additional divergence leads to an harmonic
    anomaly.
     \end{minipage}
\end{center}

\vfill
\thispagestyle{empty}

\newpage

\tableofcontents

\vfill
\newpage

\section{Introduction}
\label{sec:introduction}

The SL$(2,\mathbb Z)$ real analytic Eisenstein series,
\begin{equation}
  \label{eq:1}
  f_s(\tau,\bar\tau)=\sum_{(m,n)\in \mathbb Z^2/(0,0)}
  \frac{\tau_2^s}{\abs{m+n\tau}^{2s}}, \quad \mathfrak{Re}(s)>1, 
\end{equation}
with $\tau =\tau_1+i\tau_2\in \mathbb C$, invariant under modular
transformations
\begin{equation}
  \label{eq:2}
  \tau\to \frac{a\tau +b}{c\tau +d}, \quad a,b,c,d\in \mathbb Z,\quad
  ad-bc=1, 
\end{equation}
feature prominently in theoretical physics, for instance in the context of
string theory amplitudes, gravitational instantons and Feynman integrals, see
e.g.~\cite{Obers:1999um,Petropoulos2012,%
  Williams:2014udz,fleig_gustafsson_kleinschmidt_persson_2018,Bluemlein2019} and
references therein.

The purpose of the first part of these notes is to review how such series
directly appear in the partition function of a real massless scalar field living
on a Euclidean spacetime manifold with two small dimensions with periodic
boundary conditions. One of these small dimensions is the Euclidean time, the
other being one of the spatial dimensions, while the remaining $d-1$ spatial
dimensions are taken to be large. When $d=1$, one recovers the well-known
modular invariant result.

Casimir energies and partition functions for such configurations were
originally computed in \cite{Ambjorn:1981xw} in terms of Epstein zeta
functions (see also \cite{Plunien:1986ca,Gies:2003cv} for worldline
methods and \cite{Plunien:1987fr,doi:10.1142/2065,Bordag:2009zzd} for
further developments and reviews). This result has also been derived
in the context of conformal field theories in higher dimensions in
\cite{Cappelli:1988vw}. The aim there was an attempt to generalize
some of the successes of two dimensional conformal field theories in
general and of modular invariance, and thus of high/low temperature
duality, in particular
\cite{Belavin:1984vu,Cardy:1986ie,CAPPELLI1987445} to higher
dimensions \cite{Cardy:1991kr}. More recent investigations along these
lines include
\cite{Dolan1998,Shaghoulian:2015kta,Shaghoulian:2016gol,Horowitz:2017ifu}.

Our motivation for re-analyzing these scalar field results comes from
a recent investigation on the implications of boundary or fall-off
conditions for both the degrees of freedom and the partition functions
of gauge and gravitational theories
\cite{Barnich:2019xhd,GlennBarnich}. More concretely, we have studied
how a modular covariant partition function can be constructed for
electromagnetism in the context of the Casimir set-up, that is to say
for a slab geometry with two perfectly conducting parallel and
infinitely extended metal plates separated by a distance $a$. We have
shown how the well-known $E$ and $H$ modes (see
e.g.~\cite{Deutsch1979} for an early review) may be combined into a
single massless scalar field with periodic boundary conditions on an
interval of length $2a$. Modular covariance for electromagnetism then
follows from the scalar field result when understanding what
electromagnetic operator corresponds to scalar field momentum in the
compact dimension \cite{Alessio:2020okv}. Moreover, this analysis may
be generalized directly to the case of gravitons \cite{Alessio2021}.

In the second part of these notes, we compute the partition function on the flat
Euclidean spacetime manifold $\mathbb R^{d-q}\times \mathbb T^{q+1}$ in terms of
a SL$(q+1,\mathbb Z)$ Eisenstein series with $s=\frac{d+1}{2}$. On a skewed
torus, the periodicities of the scalar field are described by general lattice
vectors. A key point is to introduce adapted coordinates in which these
periodicities become equal to unity along each direction. In these coordinates,
the information on the torus is encoded in the coupling of the scalar field to a
flat background metric that contains the information on the geometry of the
torus. The more skewed the torus, the more observables are turned on in the
partition function. High/low temperature duality follows by two complementary
ADM parametrizations of this metric. For the partition function on $\mathbb
T^{d+1}$, the result is divergent. As understood in \cite{Obers:1999um}, using
dimensional regularization, the subtraction of the pole leads to an harmonic
anomaly.

The literature on partition functions for massless scalar fields and
its dependence on boundary conditions on the one hand, and on physical
applications of Eisenstein series on the other, is substantial. Hence,
the question on the added value of the current discussion, which
necessarily has considerable overlaps with existing literature, is
legitimate.

(i) From the viewpoint of the finite temperature Casimir effect, our
discussion and techniques follow closely the approach developed in
\cite{Lutken:1988ge,PhysRevD.40.4191,Wotzasek:1989gq,Ravndal:1990mu}
in order to derive and understand temperature inversion symmetry,
originally discovered in \cite{Brown:1969na} through the method of
images, in terms of functional methods and Epstein zeta functions. Our
addition here consists in including a chemical potential for linear
momentum in the compact dimension which brings one from Epstein zeta
functions with temperature inversion symmetry to real analytic
SL$(2,\mathbb Z)$ Eisenstein series with modular covariance.

(ii) From the viewpoint of modular transformations in higher dimensions, as
compared to the analysis in \cite{Cappelli:1988vw}, our derivation provides the
full analytic expression for the partition function in the case of the simplest
model of a free massless scalar field on the flat Euclidean manifold
$\mathbb{R}^p\times\mathbb{T}^{q+1}$ for $p+q=d$, in terms of SL$(q+1,\mathbb
Z)$ Eisenstein series, with an explicit proof of SL$(q+1,\mathbb Z)$ invariance
built in (see equation \eqref{F21.a} and remark (iii) thereafter). As compared
to \cite{Shaghoulian:2015kta}, for the model under consideration, there is full
control on finite-size corrections, and also a new formula, valid now at low
rather than at high temperature, that relates the leading contribution to the
entropy to the Casimir energy density of a massless scalar field in one
dimension lower (see equations \eqref{FF14} and \eqref{eq:65bis}). With respect
to \cite{Obers:1999um}, the high temperature/large volume discussion is
extended from a factorized torus $\mathbb T^{q}\times \mathbb S^1$ to a general
torus $\mathbb T^{q+1}$ through a dual ADM parametrization (see equations
\eqref{FF17a} and \eqref{eq:284}).

(iii) From the viewpoint of the relevance of Eisenstein series in
physical applications, we show that, in the context of quantum
statistical physics, partitions functions of massless scalar fields on
Euclidean spacetime manifolds of the form
$\mathbb{R}^p\times\mathbb{T}^{q+1}$ are among the simplest physical
observables that are directly expressed through such series. Various
approaches to computing these partition functions, such as functional
integrals, canonical quantization or proper-time/heat
kernel/world-line methods illustrate complementary aspects of
Eisenstein series. 

The plan of the paper is the following. In the next section, we briefly review
how to derive the scalar black body result, i.e., the partition function for a
massless scalar field in $d$ large spatial dimensions by canonical, path
integral and proper time methods. At the end of the section, we recall how this
black body partition function can be used to derive the Casimir energy at zero
temperature in the case of one small spatial dimension.

We then turn to the case of finite temperature with one small spatial
dimension where the partition function can be simply expressed in
terms of the Eisenstein series $f_{(d+1)/2}$ when turning on a purely
imaginary chemical potential for the momentum operator in the small
dimension. Our discussion covers the Casimir effect at finite and zero
temperature.

In the last section, we generalize the discussion to $p$ large and
$q=d-p$ small spatial dimensions. We clarify how the partition
function depends on the full geometry of the $q+1$ dimensions torus,
including the case $q=1$, and express the result in terms of an
SL$(q+1,\mathbb Z)$ invariant partition function. All results of the
previous sections may be obtained as particular cases, so that one may
start directly with this section if one is interested in the final
results and the general structure, but not necessarily the way our
understanding of these matters has evolved from studying the simpler
particular cases first.

\section{Generalities. One small spacetime dimension}
\label{sec:massl-scal-part}

\subsection{Mode expansion}
\label{sec:mode-expansion}

Consider a massless scalar field $\phi$ in $d$ spatial dimensions, whose
Lagrangian action reads
\begin{equation}
  \label{eq:3}
  S[\phi]=-\frac{1}{2}\int\D x^0\int_{V_d}\D^dx\ \partial_{\mu}\phi\partial^{\mu}\phi,
\end{equation}
while the first order Hamiltonian action is
\begin{equation}
  \label{eq:6}
  S_H[\phi,\pi]=\int\D x^0[\int_{V_d}\D^dx\ \dot\phi\pi-H],\quad
  H=\frac{1}{2}\int_{V_d}\D^dx\big(\pi^2+\partial_i\phi\partial^i\phi\big), 
\end{equation}
with associated canonical Poisson brackets
\begin{equation}
  \label{eq:10}
  \{\phi(x^0,x),\pi(x^0,x')\}=\delta^{d}(x-x').
\end{equation}
Our conventions for indices and their ranges are as follows: the spacetime and
spatial indices are $\mu=0,...,d$, and $i=1,...,d$, respectively. Besides the
Hamiltonian, the other observables that are relevant for us here are linear
momenta in the $x^i$ direction,
\begin{equation}
  \label{eq:8}
  P_i=-\int_{V_d}\D^dx\ \pi\partial_i\phi.
\end{equation}
Our aim is to discuss analytic expressions in terms of Eisenstein series and
associated modular properties of partition functions of the type
\begin{equation}
  \label{eq:9}
  Z_d(\beta,\mu^j)={\rm Tr}\ e^{-\beta (\hat H -i\mu^j\hat P_j)},
\end{equation}
and generalizations thereof. In order to take advantage of techniques
in complex analysis, we use purely imaginary chemical potentials
$i\mu^j,\mu^j\in\mathbb R$, as in 2-dimensional conformal field theory
and also for instance in
\cite{ROBERGE1986734,Alford:1998sd,PhysRevD.75.025003}.

In order to evaluate such partition functions, a standard procedure is to put
the system in a hyperrectangular box of volume $V_d=\prod^d_{i=1} L_i$ and to
choose periodic boundary conditions in all spatial dimensions,
\begin{equation}
  \phi(x^0,x^1,...,x^i,...,x^d)=\phi(x^0,x^1,...,x^i+L_i,...,x^d)
  \label{eq:21}.
\end{equation}
The appropriate orthonormal basis vectors $\{e_{k_i}\}$ are plane waves,
\begin{equation}
  \label{eq:4}
  e_{k_i}(x)=\frac{1}{\sqrt{V_d}}e^{ik_ix^i},\quad
  k_i=\frac{2\pi n_i}{L_i},\quad
  \left(e_{k_i},e_{k'_i}\right)=\int_{V_d}\D^dx\
  e^*_{k_i}(x)e_{k'_i}(x)=\prod_{i}\delta_{n_i,n'_i}.  
\end{equation}
The fields $\phi,\pi$ admit the mode decomposition
\begin{equation}
  \label{eq:5}
  \phi(x)=\sum_{n_i\in\mathbb{Z}^d}\phi_{k_i}e_{k_i}(x),\quad
  \pi(x)=\sum_{n_i\in\mathbb{Z}^d}\pi_{k_i}(x)e_{k_i},
\end{equation}
with reality conditions
$\phi_{k_i}=\phi^*_{-k_i},\pi_{k_i}=\pi^*_{-k_i} $.  In terms of the
Fourier components, the observables are
\begin{equation}
  \label{eq:11}
  H=\frac{1}{2}\sum_{n_i\in\mathbb{Z}^d}(\pi_{k_i}\pi^*_{k_i}
  +\omega^2_{k_i}\phi_{k_i}\phi^{*}_{k_i}),
  \quad P_i=-i\sum_{n_i\in\mathbb{Z}^d} k_i \phi_{k_i}\pi^*_{k_i}, 
\end{equation}
where $\omega_{k_i}=\sqrt{k_{j}k^j}$, while the non-vanishing canonical
Poisson brackets become
\begin{equation}
  \label{eq:7}
 \{\phi_{k_i},\pi^*_{k'_i}\}=\prod_{i}\delta_{n_i,n'_i}.
\end{equation}

The zero mode is denoted by $\phi_{0}=q,\pi_{0}=p$ and, for the Fourier
components with $n_i\neq (0,\dots,0)$, one defines standard oscillator
variables,
\begin{equation}
  \label{eq:12}
  a_{k_i}=\sqrt{\frac{\omega_{k_i}}{2}}\big[\phi_{k_i}+\frac{i}{\omega_{k_i}}\pi_{k_i}\big].
\end{equation}
In these terms,
\begin{equation}
  \label{eq:140}
  \begin{split}
    \phi&={V_d}^{-\frac{d-1}{2d}}q+\frac{1}{\sqrt{V_d}}{\sum_{n_i\in\mathbb Z^d}}'
    \frac{1}{\sqrt{2\omega_{k_i}}}(a_{k_i}e^{ik_{j}x^j}+{\rm c.c.}),\\
    \pi&={V_d}^{-\frac{d+1}{2d}}p-\frac{i}{\sqrt{V_d}}{\sum_{n_i\in\mathbb Z^d}}'
    \sqrt{\frac{\omega_{k_i}}{2}}(a_{k_i}e^{ik_{j}x^j}-{\rm c.c.}),
  \end{split}
\end{equation}
where the prime indicates that the term with $n_i=(0,\dots,0)$ is excluded from
the sum and $q,p$ are dimensionless. The observables are given by
\begin{equation}
  \label{eq:13}
  H={V_d}^{-\frac{1}{d}}\frac{p^2}{2}
  +\frac{1}{2}\sideset{}{'}\sum_{n_i\in\mathbb{Z}^d}\omega_{k_i}(a^{*}_{k_i}a_{k_i}+a_{k_i}a^{*}_{k_i}),
  \quad
  P_i=\frac{1}{2}\sideset{}{'}\sum_{n_i\in\mathbb{Z}^d}k_i(a^*_{k_i}a_{k_i}+a_{k_i}a^*_{k_i}),
\end{equation}
 while the non-vanishing Poisson brackets become
\begin{equation}
  \label{eq:14}
  \{q,p\}=1,\quad \{a_{k_i},a^{*}_{k_i}\}=-i\prod_{i}\delta_{n_i,n'_i}.
\end{equation}

\subsection{Canonical approach to partition function in large volume
  limit}
\label{sec:canon-appr-large}

Throughout this section, where we will consider all spatial dimensions to be
large, we will also take the chemical potentials $\mu^j$ to vanish. How the
final result changes when they are turned on will be discussed from a more
general perspective later.

\subsubsection{Zero mode contribution}
\label{sec:zero-mode-contr}

It follows from the expression of the Hamiltonian \eqref{eq:13} that the zero
mode sector of the theory corresponds to a free particle with Hamiltonian
$H_0={V_d}^{-\frac{1}{d}}\frac 12 p^2$. Therefore one may quantize the full theory in a Fock space
for which the vacua $|p\rangle $ are labeled by the continuous eigenvalues $p$
of $\hat p$,
\begin{equation}
  \label{eq:141}
  \hat p |p\rangle =p|p\rangle,\quad \hat a_{k_i}|p\rangle=0. 
\end{equation}
It follows that 
\begin{equation}
  \label{eq:145}
  Z_0(\beta)={\rm Tr}\, e^{-\beta\hat H_0}=\int^{+\infty}_{-\infty}\text{d}p\,
  e^{-\frac{b}{2}p^2}\langle p|p \rangle=\sqrt{\frac{2\pi}{b}}\delta(0),
  \quad b={V_d}^{-\frac{1}{d}}\beta. 
\end{equation}

\subsubsection{Oscillator contribution}
\label{sec:oscill-contr}

For the oscillator modes quantized in Fock space, one may directly evaluate the
trace using the normal ordered Hamiltonian $:\hat H':$ so that no divergences
occur. If $N_{k_i}$ denotes the occupation number of the oscillator associated
to $k_i$, we have, 
\begin{equation}
  \label{eq:15}
  Z'_d(\beta)={\rm Tr}\ e^{-\beta
    :\hat H':}=\sideset{}{'}\prod_{n_i\in\mathbb Z^d}\sum_{N_{k_i}\in \mathbb N} e^{-\beta \omega_{k_i}
    N_{k_i}}=\sideset{}{'}\prod_{n_i\in\mathbb Z^d}\frac{1}{1-e^{-\beta \omega_{k_i}}}, 
\end{equation}
and thus
\begin{equation}
  \label{eq:16}
  \ln Z'_d(\beta)=-\sideset{}{'}\sum_{n_i\in\mathbb Z^d}\ln (1-e^{-\beta \omega_{k_i}}). 
\end{equation}

For large $L_i$, the sums can be approximated by integrals using
$\text{d}k_i=\frac{2\pi}{L_i} \text{d}n_i$. After performing the integrals over
the angles in hyperspherical coordinates, one finds
\begin{multline}
  \label{eq:17}
  \ln Z'_d(\beta)= -\frac{V_d}{(2\pi)^d}\int \D^dk \ln (1-e^{-\beta
    \omega_{k_i}})\\=-\frac{V_d}{(2\pi)^d}{\rm Vol}(\mathbb
  S^{d-1})\int^\infty_0\D k k^{d-1}\ln (1-e^{-\beta k}). 
\end{multline}
Since the finite contribution of the zero mode $ \ln Z_0(\beta)\sim \ln V_d$ is
subdominant in the large volume limit, it may be neglected in most of the paper,
unless otherwise specified.

On the one hand, ${\rm Vol}(\mathbb
S^{d-1})={2\pi^{\frac{d}{2}}}/{\Gamma(\frac{d}{2})}$ while, on the other, the
integral becomes after the change of variables $x=\beta k$ and a standard
integrations by parts,
\begin{equation}
\int^\infty_0\D k k^{d-1}\ln (1-e^{-\beta k})=-\frac{1}{d\beta^d}
\int^\infty_0
\D x\frac{x^d}{e^x-1}=-\frac{1}{d\beta^d}\Gamma(d+1)\zeta(d+1).\label{eq:19}
\end{equation}
One thus ends up with the ``scalar black body'' partition function
\begin{equation}
  \label{eq:121}
  \ln Z_d(\beta)=
  \frac{\Gamma(d)\zeta(d+1)}{2^{d-1}\pi^{\frac{d}{2}}\Gamma(\frac{d}{2})}\frac{V_d}{\beta^d}.
\end{equation}
Furthermore, using the reduplication formula
\begin{equation}
  \label{eq:57}
  \Gamma(z)\sqrt{\pi}=2^{z-1}\Gamma(\frac{z}{2})\Gamma(\frac{z+1}{2}),
\end{equation}
at $z=d$, and the definition of the completion of the zeta function,
\begin{equation}
  \label{eq:73}
  \xi(z)=\frac{\Gamma(\frac{z}{2})\zeta(z)}{\pi^{\frac{z}{2}}},
\end{equation}
at $z=d+1$, the result can be written as
\begin{equation}
     \label{eq:18}
     \boxed{\ln Z_d(\beta)=\xi(d+1)\frac{V_d}{\beta^d}}=
     \left\{\begin{array}{ll} \frac{\pi}{6}\frac{L_1}{\beta}  & {\rm for}\ d=1\\
             \frac{\zeta(3)}{2\pi}\frac{L_1L_2}{\beta^2} \
             & {\rm for}\ d=2\\
             \frac{\pi^2}{90}\frac{L_1L_2L_3}{\beta^3} \
             & {\rm for}\ d=3\\
             & \vdots
           \end{array}
           \right..   
\end{equation}

{\bf Remarks:}

(i) This result for the partition function is independent of the choice of
boundary conditions in the large volume limit. For instance, for Dirichlet
conditions $\phi(t,x^1,...,0,...,x^d)=0=\phi(t,x^1,...,L_i,...,x^d)$, the
appropriate basis is $e_{k_i}=\sqrt{\frac{2^d}{V_d}}\prod_i\sin k_i
x=e^*_{k_i}$, $k_i=\frac{\pi n_i}{L_i}$. In particular, there is no mode at
$n_i=(0,\dots,0)$ that has to be dealt with. All mode expansions are the same
except that sums and products are restricted to $n_i>0$. In the evaluation of
the partition function after \eqref{eq:16}, this is compensated by the fact that
$\text{d}k_i=\frac{\pi}{L_i} \text{d}n_i$. Equivalently, the sums can be extended to $\mathbb
Z^d/(0,\dots,0)$ while dividing by $2^d$, because the summands are even functions
of the $n_i$, and the computation proceeds as before.

(ii) For $d=3$, the standard argument to get the partition functions for the
electromagnetic field is to multiply the scalar field result by $2$ in order to
take into account the $2$ independent polarizations of the photon. On account of
the equivalence established in \cite{Alessio:2020okv} of the electromagnetic
field with Casimir boundary conditions and a scalar field on the double interval
with periodic boundary conditions in the direction $x^3$, together with the
independence of the results on boundary conditions in the large volume limit,
this can also be obtained by the replacement $L_3= 2 a$, where $a$
is the distance between the two parallel plates. In the case where the $x^3$
dimension is ``small'', the former argument no longer applies but the latter
still does.

\subsection{Functional approach and zeta function}
\label{sec:funct-appr-zeta}

The functional evaluation of partition functions \cite{Hawking:1976ja} (see also
e.g.~\cite{DeWitt:1975ys,Dowker:1975tf} for earlier connected work,
\cite{Kapusta:1981aa} for the inclusion of chemical potentials and
\cite{doi:10.1142/2065} for a review) starts from the Hamiltonian path integral
representation
\begin{equation}
  \label{eq:35}
  Z_d(\beta,\mu^j)=\int \prod_{x^a} \text{d}\phi(x^a) \prod_{x^a}
  \frac{\text{d}\pi(x^a)}{2\pi}\ e^{-S^E_H},
\end{equation}
where $x^a=(x^1,\dots,x^{d+1})$ and the first order Euclidean action is
\begin{equation}
  \label{eq:36}
  S^E_H[\phi,\pi]=\int^{\beta}_0\D x^{d+1}\big[\int_{V_d}\D^dx
  (-i\pi\partial_{d+1}\phi) +(H-i\mu^jP_j)\big]. 
\end{equation}
The sum is over periodic phase space paths of period $\beta=L_{d+1}$ in
Euclidean time $x^{d+1}$,
\begin{equation}
  \phi(x^i,x^{d+1}+\beta)=\phi(x^i,x^{d+1}),\quad \pi(x^i,x^{d+1}+\beta)=\pi(x^i,x^{d+1})
  \label{eq:20}. 
\end{equation}
After integration over the momenta, this leads to
\begin{equation}
  \label{eq:37}
  Z_d(\beta,\mu^j)=\big({\rm det}[2\pi \delta^{d+1}(x',x)]\big)^{-\frac{1}{2}}
  \int \prod_{x^a} d\phi(x^a)\   e^{-S^E_L}
\end{equation}
with
\begin{equation}
  \label{eq:38}
  S^E_L[\phi]=\frac{1}{2}\int_{V_{d+1}}\D^{d+1} x 
   [(\partial_{d+1}\phi-\mu^j\partial_j
  \phi)^2+\partial_j\phi\partial^j\phi],
\end{equation}
where the integration is now over a $(d+1)$-dimensional hyperrectangle in
Euclidean space of volume $V_{d+1}=V_d\beta=\prod^{d+1}_{a=1}L_a$. Except for the
replacement $\partial_{d+1}\to \partial_{d+1}-\mu^j\partial_j$, the operator in
the action is minus the Laplacian in $d+1$ dimensions with periodic boundary
conditions in all dimensions. Since the eigenfunctions are
$\frac{1}{\sqrt{V_{d+1}}}e^{ik_a x^a}$ with $k_{d+1}=\frac{2\pi n_{d+1}}{\beta}$, the
eigenvalues are
\begin{equation}
  \label{eq:39}
  \lambda_{n_a}=  \Big[\Big(\frac{2\pi n_{d+1}}{\beta}-\sum_j
  \mu^j\frac{2\pi n_j}{L_j}\Big)^2
  +\sum_i\Big(\frac{2\pi n_i}{L_i}\Big)^2\Big]. 
\end{equation}
In order to define the partition function, formally given by 
\begin{equation}
  \label{eq:153}
  Z_{d}(\beta,\mu^j)=\sideset{}{'}\prod_{n_a\in \mathbb Z^{d+1}}
  \sqrt{\frac{2\pi}{\lambda_{n_a}}}=\Big[{\rm det}
  (\frac{-\Delta_\mu}{2\pi})\Big]^{-\frac 12}, 
\end{equation}
one introduces a parameter $\nu$ of dimension inverse length and
considers the dimensionless operators $A_\mu=-\Delta_\mu/2\pi\nu^2$. The
associated zeta function is defined by
\begin{equation}
  \zeta_{A_\mu}(s)=\sideset{}{'}\sum_{n_a\in\mathbb Z^{d+1}}\lambda_{n_a}^{-s}(2\pi\nu^2)^s
  \label{eq:42}. 
\end{equation}
In these terms, the partition function is given by
\begin{equation}
  \label{eq:22}
  \ln Z_d(\beta,\mu^j)=\frac 12 \zeta'_{A_\mu}(0)
  =\frac{1}{2}\zeta'_{-\Delta_\mu}(0)+\frac{1}{2}\ln
  (2\pi\nu^2)\zeta_{-\Delta_\mu}(0).  
\end{equation}

\subsection{Functional approach to partition function in large volume limit}
\label{sec:funct-appr-large}

Let us recover the canonical result of section \ref{sec:canon-appr-large}
through the functional approach. In this case, in the absence of chemical
potentials, $\mu^j=0$, $A_0=-\frac{\Delta_0}{2\pi\nu^2}$ we again turn the
sums over the $n_i$ into integrals,
\begin{equation}
  \label{eq:40}
  \zeta_{A_0}(s)=(2\pi\nu^2)^s\frac{V_d} {(2\pi)^d}\int\D^d k\
  \sum_{n_{d+1}\in\mathbb Z}[(\frac{2\pi
    n_{d+1}}{\beta})^2+k_i k^i]^{-s}, 
\end{equation}
From the Euclidean spacetime point of view, there is thus only one small dimension
$x^{d+1}$ of length $\beta$. After performing the integrals over the angles in
hyperspherical coordinates, we have
\begin{equation}
  \label{eq:23}
  \zeta_{A_0}(s)=(2\pi\nu^2)^s\frac{V_d2\pi^{\frac{d}{2}}}{(2\pi)^d
    \Gamma(\frac{d}{2})}\int^\infty_0\D k\
  k^{d-1}\sum_{n_{d+1}\in\mathbb Z}[(\frac{2\pi n_{d+1}}{\beta})^2+k^2]^{-s}
\end{equation}
The divergent integral contained in the term at $n_{d+1}=0$ is regulated through
an infrared cut-off $\epsilon$,
\begin{equation}
  \label{eq:24}
  \int^\infty_\epsilon\D k\ k^{d-1-2s}=-\frac{\epsilon^{d-2s}}{d-2s},\quad \mathfrak{Re}(s)>\frac{d}{2}.
\end{equation}
This expression and its derivative with respect to $s$ both vanish at
$s=0$ in the limit when $\epsilon\to 0$ and can thus be discarded. In
the remaining terms, the integrals converge for
$\mathfrak{Re}(s)>\frac{d}{2}$. After the change of variables,
$k=\frac{2\pi}{\beta}y$, we are left with
\begin{equation}
  \label{eq:25}
  \zeta_{A_0}(s)=(2\pi\nu^2)^s\frac{V_d 2
    \pi^{\frac{d}{2}}}{(2\pi)^d\Gamma(\frac{d}{2})}(\frac{2\pi}{\beta})^{d-2s}\int^\infty_0\D y\
  y^{d-1}\sideset{}{'}\sum_{n_{d+1}\in\mathbb Z}[(n_{d+1})^2+y^2]^{-s}.
\end{equation}
The additional change of variables $y=n_{d+1}\sinh x$ together with
\begin{equation}
\int^\infty_0 \text{d}x
\frac{\sinh^{d-1}(x)}{\cosh^{2s-1}(x)}=\frac{1}{2}
\frac{\Gamma(\frac{d}{2})\Gamma(s-\frac{d}{2})}{\Gamma(s)},
\label{eq:26}
\end{equation}
then yields
\begin{equation}
  \label{eq:27}
  \zeta_{A_0}(s)=(2\pi\nu^2)^s\frac{V_d 2
    \pi^{\frac{d}{2}}}{(2\pi)^d}(\frac{2\pi}{\beta})^{d-2s}
  \frac{\Gamma(s-\frac{d}{2})\zeta(2s-d)}{\Gamma(s)}, 
\end{equation}
which can be written more compactly as
\begin{equation}
  \label{eq:154}
  \zeta_{A_0}(s)=\frac{\nu^{2s}V_d}{\beta^{d-2s}} 2^{1-s}
  \frac{\xi(2s-d)}{\Gamma(s)}. 
\end{equation}
Since $\frac{1}{\Gamma(s)}=s+{O}(s^2)$, it follows that
$\zeta_{A_0}(0)=0$ and
\begin{equation}
  \label{eq:155}
\ln Z_d(\beta)=\xi(-d)
    \frac{V_d}{\beta^d}.
\end{equation}
When using the reflection formula
\begin{equation}
  \label{eq:28}
  \pi^{-\frac z2}\Gamma(\frac{z}{2})\zeta(z)=\pi^{\frac{z-1}{2}}
  \Gamma(\frac{1-z}{2})\zeta(1-z) \iff \xi(z)=\xi(1-z). 
\end{equation}
at $z=-d$, this agrees with \eqref{eq:18}.

\subsection{Schwinger proper time, heat kernel and worldline approach}
\label{sec:world-line-approach}

The quadratic Lagrangian path integral \eqref{eq:37} is formally solved by the
one-loop contribution
\begin{equation}
  \label{eq:69}
  \ln Z_d(\beta,\mu^j)=-\frac{1}{2}{\rm Tr}\ \ln \frac{\delta^2 S^E_L}{\delta \phi\delta\phi},
\end{equation}
which can be written as an integral over Schwinger proper time $t$ as
\begin{equation}
  \label{eq:111}
  \ln Z_d(\beta,\mu^j)=\frac{1}{2}\int^\infty_0
  \frac{\text{d}t}{t}\, {\rm Tr}
  \ e^{-t \hat {\rm H}_{\mu}},
\end{equation}
after having removed the zero
modes. Here
\begin{equation}
  \label{eq:112}
  \hat {\rm H}_{\mu}=(\hat p_{d+1} -\mu^j\hat p_j)^2+\hat p_j\hat p^j,
\end{equation}
is not the field theory Hamiltonian but the one of a free particle of mass
$\frac{1}{2}$ in $d+1$ (Euclidean) dimensions described by the canonical
variables $(x^a,p_a)$. In terms of the operator
\begin{equation}
  \label{eq:113}
  \hat T_\mu(t)=e^{-t \hat {\rm H}_{\mu}},
  \quad -\frac{\text{d}}{\text{d}t}
  \hat T_\mu(t)
  =\hat {\rm H}_{\mu}\hat T_\mu(t), 
\end{equation}
the relation to the zeta function is
\begin{equation}
  \label{eq:117}
  \zeta_{\hat {\rm H}_{\mu}}(s)=\frac{1}{\Gamma(s)}
  \int_0^\infty \text{d}t\ t^{s-1}
  {\rm Tr}\ (\hat T_\mu(t)-\hat \Pi),  
\end{equation}
where $\hat \Pi$ denotes the orthogonal projector onto ${\rm Ker}\ \hat
{\rm H}_{\mu}$, ${\rm Tr}\,\hat \Pi={\rm dim}\, {\rm Ker}\, \hat
{\rm H}_{\mu}$.

{\bf Remarks:}

(i) If one takes vanishing chemical potentials $\mu^j=0$ and considers
infinite Euclidean spacetime, which means in particular zero temperature,
$\beta=L_{d+1}\to \infty$, the matrix element
\begin{equation}
  \label{eq:114}
  K_{-\Delta_0}(x',x;t)=\langle  x' |\hat T(t)
  |x \rangle
\end{equation}
is called the heat kernel since it satisfies the heat equation
\begin{equation}
  \label{eq:115}
  \frac{\text{d}}{\text{d}t}K_{-\Delta_0}(x',x;t)=\Delta_0
  K_{-\Delta_0}(x',x;t),
  \quad K_{-\Delta_0}(x',x;0)=\delta^{d+1}(x'-x). 
\end{equation}
The solution is well-known and can be obtained for instance by working out the
(Euclidean) path integral for the transition amplitude of a particle of mass
$m=\frac{1}{2}$ in $d+1$ dimensions:
\begin{equation}
  \label{eq:116}
  K_{-\Delta_0}(x',x;t)=\frac{1}{(4\pi t)^{\frac{d+1}{2}}}
  e^{-\frac{1}{4t}(x'_a-x_a)(x^{\prime a}- x^a)}. 
\end{equation}

(ii) If one takes vanishing chemical potentials $\mu^j$ and only one spacetime
dimension with finite $\beta$ instead, the periodic boundary condition,
$\phi(0)=\phi(\beta)$, requires one to solve the heat kernel for a particle on a
circle. The first form of the solution may be obtained by inserting a resolution
of the identity in terms of eigenstates of the Laplacian.
\begin{equation}
  \label{eq:178}
  \varphi_n(x)=\langle x|n \rangle=\frac{1}{\sqrt \beta}e^{i\omega_n x},
  \quad \omega_n=\frac{2\pi}{\beta}n, 
\end{equation}
so that
\begin{equation}
  \label{eq:179}
  K_{-\Delta_P}(x',x; t)=\sum_{n\in\mathbb Z}
  e^{- t \omega^2_n}\varphi_n(x')\varphi_n^*(x)
\end{equation}
and may be expressed in
terms of the Jacobi theta function defined by
\begin{equation}
  \label{eq:147}
  \vartheta_3(z|\tau)=\sum_{n\in\mathbb Z} e^{\pi i n^2\tau+2\pi i nz}, 
\end{equation}
as follows 
\begin{equation}
  \label{eq:118}
  K_{-\Delta_P}(x',x;t)=\frac{1}{\beta}\sum_{n\in\mathbb Z}
  e^{-t(\frac{2\pi n}{\beta})^2+ 2\pi i n\frac{x'-x}{\beta}}=\frac{1}{\beta}
  \vartheta_3(\frac{x'-x}{\beta}|i\frac{4\pi}{\beta^2}t).  
\end{equation}
The second form of the solution follows by considering a path integral including
paths that wind $m$ times around the circle of length $\beta$, 
\begin{equation}
  \label{eq:148}
  K_{-\Delta_P}(x',x;t)=\frac{1}{\sqrt{4\pi t}}\sum_{m\in\mathbb Z}
  e^{-\frac{1}{4t}(x'-x+\beta m)^2}.
\end{equation}
Both expressions are related through the Poisson summation formula, 
\begin{multline}
  \label{W11}
  \sum_{m^\alpha\in\mathbb{Z}^n}e^{-\pi
    (m^{\alpha}+y^{\alpha})A_{\alpha\beta}(m^{\beta}+y^{\beta})
    +2\pi i m^{\alpha}z_{\alpha}}
  \\=\frac{1}{\sqrt{\mathrm{det}\, A}}
  \sum_{n_\alpha\in\mathbb{Z}^n}e^{-\pi (n_{\alpha}+z_{\alpha})
    (A^{-1})^{\alpha\beta}(n_{\beta}+z_{\beta})
    -2\pi i (n_{\alpha}+z_{\alpha})y^{\alpha}},
\end{multline}
in the simple case where $n=1$. For later use, one also defines
\begin{equation}
  \label{eq:156}
  \theta(t)=\vartheta_3(0|it)=\sum_{n\in\mathbb Z}e^{-t\pi n^2},
\end{equation}
with modular transformation
\begin{equation}
  \label{eq:164}
  \vartheta_3(0|-\frac 1\tau)=\sqrt{-i\tau}\vartheta_3(0,\tau)
  \Longrightarrow\theta(\frac 1t)=\sqrt t\theta(t),
\end{equation}
in terms of which both expressions for the trace of the heat kernel,
\begin{equation}
  \label{eq:166}
  K_{-\Delta_P}(t)=\int^\beta_0\text{d}x\, K_{-\Delta_P}(x,x;t)
\end{equation}
of a particle on the circle may be written compactly as
\begin{equation}
  \label{eq:174}
  K_{-\Delta_P}(t)=\theta(\frac{4\pi}{\beta^2} t)
  =\frac{\beta}{\sqrt{4\pi t}}\theta(\frac{\beta^2}{4\pi t}). 
\end{equation}

\subsection{Worldline approach to partition function in the large volume limit}
\label{sec:world-line-approach-1}

The full heat kernel in the large volume limit is obtained by combining the
heat kernel in \eqref{eq:116} with $d\to d-1$ for the large spatial dimensions
with that in \eqref{eq:118},
\begin{equation}
  \label{eq:149}
  \langle x'|\hat T_0(t)|x\rangle=
  \frac{1}{\beta(4\pi t)^{\frac{d}{2}}}
  e^{-\frac{1}{4t}(x'_i-x_i)(x^{\prime i}- x^i)}
  \vartheta_3(\frac{x'_{d+1}-x_{d+1}}{\beta};i\frac{4\pi}{\beta^2}t). 
\end{equation}
According to \eqref{eq:111}, the partition function in the large volume limit is 
\begin{equation}
  \label{eq:119}
  \ln Z_d(\beta)= \frac{V_d}{2^{d+1}\pi^{\frac{d}{2}}}
  \sideset{}{'}\sum_{n_{d+1}\in\mathbb Z}\int^\infty_0 \text{d}t\ t^{-\frac{d}{2}-1}
     e^{-t (\frac{2\pi n_{d+1}}{\beta})^2}.
\end{equation}
After the change of variables $t\to (\frac{2\pi}{\beta})^2t$, this becomes
\begin{equation}
  \label{eq:30}
  \ln Z_d(\beta)= \frac{V_d}{\beta^d}\frac{\pi^{\frac d2}}{2}
  \sideset{}{'}\sum_{n_{d+1}\in\mathbb Z}\int^\infty_0 \text{d}{t}\
  {t}^{-\frac{d}{2}-1}
  e^{- tn_{d+1}^2}, 
\end{equation}
which may be written in terms of the theta function as
\begin{equation}
  \label{eq:165}
  \ln Z_d(\beta)= \frac{V_d}{\beta^d}
  \int^\infty_0 \text{d}{t}\ {t}^{-\frac{d}{2}-1}\frac 12 (\theta(t)-1). 
\end{equation}
Going back to \eqref{eq:30} and assuming that $d<-2$, one may use 
\begin{equation}
    \label{eq:125}
    \frac{1}{\Gamma(s)}\int^\infty_0 \text{d}{t}\ {t}^{s-1} e^{-{t} \lambda}
    =\frac{1}{\lambda^s},\quad \mathfrak{Re}(s)>0,
\end{equation}
with $\lambda>0$, so that
\begin{equation}
    \label{eq:120}
    \frac{\pi^{\frac d2}}{2}
    \sideset{}{'}\sum_{n_{d+1}\in \mathbb Z}\int^\infty_0 \text{d}{t}\ {t}^{-\frac{d}{2}-1}
    e^{-{t} n_{d+1}^2}
    =\pi^{\frac d2}\Gamma(-\frac{d}{2})\zeta(-d)=\xi(-d). 
\end{equation}
It follows that \eqref{eq:30} agrees with \eqref{eq:155}. Again, the result for
positive integer $d$ is obtained after using the reflection formula.
     
When using \eqref{eq:148} instead of \eqref{eq:118}, the full heat kernel in the
large volume limit is instead
\begin{equation}
  \label{eq:159}
  \langle x'|\hat T_0(t)|x\rangle=\frac{1}{(4\pi t)^{\frac{d+1}{2}}}
  e^{-\frac{1}{4t}(x'_i-x_i)(x^{\prime i}- x^i)}
  \sum_{m^{d+1}\in \mathbb Z}e^{-\frac{1}{4t}(x'-x +\beta m^{d+1})^2}. 
\end{equation}
For the partition function, after dropping the $m^{d+1}=0$ term, this gives
\begin{equation}
  \label{eq:150}
  \ln Z_d(\beta)=\frac{V_d\beta}{2^{d+2}\pi^{\frac{d+1}{2}}}
    \sideset{}{'}\sum_{m^{d+1}\in\mathbb Z}\int^\infty_0 \text{d}t\ t^{-\frac{d+3}{2}}
    e^{-\frac{1}{4t}(m^{d+1}\beta)^2}.
\end{equation}
After the change of variables $t\rightarrow\frac{\beta^2}{4t}$, 
one now gets directly
\begin{equation}
  \label{eq:160}
  \ln Z_d(\beta)=\frac{V_d}{\beta^d}\frac{1}{2\pi^{\frac{d+1}{2}}}
  \sideset{}{'}\sum_{m^{d+1}\in \mathbb Z}\int^\infty_0 \text{d}{t}\ {t}^{\frac{d+1}{2}-1}
  e^{-t(m^{d+1})^2}=\frac{V_d}{\beta^d}\xi(d+1),
\end{equation}
in agreement with the result obtained from canonical quantization in
\eqref{eq:18}.

In terms of the theta function, the first expression for the partition function in
\eqref{eq:160} is
\begin{equation}
  \label{eq:157}
  \boxed{\ln Z_d(\beta)= \frac{V_d}{\beta^d}\int^\infty_0\text{d}t\,
    t^{\frac {d+1}{2}-1}\frac 12 \big(\theta(t)-1\big)}. 
\end{equation}
In other words, equivalence of the canonical and worldline approach for the
partition function shows for $z=d+1$ that the completion of the zeta
function is the Mellin transform of $\frac 12 (\theta(t)-1)$,
\begin{equation}
  \label{eq:158}
  \xi(z)=\frac 12\int^\infty_0\text{d}t\, t^{\frac z2 -1} \big(\theta(t)-1\big). 
\end{equation}

\subsection{Casimir energy on $\mathbb R^{d-1}\times \mathbb
  S^1$}
\label{sec:canon-appr-casim}

Suppose now that the spatial dimension $x^d$ is small, while all other spatial
dimensions $x^I$, for $I=1,\dots,d-1$ are large, $L_I\gg L_d$. In other words,
one considers a slab geometry with two infinite parallel hyperplanes separated
by a distance $L_d$, or because of the periodic boundary conditions, a scalar
field on the spatial manifold $\mathbb R^{d-1}\times \mathbb S^1_{L_d}$. Suppose
also that one uses symmetric instead of normal ordering, so that
\begin{equation}
  \label{eq:29}
  \hat H=\frac{1}{2}\sideset{}{'}\sum_{n_i\in\mathbb{Z}^d}\omega_{k_i}(\hat
  a^{\dagger}_{k_i}\hat a_{k_i}+\hat a_{k_i}\hat a^{\dagger}_{k_i})
  =\sideset{}{'}\sum_{n_i\in\mathbb{Z}^d}\omega_{k_i}(\hat
  a^{\dagger}_{k_i}\hat a_{k_i}+\frac{1}{2}).
\end{equation}
The divergent zero-point energy is 
\begin{equation}
  \label{eq:32}
  E^d_0=\langle 0 |\hat{H}|0\rangle
  =\frac{1}{2}\sideset{}{'}\sum_{n_i\in\mathbb{Z}^d}\omega_{k_i}.
\end{equation}
Transforming $d-1$ sums into integrals, and using $\zeta$ function
regularization to give a meaning to this divergent quantity, we need to evaluate
\begin{equation}
  \label{eq:34}
  E^d_0(s)= \frac{1}{2}\nu^{2s}\frac{V_{d-1}}{(2\pi)^{d-1}}\sum_{n_d\in \mathbb Z}
  \int{\D^{d-1}k}\ [(\frac{2\pi
    n_d}{L_d})^2+k_Ik^I]^{-s+\frac{1}{2}},
\end{equation}
where $V_{d-1}=\prod_{I=1}^{d-1}L_I$. This is the same computation as in
\eqref{eq:40}, up an overall factor of $(2\pi\nu^2)^s$ versus
$\frac{1}{2}\nu^{2s}$ and the replacements $d\to d-1$, $s\to s-\frac{1}{2}$,
$\beta=L_{d+1}\to L_d$. The result can thus be read off \eqref{eq:27},
\begin{equation}
  \label{eq:41}
  E^d_0(s)=\frac{1}{2}\nu^{2s}\frac{V_{d-1}}{L_d^{d-2s}}2^{2-2s}\pi^{\frac{d+1}{2}-2s}
    \frac{\Gamma(s-\frac d2)\zeta(2s-d)}{\Gamma(s-\frac{1}{2})}, 
\end{equation}
which yields at $s=0$,
\begin{equation}
  \label{eq:122}
  E_0^d(0)=-\xi(-d)\frac{V_{d-1}}{L_d^{d}},
\end{equation}
and thus, after using the reflection formula, 
\begin{equation}
    \label{eq:43}
    \boxed{E^d_0(0)=-
      \xi(d+1)\frac{V_{d-1}}{L_d^{d}}}=
    \left\{\begin{array}{ll}
             -\frac{\pi}{6}\frac{1}{L_1}\
             & {\rm for}\ d=1\\
             -\frac{\zeta(3)}{2\pi}\frac{L_1}{L_2^2} \
             & {\rm for}\ d=2\\
             -\frac{\pi^2}{90}\frac{L_1L_2}{L_3^3} \
             & {\rm for}\ d=3\\
             & \vdots
           \end{array}
           \right..  
\end{equation}

{\bf Remarks:}

(i) For later purpose and to make contact with the literature, one may write
\begin{equation}
  \label{eq:133}
  E_0^d=-\varepsilon^d_{\rm vac}\frac{V_{d-1}}{L_d^{d}},
\end{equation}
where $\varepsilon^d_{\rm vac}$ is a dimensionless number characterizing the
Casimir energy whose value is
\begin{equation}
  \label{eq:134}
  \varepsilon^d_{\rm vac}=\xi(d+1)
\end{equation}
in the current model. 

The relation between the partition function \eqref{eq:18} in the large volume
limit and the Casimir energy density (see
e.g.~\cite{BALIAN1978165,Ambjorn:1981xw,Lutken:1988ge,Ford:1988gt,%
  PhysRevD.40.4191,Wotzasek:1989gq}) may then be written as
\begin{equation}
  \label{eq:82}
  \ln Z_d(\beta)=\varepsilon^d_{\rm vac}\frac{V_d}{\beta^d}. 
\end{equation}

(ii) On account of the previously discussed equivalence, the correct
electromagnetic results with perfectly conducting plates at $x^3=0$ and
$x^3=a$ are again obtained through the replacement $L_3=2a$.

(iii) Note that, even with symmetric ordering,
\begin{equation}
  \label{eq:68}
  P^0_i=\langle 0|\hat P_i|0 \rangle=0, 
\end{equation}
irrespective of the existence or not of a small spatial dimension, because
either $\sum_{n_i\in \mathbb Z}k_i=0$ or $\int \text{d}k_i k_i=0$.

(iv) The computation of the Casimir energy can also be done using a suitable
integral representation. Starting from the representation \eqref{eq:125} at
$s=-\frac 12$ (to be understood in the sense of analytic continuation), the zero
point energy in \eqref{eq:32} is written as
\begin{equation}
  \label{eq:31}
  E^d_0=\frac{1}{2\Gamma(-\frac 12)}\sideset{}{'}\sum_{n_i\in\mathbb{Z}^d}\int^\infty_0 \text{d}t\,
  t^{-\frac 32}e^{-t\omega^2_{k_i}}
\end{equation}
After changing sums to integrals for the $d-1$ large dimensions and integrating
over the angles, one gets
\begin{equation}
  \label{eq:151}
  E^d_0=-\frac{1}{4\sqrt \pi}\frac{V_{d-1}}{(2\pi)^{d-1}}\frac{2\pi^{\frac{d-1}{2}}}{\Gamma(\frac{d-1}{2})}
  \sum_{n_d\in\mathbb Z}\int^\infty_0	\text{d}t\, t^{-\frac 32}
    \int^\infty_0\text{d}k k^{d-2} e^{-t\big[(\frac{2\pi}{L_d}n_d)^2+k^2\big]}.
\end{equation}
The change of variables $k=\frac{2\pi}{L_d}\sqrt y$, $t\to t=(\frac{L_d}{2\pi})^2t$ then
leads to
\begin{equation}
  \label{eq:152}
  E^d_0=-\frac{V_{d-1}}{L_d^d}\frac{\pi^{\frac d2}}{2\Gamma(\frac{d-1}{2})}
   \sum_{n_d\in\mathbb Z}\int^\infty_0\text{d}t\, t^{-\frac 32}e^{-tn_d^2}\int^\infty_0\text{d}y y^{\frac{d-1}{2}-1}e^{-ty}. 
\end{equation}
Performing the integral over $y$ using \eqref{eq:125} again gives
\begin{equation}
  \label{eq:161}
  E^d_0=-\frac{V_{d-1}}{L_d^d}\frac{\pi^{\frac d2}}{2}
  \sum_{n_d\in\mathbb Z}\int^\infty_0\text{d}t\, t^{-\frac d2 -1}e^{-tn_d^2}
\end{equation}
The integral term with $n_d=0$ is divergent, $
\int^\infty_\epsilon \text{d}t\, t^{-\frac d2 -1}=\frac{2\epsilon^{-\frac
    d2}}{d}$ and neglected, while the terms with $n_d\neq 0$ are evaluated using
\eqref{eq:125} to yield
\begin{equation}
  \label{eq:162}
  E^d_0=-\frac{V_{d-1}}{L_d^d}{\pi^{\frac d2}}\Gamma(-\frac d2)\zeta(-d), 
\end{equation}
in agreement with \eqref{eq:122}.

\section{Two small spacetime dimensions and Eisenstein series}
\label{sec:masl-scal-part}

\subsection{Functional approach}
\label{sec:funct-appr-extend}

We now consider the case where only the first $d-1$ spatial dimensions are
large, while $L_d$ is small, and where a chemical potential for linear momentum
in the small dimension $x^d$ is turned on. We start here with the functional
approach because it yields directly the result in terms of the Eisenstein series
in \eqref{eq:1}. 
In this case $I=1,\dots, d-1$,
$V_{d-1}=\prod_{I=1}^{d-1}L_I$ and, when taking into account \eqref{eq:39}, we
get instead of \eqref{eq:40},
\begin{equation}
  \label{eq:33}
  \zeta_{-\Delta_\mu}(s)=\frac{V_{d-1}}{(2\pi)^{d-1}}\int \D^{d-1} k\
  \sideset{}{'}\sum_{(n_{d+1},n_d)\in\mathbb{Z}^2}[(\frac{2\pi n_{d+1}}{\beta}
  -\mu\frac{2\pi n_d}{L_d})^2
  +(\frac{2\pi n_{d}}{L_d})^2+k_I k^I]^{-s}. 
\end{equation}
The computation then proceeds as in section \ref{sec:funct-appr-large}, up to
the replacements $d\to d-1$ and of the zeta function by a suitable Eisenstein
series. More precisely, instead of \eqref{eq:25}, we now get
\begin{multline}
  \label{eq:44}
  \zeta_{-\Delta_\mu}(s)=\frac{V_{d-1}}{(2\pi)^{d-1}}
  \frac{2\pi^{\frac{d-1}{2}}}{\Gamma(\frac{d-1}{2})}
  (\frac{2\pi}{\beta})^{d-1-2s}\\
  \sideset{}{'}\sum_{(n_{d+1},n_d)\in\mathbb{Z}^2}\int_0^{\infty}\D y\ y^{d-2}
  [(n_{d+1}-n_d\frac{\beta\mu}{L_d})^2
  +n_d^2(\frac{\beta}{L_d})^2+y^2]^{-s},
\end{multline}
For later convenience, we also change parametrization and use $\alpha=\beta\mu$
so that we are now computing
\begin{equation}
  \label{eq:45}
  {\rm Z}_{d,1}(\beta,\alpha)={\rm Tr}\ e^{-\beta \hat H+i\alpha \hat P_d}. 
\end{equation}
The additional subscript $1$ in ${\rm Z}_{d,1}$ refers to the number of small
spatial dimensions. In section \ref{sec:high-dimens-tori} below, we will consider $q$ small spatial
dimensions and compute ${\rm Z}_{d,q}$. 
Introducing the modular parameter
\begin{equation}
  \label{eq:46}
  \tau=\frac{\alpha+i\beta}{L_d},
\end{equation}
the sum simplifies to
\begin{equation}
  \label{eq:47}
  \sideset{}{'}\sum_{(n_{d+1},n_d)\in\mathbb{Z}^2}\int_0^{\infty}\D y\ y^{d-2}
  [\abs{n_{d+1}+n_d\tau}^2+y^2]^{-s}. 
\end{equation}
The integral is convergent for $\mathfrak{Re}(s)>\frac{d-1}{2}$ and
can be performed through the change of variables
$y=\abs{n_{d+1}+n_d\tau}^2\sinh x$ which gives
\begin{equation}
  \label{eq:48}
  \zeta_{-\Delta_\alpha}(s)=\frac{V_{d-1}}{(2\pi)^{d-1}}
  \frac{\pi^{\frac{d-1}{2}}\Gamma(s-\frac{d-1}{2})}{\Gamma(s)}
  (\frac{2\pi}{\beta})^{d-1-2s}
  \sideset{}{'}\sum_{(n_{d+1},n_d)\in\mathbb{Z}^2}\abs{n_{d+1}+n_d\tau}^{d-1-2s}.  
\end{equation}
In terms of the real analytic Eisenstein series \eqref{eq:1}, this can be
written as 
\begin{equation}
  \label{eq:49}
  \zeta_{-\Delta_\alpha}(s)=\frac{V_{d-1}}{(2\pi)^{d-1}}\frac{\pi^{\frac{d-1}{2}}
    \Gamma(s-\frac{d-1}{2})}{\Gamma(s)\tau_2^{s-\frac{d-1}{2}}}
  (\frac{2\pi}{\beta})^{d-1-2s}f_{s-\frac{d-1}{2}}(\tau,\bar\tau),
\end{equation}
When using the functional relation for the analytically continued Eisenstein
series, 
\begin{equation}
\pi^{-z}\Gamma(z)f_z(\tau,\bar\tau)=\pi^{z-1}\Gamma(1-z)f_{1-z}(\tau,\bar\tau).\label{eq:50}
\end{equation}
for $z=s-\frac{d-1}{2}$, and also $\tau_2=\frac{\beta}{L_d}$, we
end up with
\begin{equation}
  \zeta_{-\Delta_{\alpha}}(s)=\frac{\Gamma(\frac{d+1}{2}-s)}{2^{2s}\pi^{\frac{d+1}{2}}\Gamma(s)}
  \frac{V_{d-1}}{L_d^{d-1-2s}}
  \frac{f_{\frac{d+1}{2}-s}(\tau,\bar\tau)}{\tau_2^{\frac{d-1}{2}-s}}.\label{eq:51}
\end{equation}
Again, since $\Gamma(s)\approx \frac{1}{s}+{O}(s^0)$,
$\zeta_{-\Delta_{\alpha}}(0)$, if $\mathcal Z_{d,1}(\tau,\bar\tau)={\rm Z}_{d,1}(\beta,{\alpha})$, the
partition function is
\begin{equation}
  \ln{\mathcal Z}_{d,1}(\tau,\bar\tau)=
  \frac{\Gamma(\frac{d+1}{2})V_{d-1}}{2\pi^{\frac{d+1}{2}}L_d^{d-1}\tau_2^{\frac{d-1}{2}}}
  f_{\frac{d+1}{2}}(\tau,\bar\tau).
  \label{eq:52}
\end{equation}
In terms of the completion of the real analytic Eisenstein series, 
\begin{equation}
\label{eq:52.1}
\xi(z;\tau,\bar{\tau})=\frac{\Gamma(\frac{z}{2})f_{\frac{z}{2}}(\tau,\bar{\tau})}{2\pi^{\frac{z}{2}}},
\end{equation}
which satisfies the reflection formula
\begin{equation}
  \label{eq:138}
  \xi(z;\tau,\bar{\tau})=\xi(2-z;\tau,\bar{\tau}),
\end{equation}
this gives
\begin{equation}
\label{eq:52.2}
\boxed{\ln{\mathcal Z}_{d,1}(\tau,\bar\tau)=
  \frac{V_{d-1}}{L_d^{d-1}\tau_2^{\frac{d-1}{2}}}
  \xi(d+1;\tau,\bar{\tau})}.
\end{equation}
Since $\xi(z;\tau,\bar\tau)$ is invariant under modular transformations
\eqref{eq:2} while $\tau_2$ transforms as
\begin{equation}
\tau_2\rightarrow\frac{\tau_2}{\abs{c\tau+d}^2},\label{eq:54}
\end{equation}
the partition function \eqref{eq:52.1} transforms as
 \begin{equation}
   \boxed{\ln{\mathcal Z}_{d,1}(\tau',\bar\tau')=\abs{c\tau+d}^{d-1}
     \ln{\mathcal Z}_{d,1}(\tau,\bar \tau)}.\label{eq:55}
\end{equation}

{\bf Remarks:}

(i) The above result holds for $d>1$ because if $d=1$, i.e., for the free boson
on $\mathbb T^2$, $\ln {\mathcal
  Z}_{1,1}(\tau,\bar\tau)=\frac{1}{2\pi}f_{1}(\tau,\bar\tau)$, is not convergent
and more care is needed to derive the correct result. This is reviewed in
section \ref{sec:torus-part-funct}.

(ii) As discussed in the introduction, for $d=3$, the result is directly
relevant for the electromagnetic field with Casimir boundary conditions after
the replacement $L_3= 2 a$.

(iii) If the chemical potential vanishes, ${\alpha}=0$, $\tau=i\tau_2$, with
$\tau_2=\frac{\beta}{L_d}$ and $\mathcal Z_{d,1}(i\tau_2,-i\tau_2)=\mathcal
Z_{d,1}(\tau_2)= Z_{d,1}(\beta)$, the result can be written in terms of an Epstein zeta
function,
\begin{equation}
  \label{eq:59}
  \zeta(s;1,\tau_2^2)=\sideset{}{'}\sum_{(m^{d+1},m^d)\in\mathbb{Z}^2}
  \frac{1}{[(m^d)^2+(m^{d+1})^2\tau_2^2]^s},
\end{equation}
as 
\begin{equation}
  \label{eq:58}
  \ln{\mathcal Z}_{d,1}(\tau_2)=\frac{\Gamma(\frac{d+1}{2})}{2\pi^{\frac{d+1}{2}}}
  \frac{V_{d-1}\tau_2}{L_d^{d-1}}\zeta(\frac{d+1}{2};1,\tau_2^2).
\end{equation}
If $a=0=d$, $b=1=-c$, the modular transformation reduces to temperature
inversion $\tau_2\to \frac{1}{\tau_2}$ with
\begin{equation}
  \label{eq:60}
  \ln {\mathcal Z}_{d,1}(\frac{1}{\tau_2})=\tau_2^{d-1}\ln {\mathcal Z}_{d,1}(\tau_2). 
\end{equation}
Some properties and a trigonometric expansion for the Epstein zeta function are
reviewed in Appendix \ref{AppB}.

(iv) Setting $m^d=0$ in \eqref{eq:58} gives
\begin{equation}
  \ln {\mathcal Z}^{\rm high}_{d,1}(\tau_2)= \xi(d+1)\frac{V_{d-1}}{L_d^{d-1} \tau_2^d},\label{eq:61}
\end{equation}
which is the black body result \eqref{eq:18}. From the analysis of section
\ref{sec:massl-scal-part}, we know that this is the dominant contribution in the
high temperature/large $L_d$ limit $\tau_2\ll 1$. This justifies a posteriori
the use of a normal ordered Hamiltonian in section \ref{sec:canon-appr-large}.

(v) From equation \eqref{eq:60}, it follows that in the low temperature/small
$L_d$ limit $\tau_2\gg 1$, the dominant contribution is given by
\begin{equation}
  \label{eq:62}
  \ln {\mathcal Z}^{\rm low}_{d,1}(\tau_2)=\xi(d+1)\frac{V_{d-1}\tau_2}{L_d^{d-1}}.
\end{equation}
The result for the Casimir energy with one small spatial dimension in
\eqref{eq:43}, then follows by taking $-\frac{\partial}{\partial \beta}$ of this
dominant contribution.

(vi) Without necessarily implying the existence of an underlying
infinite-dimensional Virasoro algebra, one may write the partition function in a
form suggested by conformal field theories in $2$ spacetime dimensions as
\begin{equation}
  \label{eq:181}
  {\mathcal Z}_{d,1}(\tau,\bar\tau)={\rm Tr}\ e^{2\pi i\tau(\hat L_0-\frac{c_d}{24})
    -2\pi i\bar\tau(\hat{\bar{L}}_0-\frac{c_d}{24})}.
\end{equation}
When taking the original expression of the partition function \eqref{eq:45} and
the expression of the modular parameter \eqref{eq:46} into account, it follows
that
\begin{equation}
  \label{eq:183}
  \hat L_0=\frac{L_d}{4\pi}(\hat H+\hat P_d)+\frac{c_d}{24},\quad \hat{\bar{L}}_0=
  \frac{L_d}{4\pi}(\hat H-\hat P_d)+\frac{c_d}{24},
\end{equation}
or, equivalently, 
\begin{equation}
  \label{eq:182}
  \hat H=\frac{2\pi}{L_d}(\hat L_0+\hat{\bar{L}}_0-\frac{c_d}{12}),\quad
  \hat P_d=\frac{2\pi}{L_d}(\hat L_0-\hat{\bar{L}}_0).
\end{equation}
Note that in terms of the energy-momentum tensor,
\begin{equation}
  \label{eq:163}
  T_{\mu\nu}=\partial_\mu\phi\partial_\nu\phi-
  \frac{1}{2} \eta_{\mu\nu}\partial_\rho\phi\partial^\rho\phi, 
\end{equation}
these observables are given by 
\begin{equation}
  \label{eq:261}
  L_0=\frac{L_d}{2\pi}\int_{V_d}\text{d}^dx(T_{--}+T_{+-}),\quad
  \bar L_0=\frac{L_d}{2\pi}\int_{V_d}\text{d}^dx(T_{++}+T_{+-}),
\end{equation}
where $x^\pm=x^0\pm x^d$, $T_{\pm\pm}=\partial_\pm\phi\partial_\pm\phi$ and
$T_{+-}=\frac{1}{4}\partial_I\phi\partial^I\phi$ no longer vanishes for $d>1$. As follows
from Remark (v) and will be worked out in detail in section
\ref{sec:canon-appr-extend}, in a low temperature expansion, the leading
contribution to the partition function is
\begin{equation}
  \label{eq:184}
  \ln {\rm Z}^{\rm low}_{d,1}(\beta,0)= -\beta E^d_0,
\end{equation}
with $E_0^d$ given in \eqref{eq:43}. Identifying this contribution with that
involving the central charge $c_d$ in \eqref{eq:181}, it follows that
\begin{equation}
  \label{eq:186}
  E_0^d=-\frac{\pi c_d}{6L_d},
\end{equation}
and then, when taking \eqref{eq:43} into account, that 
\begin{equation}
  \label{eq:185}
  \boxed{c_d=\frac{6\xi(d+1)}{\pi}\frac{V_{d-1}}{L_d^{d-1}}}. 
\end{equation}
In terms of the central charge, the partition function becomes 
\begin{equation}
  \label{eq:193}
  \ln{\mathcal Z}_{d,1}(\tau,\bar\tau)=\frac{\pi c_d}{12\zeta(d+1)}
  \tau_2^{-\frac{d-1}{2}}f_{\frac{d+1}{2}}(\tau,\bar\tau),
\end{equation}
while
\begin{equation}
  \label{eq:187}
  \varepsilon^d_{\rm vac}=\frac{\pi c_d}{6}\frac{L_d^d}{V_d}.
\end{equation}
It also follows from Remark (iv) and will be worked out in detail in section
\ref{sec:canon-appr-extend}, that the leading contribution at high temperature/
large $L_d$ is
\begin{equation}
  \label{eq:190}
  \ln {\rm Z}^{\rm high}_{d,1}(\beta,0)= \frac{\pi c_d L^d_d}{6\beta^d}. 
\end{equation}
Following \cite{Shaghoulian:2015kta}, the associated asymptotic micro-canonical
density of states is then
\begin{equation}
  \label{eq:191}
  \rho_d(E)=\int^\infty_0\text{d}\beta\, Z^{\rm high}_{d,1}(\beta,0)e^{\beta E}
  = \int_0^\beta \text{d}\beta\, e^{\frac{\pi c_d L^d_d}{6\beta^d}+\beta E}, 
\end{equation}
which yields in the saddle point approximation, 
\begin{equation}
  \label{eq:192}
  \ln \rho_d(E)
  \approx (d+1)(\frac{E L_d}{d})^{\frac{d}{d+1}}(\frac{\pi c_d}{6})^{\frac{1}{d+1}}. 
\end{equation}

(vii) Real analytic Eisenstein series are eigenfunctions of the Laplace-Beltrami
operator on the upper half-plane $\mathbb H$ (see
e.g.~\cite{fleig_gustafsson_kleinschmidt_persson_2018}),
\begin{equation}
  \label{eq:194}
  \Delta_{\mathbb H}=\tau_2^2(\partial_{\tau_1}^2+\partial^2_{\tau_2}),
  \quad \Delta_{\mathbb H} f_s(\tau,\bar\tau)= s(s-1) f_s(\tau,\bar\tau). 
\end{equation}
In terms of $\ln {\mathcal Z}_{d,1}(\tau,\bar\tau)$, one finds
\begin{equation}
  \label{eq:195}
  \Delta_{\mathbb H} \ln {\mathcal Z}_{d,1}
  =(d-1)\Big[\ln {\mathcal Z}_{d,1}-\tau_2\partial_{\tau_2}
  \ln {\mathcal Z}_{d,1}\Big].
\end{equation}
When taking into account that
\begin{equation}
  \label{eq:196}
  \langle \hat H  \rangle=-\partial_\beta \ln{\rm Z}_{d,1}(\beta,\alpha),\quad
  \langle \hat P_d  \rangle=-i \partial_\alpha \ln{\rm Z}_{d,1}(\beta,\alpha),
\end{equation}
as well as 
\begin{equation}
  \label{eq:198}
  \Delta \hat O=\hat O-\langle  \hat O \rangle,
  \quad \langle  (\Delta \hat O)^2 \rangle=\langle \hat O^2
  \rangle-(\langle \hat O \rangle)^2, 
\end{equation}
the interpretation of equation \eqref{eq:195} in quantum statistical terms is
\begin{equation}
  \label{eq:197}
  \boxed{\beta^2\Big[\langle (\Delta\hat H)^2-(\Delta\hat P_d)^2 \rangle\Big]=
    (d-1)\Big[\ln {\rm Z}_{d,1}+\beta\langle \hat H \rangle
  \Big]}. 
\end{equation}
Alternatively, using $\Delta_{\mathbb
  H}=-(\tau-\bar\tau)^2\partial_\tau\partial_{\bar\tau}$, this may be written as
\begin{multline}
  \label{eq:199}
  \boxed{-(2\pi(\tau-\bar\tau))^2\Big[\langle \hat L_0\hat{\bar{L}}_0
  \rangle-\langle  \hat L_0 \rangle\langle  \hat{\bar{L}}_0 \rangle\Big]}\\=
\boxed{(d-1)\Big[\ln {\mathcal Z}_{d,1}-
  i\pi(\tau-\bar\tau)\langle \hat L_0+\hat{\bar{L}}_0-\frac{c_d}{12} \rangle\Big]}. 
\end{multline}

(viii) As in section \ref{sec:canon-appr-casim}, the regularized
Casimir energy on the spatial $\mathbb R^{d-2} \times \mathbb T^2$, where
$\mathbb T^2$ is rectangular, can be obtained from the zeta function
in \eqref{eq:51}, by multiplying the result by $\frac{1}{2}\nu^{2s}$ and
the replacements $\tau \to i\frac{L_{d}}{L_{d-1}}$, $d\to d-1$,
$s\to s-\frac{1}{2}$. If $x^I$, for $I=1\dots d-2$, denote the large
dimensions and $V_{d-2}=\prod_{I=1}^{d-2}L_I$,
\begin{equation}
    \label{eq:63}
    E_0(s)=\frac{1}{2}\nu^{2s}
    \frac{\Gamma(\frac{d+1}{2}-s)}{2^{2s-1}\pi^{\frac{d}{2}}\Gamma(s-\frac{1}{2})}
    \frac{V_{d-2} L_d}{L_{d-1}^{d-2s}}
  \zeta(\frac{d+1}{2}-s;(\frac{L_{d}}{L_{d-1}})^2,1), 
\end{equation}
so that
\begin{equation}
  \label{eq:64}
  E_0(0)=-\frac{1}{2}\frac{\Gamma(\frac{d+1}{2})}{\pi^{\frac{d+1}{2}}}
  \frac{V_{d-2} L_d}{L_{d-1}^{d}}
  \zeta(\frac{d+1}{2};(\frac{L_{d}}{L_{d-1}})^2,1). 
\end{equation}

\subsection{Canonical approach}
\label{sec:canon-appr-extend}

In this section, we evaluate the partition function
\begin{equation}
  \label{57}
  {\rm Z}_{d,1}(\beta,\alpha)=\mathrm{Tr}\, e^{-\beta\hat{H}+i\alpha\hat{P}_d}, 
\end{equation}
directly in the operator formalism where $\hat{H}$ and $\hat{P}_d$ are given by
\begin{equation}
\label{58}
  \hat{H}=\sum_{n_i\in\mathbb{Z}^d}\omega_k\hat a^{\dagger}_{k_i}\hat a_{k_i}+E_0^d,
  \qquad \hat{P}_d=\sum_{n_i\in\mathbb{Z}^d}k_d\hat a^{\dagger}_{k_i}\hat a_{k_i},
\end{equation}
and $E_0^d$ is given in \eqref{eq:43}. It follows that
\begin{equation}
  \label{eq:56}
  {\rm Z}_{d,1}(\beta,\alpha)=e^{-\beta E_0^d}\sideset{}{'}\prod_{n_i\in\mathbb{Z}^d}
  \sum_{N_{k_i}\in\mathbb{N}}e^{(-\beta\omega_k+i\alpha k_d)N_{k_i}}=e^{-\beta E_0^d}
  \sideset{}{'}\prod_{n_i\in\mathbb{Z}^d}
  \frac{1}{1-e^{-\beta\omega_k+i\alpha k_d}}.  
\end{equation}
Turning again the sums into integrals in the large dimensions gives
\begin{equation}
  \label{eq:53}
  \begin{split}
    \ln {\rm Z}_{d,1}(\beta,\alpha)&=-\beta E_0^d-\frac{V_{d-1}}{(2\pi)^{d-1}}
    \int_{-\infty}^{\infty}\D^{d-1}k\sum_{n_d\in\mathbb{Z}}
    \ln \left[1-e^{-\beta\omega_k+i\alpha k_d}\right]\\
    &=-\beta E_0^d-\frac{V_{d-1}}{(2\pi)^{d-1}}
    \frac{2\pi^{\frac{d-1}{2}}}{\Gamma(\frac{d-1}{2})}\sum_{n_d\in\mathbb{Z}}
    \int_{0}^{\infty}\D k\, k^{d-2}
    \ln [1-e^{-\beta\sqrt{k^2+k^2_d}+i\alpha k_d}],
  \end{split}
\end{equation}
where $k=k_Ik^I$, the range of $I$ is from $1$ to $d-1$ and
$V_{d-1}=\prod_{I=1}^{d-1}L_I$. For the sum over $n_d$, it is convenient to
split $n_d=0$ from the other terms. When using \eqref{eq:19} for $d\to d-1$ for
the former term, we get
\begin{equation}
  \label{eq:76}
  \ln {\rm Z}_{d,1}(\beta,\alpha)=-\beta E_0^d+
  \frac{\Gamma(d-1)\zeta(d)}{2^{d-2}\pi^{\frac{d-1}{2}}
    \Gamma(\frac{d-1}{2})}\frac{V_{d-1}}{\beta^{d-1}}
  -\frac{V_{d-1}}{2^{d-2}\pi^{\frac{d-1}{2}}\Gamma(\frac{d-1}{2})}
\sideset{}{'}\sum_{n_d\in\mathbb{Z}}
   I_d(\beta,\alpha;n_d),
\end{equation}
where
\begin{equation}
    \label{eq:77}
 I_d(\beta,\alpha;n_d)= \int_{0}^{\infty}\D k\, k^{d-2}
    \ln [1-e^{-\beta\sqrt{k^2+k^2_d}+i\alpha k_d}].
\end{equation}
Introducing the variable $z$ as
\begin{align}
\label{61}
  z=\beta\sqrt{k^2+k^2_d}, \quad
  k^{d-2}\D k=\frac{1}{\beta^2\beta^{d-3}}(z^2-\beta^2k^2_d)^{\frac{d-3}{2}}z\D z,
\end{align}
we get 
\begin{equation}
I_d(\beta,\alpha;n_d)=  \frac{1}{\beta^{d-1}}\int_{\beta \abs{k_d}}^{\infty}\D z\,
  z(z^2-\beta^2k^2_d)^{\frac{d-3}{2}}\ln\left[1-e^{-z+i\alpha k_d}\right].
\end{equation}
Expanding the logarithm as
\begin{equation}
  \label{eq:66}
\ln\left[1-e^{-z+i\alpha k_d}\right]=-\sum_{l\in\mathbb{N}^*}\frac{e^{-lz+il\alpha k_d}}{l},  
\end{equation}
we get, after the change of variables $z'=lz$,
\begin{equation}
  \label{eq:67}
  I_d(\beta,\alpha;n_d)= -\frac{1}{\beta^{d-1}}  \sum_{l\in\mathbb{N}^*}\frac{e^{il\alpha
      k_d}}{l^{d}}
  \int_{l\beta \abs{k_d}}^{\infty}\D z z(z^2-l^2\beta^2k^2_d)^{\frac{d-3}{2}}e^{-z},
\end{equation}
The integrals are given in terms of a modified Bessel function of the
second kind as
\begin{equation}
  \label{eq:70}
  \int_{l\beta \abs{k_d}}^{\infty}\D z
  z(z^2-l^2\beta^2k^2_d)^{\frac{d-3}{2}}e^{-z}
  =  2^{\frac{d-2}{2}}\frac{\Gamma(\frac{d-1}{2})}{\sqrt{\pi}}(l\beta\abs{k_d})^{\frac{d}{2}}
  K_{\frac{d}{2}}(l\beta\abs{k_d}),
\end{equation}
so that
\begin{multline}
  \label{eq:71}
  \ln{\rm Z}_{d,1}(\beta,\alpha)=-\beta
  E_0^d+\frac{\Gamma(d-1)\zeta(d)}{2^{d-2}\pi^{\frac{d-1}{2}}
    \Gamma(\frac{d-1}{2})}\frac{V_{d-1}}{\beta^{d-1}}\\
  +2\frac{V_{d-1}}{L_d^{\frac{d}{2}}\beta^{\frac{d-2}{2}}}
  \sideset{}{'}\sum_{n_d\in\mathbb{Z}}\sum_{l\in\mathbb{N}^*}(\frac{\abs{n_d}}{l})^{\frac{d}{2}}
  K_{\frac{d}{2}}(2\pi 
      l\abs{n_d}\frac{\beta}{L_d})e^{2\pi i l n_d \frac{ \alpha }{L_d}}.
\end{multline}

{\bf Remarks:}

(i) At low temperature/small distance $\frac{\beta}{L_d}\gg 1$, the
leading term in the expansion of the partition function is directly
related to the Casimir energy. The leading correction is the
contribution of the modes with spatial frequencies $n_d=0$. It
coincides with the black body result \eqref{eq:18} of a massless
scalar field in $d-1$ spatial dimensions. On account of the
equivalence of this expression with \eqref{eq:31}, it can also be
written more compactly as $\xi(d)\frac{V_{d-1}}{\beta^{d-1}}$.  It is
independent of $L_d$ and thus does not contribute to the Casimir
pressure,
\begin{equation}
  \label{eq:278}
  {\rm p}_{d,1}(\beta,\alpha)
  =\frac{1}{V_{d-1}}\frac{\partial (\beta^{-1}\ln {\rm Z}_{d,1}(\beta,\alpha))}{\partial L_d}. 
\end{equation}
The asymptotic expansion
\begin{equation}
  \label{eq:78}
  K_\nu(x)=\sqrt{\frac{\pi}{2x}}e^{-x}(1+O(x^{-1})),
\end{equation}
for large $x$, implies that all other terms are exponentially suppressed. It
follows that low-temperature/small distance expansion of the Casimir pressure is
\begin{equation}
  \label{eq:87}
  \boxed{{\rm p}^{\rm low}_{d,1}(\beta,\alpha)=-d\xi(d+1)\frac{1}{L^{d+1}_{d}}+\dots},
\end{equation}
where the dots denote exponentially suppressed terms.
In the low-temperature/small distance expansion of the entropy,
\begin{equation}
  \label{eq:91}
  {\rm S}_{d,1}(\beta,\alpha)=(1-\beta\partial_\beta)\ln {\rm Z}_{d,1}(\beta,\alpha),
\end{equation}
on the other hand, the first term in \eqref{eq:71} proportional to the Casimir
energy in $d$ spatial dimensions drops out since it is linear in $\beta$ and the
leading term now comes from the lower dimensional scalar field, i.e., the modes
with $n_d=0$,
\begin{equation}
  \label{eq:65}
  \boxed{{\rm S}^{\rm low}_{d,1}(\beta,\alpha)=d\xi(d)\frac{V_{d-1}}{\beta^{d-1}}+\dots}. 
\end{equation}
It may be written in terms of the dimensionless number characterizing the
Casimir energy density defined in \eqref{eq:133} in $d-1$ spatial dimensions as
\begin{equation}
  \label{eq:75}
  {\rm S}^{\rm low}_{d,1}(\beta,\alpha)=d\varepsilon_{\rm vac}^{d-1}\frac{V_{d-1}}{\beta^{d-1}}+\dots.
\end{equation}
This is a novel low-temperature formula for the model under
consideration, that relates the leading contribution to the entropy to
the Casimir energy density of a massless scalar field in one dimension
lower. Even though it looks similar, it is not the higher dimensional
generalization of the Cardy formula \cite{Cardy:1986ie,Cardy:1991kr}
discussed in \cite{Shaghoulian:2015kta}, which is valid at high
temperature (see below). Since the two leading terms in the low
temperature/small intervall expansion of \eqref{eq:71} do not depend
on $\alpha$ they are the same when starting from $\ln
Z_{d,1}(\beta,\mu)$. It follows that the leading terms in the
expansion of
\begin{equation}
p_{d,1}(\beta,\mu) =\frac{1}{V_{d-1}}\frac{\partial (\beta^{-1}\ln
  {Z}_{d,1}(\beta,\mu))}{\partial L_d},\quad
S_{d,1}(\beta,\mu)=(1-\beta\partial_\beta) \ln
{Z}_{d,1}(\beta,\mu),\label{eq:285}
\end{equation}
are still given by the right hand sides of \eqref{eq:87} and of
\eqref{eq:65}.

(ii) When changing the double sum in the last term to a sum over $m=ln_d\in
\mathbb Z^*$ and introducing the divisor sum
\begin{equation}
  \label{eq:79}
  \sigma_{s}(m)=\sum_{n|m}n^{s},
\end{equation}
the exponentially suppressed terms, i.e., the last line of \eqref{eq:71}, may be
written as 
\begin{equation}
  \label{eq:80}
  2\frac{V_{d-1}}{L_d^{\frac{d}{2}}\beta^{\frac{d-2}{2}}}
  \sideset{}{'}\sum_{m\in\mathbb{Z}}\sigma_{-d}(m)\abs{m}^{\frac{d}{2}}
  K_{\frac{d}{2}}(2\pi m \frac{\beta}{L_d})e^{2\pi i m \frac{\alpha}{L_d}}. 
\end{equation}
When taking into account the explicit expression for the vacuum energy $E_0^d$ in
\eqref{eq:43}, the partition function may be re-written as
\begin{multline}
  \label{eq:81}
  \boxed{\ln{\rm Z}_{d,1}(\beta,\alpha)=\frac{V_{d-1}}{L_d^{d-1}}\Big[
  \xi(d+1)\frac{\beta}{L_d}
  +\xi(d)(\frac{L_d}{\beta})^{d-1}}\\
  \boxed{+2(\frac{L_d}{\beta})^{\frac{d-2}{2}}
   \sideset{}{'}\sum_{m\in\mathbb{Z}}\sigma_{-d}(m)\abs{m}^{\frac{d}{2}}
  K_{\frac{d}{2}}(2\pi m \frac{\beta}{L_d})e^{2\pi i m \frac{\alpha}{L_d}}\Big]}. 
\end{multline}
The equivalence of the functional Lagrangian and the canonical approaches to
computing the partition function then implies that this last result is the same
than \eqref{eq:52}. For $s=\frac{d+1}{2}$, this shows
\begin{multline}
  \label{eq:72}
  f_{s}(\tau,\bar\tau)=2\zeta(2s)
  \Big[\tau_2^{s}
  +\frac{\xi(2s-1)}{\xi(2s)}\tau_2^{1-s}
  \\
  +\frac{2}{\xi(2s)}\tau_2^{\frac{1}{2}}
   \sideset{}{'}\sum_{m}\sigma_{1-2s}(m)\abs{m}^{\frac{2s-1}{2}}
  K_{\frac{2s-1}{2}}(2\pi m \tau_2)e^{2\pi i m \tau_1}\Big], 
\end{multline}
which is the Fourier expansion of $f_{s}(\tau,\bar\tau)$, traditionally derived
using Poisson resummation (see
e.g.~\cite{fleig_gustafsson_kleinschmidt_persson_2018}, Appendix A).

(iii) When expressed in terms of inverse temperature and chemical potential, the
generating set of transformations of the modular group become
\begin{equation}
  \label{eq:83}
  \tau'=\tau+1\iff \left\{\begin{array}{l} (\frac{\beta}{L_d})'=\frac{\beta}{L_d}\\
                              (\frac{\alpha}{L_d})'=\frac{\alpha}{L_d}+1
                            \end{array}
                          \right.\quad
                          \tau'=-\frac{1}{\tau}\iff  \left\{\begin{array}{l}
(\frac{\beta}{L_d})'=\frac{L_d\beta}{\alpha^2+\beta^2}\\
(\frac{\alpha}{L_d})'= -\frac{L_d\alpha}{\alpha^2+\beta^2}
                            \end{array}
                          \right..                    
\end{equation}
Under the first of these transformations, 
$\ln {\rm Z}_{d,1}(\beta,\alpha)$ in \eqref{eq:81} is manifestly invariant, as required by
\eqref{eq:55} for $c=0,a=b=d=1$. For the second of these transformations, we get
\begin{equation}
  \ln {\rm Z}_{d,1}(\beta',\alpha')=\frac{(\alpha^2+\beta^2)^{\frac{d-1}{2}}}{L_d^{d-1}}
  \ln {\rm Z}_{d,1}(\beta,\alpha).\label{eq:85}
\end{equation}
When transposing and using the explicit expression in \eqref{eq:81} for the LHS,
this gives
\begin{multline}
  \label{eq:84}
  \ln {\rm Z}_{d,1}(\beta,\alpha)=\frac{V_{d-1}}{(\alpha^2+\beta^2)^{\frac{d-1}{2}}}
\Big[
  \xi(d+1)\frac{L_d\beta}{\alpha^2+\beta^2}
  +\xi(d)(\frac{\alpha^2+\beta^2}{L_d\beta})^{d-1}\\
  +2(\frac{\alpha^2+\beta^2}{L_d\beta})^{\frac{d-2}{2}}
   \sideset{}{'}\sum_{m\in\mathbb{Z}}\sigma_{-d}(m)\abs{m}^{\frac{d}{2}}
  K_{\frac{d}{2}}(2\pi m\frac{L_d\beta}{\alpha^2+\beta^2})e^{-2\pi i m
    \frac{L_d\alpha}{\alpha^2+\beta^2}}\Big].  
\end{multline}
For a high temperature/large interval expansion, it is more convenient
to use $\mu$ rather than $\alpha$, in terms of which the previous
expression becomes
\begin{multline}
  \label{eq:84a}
  \ln {Z}_{d,1}(\beta,\mu)=\frac{V_{d-1}}{L_d^{d-1}(1+\mu^2)^{\frac{d-1}{2}}}
\Big[\frac{\xi(d+1)}{1+\mu^2}(\frac{\beta}{L_d})^{-d}
  +\xi(d)(1+\mu^2)^{d-1}+\\
  +2(\frac{\beta}{L_d})^{-\frac{d}{2}}(1+\mu^2)^{\frac{d-2}{2}}
   \sideset{}{'}\sum_{m\in\mathbb{Z}}\sigma_{-d}(m)\abs{m}^{\frac{d}{2}}
  K_{\frac{d}{2}}(2\pi m\frac{L_d}{\beta}\frac{1}{1+\mu^2})e^{-2\pi i m
    \frac{L_d}{\beta}\frac{1}{1+\mu^2}}\Big].  
\end{multline}
The high temperature/large interval expansion is then defined by
$\frac{\beta}{L_d}\ll 1$ at fixed $\mu$. The leading contributions are
given by the first two terms, while the others are exponentially
suppressed,
\begin{equation}
  \label{eq:188}
  \ln Z^{\rm high}_{d,1}(\beta,\mu)
  = \frac{V_{d-1}L_d}{\beta^d}\xi(d+1)(1+\mu^2)^{-\frac{d+1}{2}}
  +\frac{V_{d-1}}{L_d^{d-1}}\xi(d)(1+\mu^2)^{\frac{d-1}{2}}+\dots.
\end{equation}
Up to suitable powers of $(1+\mu^2)$, which will be explained in the general
case below, the two leading terms in the expansion here are thus mapped to the
two leading terms in the low temperature/small interval expansion \eqref{eq:84}
by an $S$ transformation $\tau\to -\frac{1}{\tau}$. The associated expansion of
the Casimir pressure is
\begin{equation}
  \label{eq:282}
  p_{d,1}^{\rm high}(\beta,\mu)=\frac{\xi(d+1)}{\beta^{d+1}}(1+\mu^2)^{-\frac{d+1}{2}}
  -\frac{(d-1)\xi(d)}{\beta L_d^{d}}(1+\mu^2)^{\frac{d-1}{2}}+\dots,
\end{equation}
while for the entropy, one finds
\begin{equation}
  \label{eq:132}
  S^{\rm high}_{d,1}(\beta,\mu)
  =\frac{V_{d-1}L_d}{\beta^d}(d+1)\xi(d+1)(1+\mu^2)^{-\frac{d+1}{2}}
   +\frac{V_{d-1}}{L_d^{d-1}}\xi(d)(1+\mu^2)^{\frac{d-1}{2}}+\dots. 
\end{equation}
This can be written as 
\begin{equation}
  \label{eq:135}
  S^{\rm high}_{d,1}(\beta,\mu)=
  {(d+1)\varepsilon^d_{\rm vac}}\frac{V_d}{\beta^d}(1+\mu^2)^{-\frac{d+1}{2}}
  +\varepsilon^{d-1}_{\rm vac}\frac{V_{d-1}}{L_d^{d-1}}(1+\mu^2)^{\frac{d-1}{2}}
  +\dots.
\end{equation}
The first term, which can also be written as,
\begin{equation}
  \label{eq:189}
  S^{\rm high}_{d,1}(\beta,\mu)\approx
  {(d+1)}\frac{\pi c_d L^d_d}{6\beta^d}(1+\mu^2)^{-\frac{d+1}{2}},
\end{equation}
generalizes formula (8) of \cite{Shaghoulian:2015kta} to the case of a
non-vanishing chemical potential. The associated asymptotic density of states becomes
\begin{equation}
  \label{eq:192a}
  \ln \rho_d(E,\mu)
  \approx (d+1)(\frac{E L_d}{d})^{\frac{d}{d+1}}(\frac{\pi c_d}{6})^{\frac{1}{d+1}}
 \frac{1}{\sqrt{1+\mu^2}}. 
\end{equation}

\subsection{Worldline approach}
\label{sec:world-line-approach-2}

Consider a particle on a torus with 
\begin{equation}
  \hat {\rm H}_{\beta,\alpha}=(\hat
  p_{d+1}-\frac{\alpha}{\beta}\hat p_d)^2+\hat p_d^2\label{eq:139}. 
\end{equation}
The associated heat kernel is
\begin{equation}
  \label{eq:123}
  \langle  x' |e^{-t \hat {\rm H}_{\beta,\alpha}}|x \rangle
  = \frac{1}{\beta L_d}\sum_{(n_{d+1},n_d)\in \mathbb Z^2}
  e^{-t f},  
\end{equation}
where
\begin{equation}
  \label{eq:130}
  f=|\frac{2\pi n_{d+1}}{\beta}+\frac{2\pi (\alpha+i\beta)n_d}{\beta L_d}|^2+
  i2\pi[n_{d+1}\frac{x'_{d+1}-x_{d+1}}{\beta}
  + {n_{d}}\frac{x'_{d}-x_{d}}{L_d}],
\end{equation}
which can again be shown by inserting a resolution of identity. When combining
this heat kernel with the heat kernel \eqref{eq:116} for $d-1$ large dimensions,
one gets, after dropping the $n_{d+1}=0=n_d$ term,
\begin{equation}
  \label{eq:124}
  \ln {\rm Z}_{d,1}(\beta,\alpha)=\frac{V_{d-1}}{2^{d}\pi^{\frac{d-1}{2}}}
    \sideset{}{'}\sum_{(n_{d+1},n_d)\in \mathbb Z^2}\int^\infty_0 \text{d}t\,
    t^{-\frac{d+1}{2}}e^{-t(\frac{2\pi}{\beta})^2|n_{d+1}+\tau n_d|^2}.
\end{equation}
Using again \eqref{eq:125}, 
\begin{multline}
  \frac{1}{2^{d}\pi^{\frac{d-1}{2}}}
  \sideset{}{'}\sum_{(n_{d+1},n_d)\in \mathbb Z^2}\int^\infty_0 \text{d}t\,
  t^{-\frac{d+1}{2}}e^{-t(\frac{2\pi}{\beta})^2|n_{d+1}+\tau n_d|^2}
  \\=\frac{\Gamma(\frac{1-d}{2})}{2^{d}\pi^{\frac{d-1}{2}}}
  (\frac{2\pi}{\beta})^{d-1}
  \sideset{}{'}\sum_{(n_{d+1},n_d)\in \mathbb Z^2}
  \frac{1}{|n_{d+1}+\tau n_d|^{1-d}},
  \label{eq:126}
\end{multline}
which can be expressed in terms of the Eisenstein series \eqref{eq:1} to yield
\begin{equation}
  \label{eq:127}
  \ln  \mathcal Z_{d,1}(\tau,\bar\tau)=\frac{\pi^{\frac{d-1}{2}}\Gamma(\frac{1-d}{2})}{2}
  \frac{V_{d-1}}{L_d^{d-1}}
  \frac{f_{\frac{1-d}{2}}(\tau,\bar\tau)}{\tau_2^{\frac{d-1}{2}}}.
\end{equation}
This agrees with \eqref{eq:52} after using the reflection formula for Eisenstein
series \eqref{eq:50} at $z=\frac{d+1}{2}$.

Another way to obtain the heat kernel for a particle on a torus is by a
standard path integral computation, in which case one finds instead 
\begin{equation}
  \label{eq:128}
  \langle  x' |e^{-t \hat {\rm H}_{\beta,\alpha}}|x \rangle=
  \frac{1}{4\pi t}\sum_{(m^d,m^{d+1})\in \mathbb Z^2}
  e^{-\frac{1}{4t}g},
\end{equation}
where
\begin{multline}
  \label{eq:129}
g= {(\alpha^2+\beta^2)}(\frac{x'_{d+1}-x_{d+1}}{\beta}+m^{d+1})^2
    +2\alpha L_d(\frac{x'_{d+1}-x_{d+1}}{\beta}+m^{d+1})(\frac{x'_{d}-x_{d}}{L_d}+m^d)
\\+L_d^2(\frac{x'_{d}-x_{d}}{L_d}+m^d)^2.
\end{multline}
These two expressions for the heat kernel of a particle on a torus,
\eqref{eq:123} and \eqref{eq:128} are related by the Poisson summation formula.
When combining now the expression in \eqref{eq:128} with the heat kernel in the
$d-1$ large dimensions, one gets, after dropping the $m^d=0=m^{d+1}$ term,
\begin{equation}
  \label{eq:131}
  \ln  {\rm Z}_{d,1}(\beta,\alpha)=\frac{V_{d-1} L_d\beta}{2^{d+2}\pi^{\frac{d+1}{2}}}
   \sideset{}{'}\sum_{(m^d,m^{d+1})\in \mathbb Z^2}\int^\infty_0 \text{d}t\,
  t^{-\frac{d+3}{2}}e^{-\frac{1}{4t}L_d^2|m^{d+1}\tau+m^d|^2}.
\end{equation}
In this case, the change of integration variable $t\to
\frac{1}{t}$ gives
\begin{equation}
  \label{eq:231}
  \ln  {\rm Z}_{d,1}(\beta,\alpha)=\frac{V_{d-1} L_d\beta}{2^{d+2}\pi^{\frac{d+1}{2}}}
  \sideset{}{'}\sum_{(m^d,m^{d+1})\in \mathbb Z^2}\int^\infty_0 \text{d}t\,
  t^{\frac{d+1}{2}-1}e^{-\frac{t}{4}L_d^2|m^{d+1}\tau+m^d|^2}.
\end{equation}
Using now \eqref{eq:125}, the result can be shown directly to agree with
\eqref{eq:52} without using a reflection formula. Alternatively, the equivalence
of both expressions \eqref{eq:124} and \eqref{eq:131} can be used to prove the
reflection formula \eqref{eq:50}.

{\bf Remarks:}

(i) For a particle on a torus as in \eqref{eq:139}, one notes that under the
canonical transformation
\begin{equation}
  \label{eq:74}
  \left\{ \begin{array}{l} p'_{d+1}=p_{d+1}-\frac{\alpha}{\beta}p_d,\\ p'_d=p_d
          \end{array} \right.\quad \left\{\begin{array}{l} x'_{d+1}=x_{d+1},\\
                                            x'_d=x_d+\frac{\alpha}{\beta}x_{d+1}, 
  \end{array}\right.
\end{equation}
the Hamiltonian is diagonalized, $\hat {\rm H}_{\beta,\alpha}=(\hat
p'_{d+1})^2+(\hat p'_d)^2$, while the periodicities 
\begin{equation}
x_{d+1}\sim x_{d+1}+m^{d+1}\beta,\quad x_d\sim
x_d+m^dL_d,\label{eq:136}
\end{equation}
become 
\begin{equation}
x'_{d+1}\sim x'_{d+1}+m^{d+1}\beta,\quad 
x'_d\sim x'_d+m^dL_d+ m^{d+1}\alpha\label{eq:137}. 
\end{equation}
When $\alpha=0$, the role played by the canonical pairs $(x'_{d+1},p'_{d+1})$
and $(x'_d,p'_d)$ in the computation of ${\rm Tr}\ e^{-t\hat
  {\rm H}_{\beta,0}}$ is symmetric if one exchanges at the same time the
periodicities.

(ii) The Riemann theta function is defined by 
\begin{equation}
\label{F5}
\vartheta_n(z|g)=\sum_{m^\alpha\in\mathbb{Z}^n}e^{\pi i m^\alpha g_{\alpha\beta} m^\beta+2\pi im^
  \alpha z_\alpha},\qquad \alpha,\beta=1,...,n,
\end{equation}
for $g=g_{1}+ig_{2}\in\mathbb{H}_n=\{g\in M_{n}(\mathbb{C})|g=g^T, g_2>0\}$ and
$z\in\mathbb{C}^n$. Since $\vartheta_{n=1}(z|g)|_{g=\tau}=\vartheta_3(z|\tau)$,
the Riemann theta function is the higher dimensional generalization of the
Jacobi theta function. One can also define
\begin{equation}
\label{F6}
\theta_n(g)\equiv\vartheta_n(0|ig)=\sum_{m^\alpha\in\mathbb{Z}^n}
e^{-\pi m^\alpha g_{\alpha\beta} m^\beta}.
\end{equation}
Consider now the matrix 
\begin{equation}
\label{F7}
\hat{g}_{\alpha\beta}=\frac{1}{\tau_2}\left(\begin{matrix}
1&\tau_1\\\tau_1 & \abs{\tau}^2
\end{matrix}\right),
\end{equation}
with unit determinant. After the change of variables
$\frac{\tau_2 L^2_d}{4\pi t}=t'$, the partition function in
\eqref{eq:131} can be written as
\begin{equation}
\label{F8}
\boxed{\ln\mathcal{Z}_{d,1}(\tau,\bar{\tau})=
    \frac{V_{d-1}}{2L^{d-1}_d
    \tau_2^{\frac{d-1}{2}}}\int_0^{\infty}\text{d}t\,t^{\frac{d+1}{2}-1}
  \big(\theta_2(t\hat{g})-1\big)}.
\end{equation}
Note that this suitably generalizes \eqref{eq:157} to the case where the
Euclidean spacetime manifold is $\mathbb{R}^{d-1}\times\mathbb{T}^2$. This
integral expression makes modular covariance of the partition function
transparent in the sense that the full behavior is governed by the prefactor
alone since $\theta_2(\hat g)$ is modular invariant. Indeed, since $m^\alpha
\hat g_{\alpha\beta}m^\beta=\frac{1}{\tau_2}|m^{d+1}\tau+m^d|^2$, the
transformation $\tau'=\tau+1$, can be absorbed by shifting the sum, $m^{d+1}\to
m^{d+1}-1$, the transformation $\tau'=-\frac{1}{\tau}$ results in the
replacement of $\hat g$ by $\hat g^{-1}=\epsilon^T \hat g^T\epsilon$, with
$\epsilon$ completely antisymmetric, and $\epsilon^{12}=1$, which can again be
absorbed by re-arranging the sums.

Alternatively, the equivalence of the functional and world-line approaches to
computing the partition function, i.e., the equality of the right hand sides of
\eqref{eq:52.2} and \eqref{F8}, shows that, at $z=d+1$, the completed real
analytic Eisenstein series is directly related to the Mellin transform of the Riemann
theta function minus one,
\begin{equation}
\label{F10}
  \xi(z;\tau,\bar{\tau})=\frac 12 \int_0^{\infty}\text{d}t\,t^{\frac{z}{2}-1}
  \big(\theta_2(t\hat g)-1\big),\quad \mathfrak{Re}(z)>1.
\end{equation}

\subsection{One spatial dimension}
\label{sec:torus-part-funct}

In this section, we review for completeness the well-known result
\cite{Polchinski:1985zf,ItzyksonZuber1986} (see
e.g.~\cite{Itzykson:1989sy,DiFrancesco:1997nk,henkel1999conformal} for reviews)
for a torus in Euclidean spacetime, that is to say we derive the partition
function of a massless scalar field in a one spatial dimension, $d=1$, $L_1=L$,
with periodic boundary conditions. As we have seen in section
\ref{sec:funct-appr-extend}, a naive application of the functional approach
leads one to the divergent expression
\begin{equation}
\ln {\mathcal
  Z}_1(\tau,\bar\tau)=\frac{1}{2\pi}f_{1}(\tau,\bar\tau).\label{eq:110}
\end{equation}

\subsubsection{Canonical approach}
\label{sec:canonical-approach}

In the analysis of section \ref{sec:mode-expansion}, the full Hamiltonian,
including the $n=0$ mode, which is a free particle, is now given by
\begin{equation}
  \label{eq:86}
  H=\frac{1}{L}\frac{p^2}{2}+\frac{1}{2}\sideset{}{'}\sum_{n\in\mathbb{Z}}\omega_n
  (a^{*}_{n}a_{n}+a_{n}a^{*}_{n}),\qquad\omega_n=\frac{2\pi\abs{n}}{L}.
\end{equation}
After the conventional redefinition $p=\sqrt{{4\pi}}a_0$, the
quantum Hamiltonian including the zero point energy $E^1_0$ in
\eqref{eq:43} can be written as 
\begin{equation}
  \label{eq:90}
  \hat H=-\frac{\pi}{6 L}+\frac{2\pi}{L}\big(\hat a_0^2
  + \sideset{}{'}\sum_{n\in\mathbb{Z}}\abs{n}
  \hat a^{\dagger}_{n}\hat a_{n}\big).
\end{equation}
Keeping the finite part of \eqref{eq:145}, we get
\begin{multline}
  \label{eq:88}
  {\rm Z}_1(\beta,\alpha)=\sqrt{\frac{L}{2\pi\beta}}e^{\frac{\pi\beta}{6 L}}
   \sideset{}{'}\prod_{n\in\mathbb{Z}}
  \sum_{N_n\in\mathbb{N}}e^{\frac{2\pi}{L}(-\beta\abs{n}+i\alpha n)N_n}	\\
  =\sqrt{\frac{L}{2\pi\beta}}e^{\frac{\pi\beta}{6 L}}
   \sideset{}{'}\prod_{n\in\mathbb{Z}}
  \frac{1}{1-e^{\frac{2\pi}{L}(-\beta\abs{n}+i\alpha n)}}.
\end{multline}

Defining
\begin{equation}
  \label{eq:92}
  q=e^{2\pi i \tau},\quad \bar{q}=e^{-2\pi i \bar{\tau}},
\end{equation}
where $\tau$ and $\bar{\tau}$ are given in \eqref{eq:46}, we have
\begin{equation}
  \label{eq:93}
  \ln\mathcal Z_1(\tau,\bar\tau)=-\frac{1}{2}\ln(2\pi)+\frac{\pi\tau_2}{6}
 -\frac{1}{2}\ln{(\tau_2)}
  -\sum_{n\in\mathbb{N}^*}\ln{[(1-q^n)(1-\bar q^n)]}.
\end{equation}
In terms of Dedekind's eta function, 
\begin{equation}
  \label{eq:94}
  \eta(q)=q^{\frac{1}{24}}\prod_{n\in\mathbb{N}^*}(1-q^n),
\end{equation}
and up to irrelevant numerical factors, the modular invariant partition function
can then be written compactly as
\begin{equation}
  \label{eq:95}
\boxed{\mathcal{Z}_1(\tau,\bar\tau)
  =\frac{1}{\sqrt{\tau_2}\abs{\eta(q(\tau))}^2}}.
\end{equation}

Note that in this case, since $\ln \mathcal{Z}_1(\tau,\bar\tau)$ is the sum of a
holomorphic and anti-holomorphic part, up to the contribution from the free particle,
\begin{equation}
  \label{eq:200}
  \Delta_{\mathbb H}\ln \mathcal{Z}_1(\tau,\bar\tau)=\frac{1}{2}.
\end{equation}
This is not the formal continuation of \eqref{eq:195} to $d=1$ because of the
contribution of the free particle, which gives rise to the factor $1/2$ on the
right hand side. Hence, in this case,
\begin{equation}
  \label{eq:201}
  \langle (\Delta\hat H)^2-(\Delta\hat P)^2 \rangle=\frac{1}{2\beta^2},\quad
  \langle \hat L_0\hat{\bar{L}}_0
  \rangle-\langle  \hat L_0 \rangle\langle  \hat{\bar{L}}_0 \rangle
  =-\frac{1}{8\pi^2(\tau-\bar\tau)^2}.
\end{equation}

\subsubsection{Dimensional continuation}
\label{sec:functional-approach}

A different way to give a meaning the non-convergent expression of the partition
function \eqref{eq:110} obtained in the functional or heat kernel approach is to
study the limit $s\to 1$ of the real analytic Eisenstein series
$f_s(\tau,\bar\tau)$ by starting from its Fourier expansion
\cite{fleig_gustafsson_kleinschmidt_persson_2018} given in equation \eqref{eq:72},
\begin{multline}
\label{F1}
f_{s}(\tau,\bar{\tau})=2\zeta(2s)\tau_2^{s}+2\sqrt{\pi}
\frac{\Gamma(s-\frac 12)\zeta(2s-1)}{\Gamma(s)}
\tau_2^{1-s}\\+2\frac{\pi^s}{\Gamma(s)}\tau_2^{\frac{1}{2}}\sideset{}{'}\sum_{n\in\mathbb{Z}}
\sideset{}{'}\sum_{m\in\mathbb{Z}}\abs{\frac{n}{m}}^{s-\frac{1}{2}}
K_{s-\frac{1}{2}}(2\pi\abs{nm} \tau_2)e^{2\pi i n\abs{m} \tau_1}.
\end{multline}
Only the second term diverges in the limit $s\rightarrow 1$, while the first
term on the right hand side becomes
\begin{align}
  \label{eq:202}
  \frac{\pi^2}{3}\tau_2=-\frac{\pi}{12}\ln \abs{q}^2.
\end{align}
Using $K_{\frac 12}(z)=e^{-z}\sqrt{\frac{\pi}{2z}}$, the term on the last line
becomes
\begin{multline}
  \label{eq:169}
  \pi\sideset{}{'}\sum_{n\in\mathbb{Z}}\sideset{}{'}
  \sum_{m\in\mathbb{Z}}\frac{1}{\abs{m}}e^{-2\pi\abs{nm}\tau_2+2\pi i n\abs{m}
    \tau_1}
  =2\pi\sum_{m\in\mathbb{N^*}}\sum_{n\in\mathbb{N^*}}\frac{1}{m}(e^{2\pi i n
    m\tau}+e^{-2\pi i n m\bar{\tau}})\\=
  -2\pi\sum_{n\in \mathbb N^*}\big[\ln(1-e^{2\pi i n \tau})+\ln(1-e^{-2\pi i
    n\bar{\tau}})\big]
  \\=-2\pi\ln \prod_{n\in\mathbb N^*} (1-q^n)\prod_{n\in\mathbb N^*}
  (1-\bar{q}^n). 
\end{multline}
These terms combine into
\begin{equation}
  \label{eq:203}
  -2\pi \ln \prod_{n\in\mathbb N^*} q^{\frac{1}{24}}(1-q^n)
  \prod_{n\in\mathbb N^*}\bar{q}^{\frac{1}{24}}(1-\bar{q}^n)=-2\pi \ln \abs{\eta(q)}^2.
\end{equation}
For the divergent term, we consider the expansion around $s=1$ using
\begin{equation}
\zeta(s)=\frac{1}{s-1}+\gamma+ {O}(s-1),\label{eq:205}
\end{equation}
so that
\begin{multline}
  \label{eq:204}
  2\sqrt{\pi}\frac{\Gamma(s-\frac 12)\zeta(2s-1)}{\Gamma(s)}\tau_2^{1-s}=
  \frac{\pi}{s-1}+\pi\bigg(3\gamma+  \frac{\Gamma'(\frac{1}{2})}{\Gamma(\frac{1}{2})}
  -\ln\tau_2\bigg)+{O}(s-1)\\=\frac{\pi}{s-1}+\pi\bigg(2\gamma-2 \ln 2
  -\ln\tau_2\bigg)+{O}(s-1).
\end{multline}
We thus get
\begin{equation}
\label{F4}
\lim_{s\to 1}\bigg(f_s(\tau,\bar{\tau})-\frac{\pi}{s-1}\bigg)=
2\pi\Big(\gamma-\ln 2-\ln\sqrt{\tau_2}\abs{\eta(q)}^2\Big).
\end{equation}
Therefore, once we remove the divergence, the real analytic Eisenstein series
for $s=1$ contains the result for the partition function of the $d=1$ case. Note
that the divergence in the second term of the expansion in \eqref{F1}, comes
from the $n_1=0$ term in the mode expansion of the field (at fixed time), as
shown in the canonical approach in the previous section. In $d>1$ this mode is a
scalar field in $d-1$ spatial dimensions. For $d=1$, it is a free particle,
which gives the factor $\sqrt{\tau_2}$ needed for modular invariance. Associated
with the divergence, the finite, $\tau_2$ independent term, is undetermined and
needs to be fixed by a normalization condition. 

How this computation appears from the viewpoint of the integral
representation connected to the world-line approach is discussed in
\cite{ItzyksonZuber1986}.

\section{Higher dimensional tori}
\label{sec:high-dimens-tori}

\subsection{Generalities}
\label{sec:generalities-tori}

We now consider the more general case of the Euclidean spacetime manifold
$\mathbb R^p\times \mathbb{T}^{q+1}$ \cite{Ambjorn:1981xw,Cappelli:1988vw}, with
$p+q=d$. Coordinates on $\mathbb R^p$ are denoted by $x^I$, $I=1,...,p$. The
skewed torus $\mathbb T^{q+1}$ is defined to be the quotient of
$\mathbb{R}^{q+1}$ with coordinates $x^a=(x^i,x^{d+1})$, $i=p+1,\dots d$, by the
lattice $\Lambda^{d+1}$ generated by a set of $q+1$ constant frame vectors ${
  e}\indices{^a_\alpha}\in\mathbb{R}^{d+1}$,
\begin{equation}
  \label{F423}
  \mathbb{T}^{q+1}=\mathbb{R}^{q+1}/\Lambda^{q+1},\quad
  \Lambda^{q+1}=\big\{m^{\alpha}{e}\indices{^a_\alpha}|\,
  m^{\alpha}\in\mathbb{Z}^{q+1}\big\},
\end{equation}
with $\alpha=1,\dots,q+1$.

Functions on this manifold satisfy the
periodicity conditions
\begin{equation}
  \label{eq:219}
  \phi(x^I,x^a)=\phi(x^I,x^{a}+{e}\indices{^a_\alpha}),\quad\forall\alpha   
\end{equation}
In $x^{a}$ coordinates, the information on the geometry of the torus is thus
encoded in the skewed periodicities determined by the frame vectors
${e}\indices{^a_\alpha}$ whereas the Lagrangian Euclidean action is
\begin{align}
\label{eq:219.a}
  S^E_L[\phi]=\frac 12 \int_{V_{d+1}} \text{d}^{d+1}x\,
  [\partial_I\phi\partial^I\phi+ \delta^{ab}\partial_a\phi\partial_b\phi].
\end{align}
Associated to the $q+1$ linearly independent frame vectors, there is a co-frame
${e}\indices{_{a}^{\alpha}}$ such that
\begin{equation}
  \label{eq:212}
  { e}\indices{^a_\alpha}{ e}\indices{_b^\alpha}=\delta^a_b,\quad
  { e}\indices{^a_\alpha}{ e}\indices{_a^\beta}=\delta^\beta_\alpha. 
\end{equation}
This allows one to define coordinates
$y^{\alpha}={ e}\indices{_a^{\alpha}} x^{a}$ in which
periodicities of the field become simply unity in all these directions
separately,
\begin{equation}
  \label{eq:220}
  \phi(x^I,y^{\alpha})=\phi(x^I,y^{\alpha}+\delta_\beta^\alpha),\quad\forall \beta.
\end{equation}
In turn, the Cartesian metric $\delta_{ab}$ becomes
\begin{equation}
  \label{eq:218}
   g_{\alpha\beta}={e}\indices{^a_\alpha}\delta_{ab}{e}\indices{^b_\beta},  
\end{equation}
When working in $y^\alpha$ coordinates, the information on the non-trivial
geometry of the torus gets encoded in the Lagrangian action through the coupling
of the scalar field to the metric $ g_{\alpha\beta}$,
\begin{equation}
  S^E_L[\phi; g]=\frac 12 \int_{V_{d+1}}
  \text{d}^px\text{d}^{q+1}y\, \sqrt{ g}[g^{\alpha\beta}
  \partial_\alpha\phi\partial_\beta \phi+\partial_I\phi\partial^I\phi].
\end{equation}
Note that since the metric is constant and flat, there is no need here to
consider the conformally coupled scalar field with its improved energy-momentum
tensor, as usually done in the case of a general curved background (see
e.g.~\cite{Callan:1970ze,DeWitt:1975ys,Cappelli:1988vw}). The volume of the
torus is $\text{V}_{q+1}=\mathrm{det}\, { e}\indices{^a_\alpha}=\sqrt{{g}}$.
For later use, we define the re-scaled frame, co-frame, metric and inverse metric,
\begin{equation}
  \label{eq:224}
  \begin{split}
    \hat e\indices{^a_\alpha}=(\text{V}_{q+1})^{-\frac{1}{q+1}}
    {e}\indices{^a_\alpha},\quad \hat e\indices{_a^\alpha}=
    (\text{V}_{q+1})^{\frac{1}{q+1}}{e}\indices{_a^\alpha},\\
    \hat g_{\alpha\beta}=
    (\text{V}_{q+1})^{-\frac{2}{q+1}} g_{\alpha\beta},
    \quad 
    \hat g^{\alpha\beta}=(\text{V}_{q+1})^{\frac{2}{q+1}} g^{\alpha\beta},
  \end{split}
\end{equation}
all of which have unit determinant.

Even though we will not need this construction here, we note that for the dual
lattice, the periodicities in $x_a$ coordinates are defined instead by the
vectors $e\indices{_a^\alpha}$. The coordinates in which these periodicities all
become unity separately in all directions are $y_\alpha= x_a
e\indices{^a_\alpha}$. The metric associated with the dual torus is
$g^{\alpha\beta}$, the inverse of the metric associated with the original torus.

\subsection{Functional approach}
\label{sec:functional-approach-1}

The aim is now to compute the partition function of the massless
scalar field in this flat background,
\begin{equation}
  \label{eq:217}
  Z_{d,q}(g)=\int \prod_{x^I,y^\alpha}d\phi(x^I,y^{\alpha})
 \, e^{-S^E_L[\phi;g]}. 
\end{equation}
As before, we treat $\mathbb R^p$ by first considering $\mathbb
S^1_{L_1}\times\dots \times \mathbb S^1_{L_p}$ with coordinates $x^I$,
$I=1,\dots p$, so that the fields satisfy the periodicity conditions
$\phi(x^I+L_{I},y^\alpha)=\phi(x^I,y^\alpha)$. The passage to $\mathbb R^p$ is
done by replacing sums by integrals at the expense of a factor
$(2\pi)^{-p}V_{p}$, $V_p=\prod^p_{I=1}L_I$. On $\mathbb S^1_{L_1}\times \dots
\times \mathbb S^1_{L_p}\times \mathbb T^{q+1}$, an orthonormal basis of
functions is
\begin{equation}
\label{FF1}
e_{n_I,n_{\alpha}}(x,y)=\frac{1}{\sqrt{V_{d+1}}}e^{k_I x^I+2\pi
  in_{\alpha} y^\alpha},
\end{equation}
with $k_I=\frac{2\pi n_I}{L_I}$, $n_I\in\mathbb{Z}^p$,
$n_{\alpha}\in\mathbb{Z}^{q+1}$, and $V_{d+1}=V_{d-q}\text{V}_{q+1}$. The
orthonormality condition is
\begin{multline}
\label{FF2}
(e_{n_I,n_{\alpha}},e_{n'_I,n'_{\alpha}}) =\int_{V_{d+1}}
\text{d}^{p}x\text{d}^{q+1}y\,\sqrt{g}e^{*}_{n_I,n_{\alpha}}(x,y)e_{n'_I,n'_{\alpha}}(x,y)
\\=\frac{1}{V_{d+1}}\int_{V_{d+1}}
\text{d}^px\text{d}^{q+1}y\,\sqrt{g}e^{i(k'_I-k_I)x^I}e^{2\pi
  i(n'_{\alpha}-n_{\alpha})y^{\alpha}}
=\prod_{I,\alpha}\delta_{n_I,n'_I}\delta_{n_{\alpha},n'_{\alpha}}.
\end{multline}
These basis functions are eigenfunctions of the Laplacian,
\begin{equation}
  \label{eq:222}
  \Delta
  =\partial_I\partial^I+ g^{\alpha\beta}\partial_\alpha\partial_\beta, 
\end{equation}
\begin{equation}
  \label{eq:223}
  \Delta\
  e_{n_I,n_{\alpha}}=-\lambda^2_{n_I,n_{\alpha}}e_{n_I,n_{\alpha}},
  \quad
  \lambda^2_{n_I,n_\alpha}=k_Ik^I+(2\pi)^2n_\alpha
   g^{\alpha\beta} n_\beta. 
 \end{equation}
When taking into account that the $p$ first dimensions become large,
the associated zeta function is
\begin{multline}
  \label{eq:33a}
  \zeta_{-\Delta}(s)=\frac{V_{p}}{(2\pi)^{p}}\int \D^{p} k\
  \sideset{}{'}\sum_{n_\alpha\in \mathbb Z^{q+1}}\big[(2\pi)^2n_\alpha
   g^{\alpha\beta} n_\beta+k_I k^I\big]^{-s}=
\frac{V_{p}}{(2\pi)^p}\frac{2\pi^{\frac{p}{2}}}{\Gamma(\frac{p}{2})}\\
\sideset{}{'}\sum_{n_{\alpha}\in\mathbb{Z}^{q+1}}
\int_0^{\infty}\D k\, k^{p-1}
 \big[(2\pi)^2n_\alpha g^{\alpha\beta} n_\beta+k^2\big]^{-s} = \frac{V_{p}\pi^{\frac{p}{2}-2s}
\Gamma(s-\frac{p}{2})}{2^{2s}\Gamma(s)}\sideset{}{'}\sum_{n_{\alpha}\in\mathbb{Z}^{q+1}}
\big[n_\alpha g^{\alpha\beta} n_\beta\big]^{\frac{p}{2}-s}
  \\ =\frac{V_{p}\pi^{\frac{p}{2}-2s}\Gamma(s-\frac{p}{2})}{
2^{2s}\Gamma(s)(\text{V}_{q+1})^{\frac{p-2s}{q+1}}}
\sideset{}{'}\sum_{n_{\alpha}\in\mathbb{Z}^{q+1}}
\big[n_{\alpha}\,\hat{g}^{\alpha\beta}n_{\beta}\big]^{\frac{p}{2}-s}
=\frac{V_p\pi^{\frac{p}{2}-2s}\Gamma(s-\frac{p}{2})}{
  2^{2s}
  \Gamma(s)(\text{V}_{q+1})^{\frac{p-2s}{q+1}}}f_{s-\frac{p}{2}}(q+1;\hat
g^{-1}).
\end{multline}
In the last line, we have identified the sum in terms of the SL$(n,\mathbb{Z})$
Eisenstein series in equation (2.6) of \cite{Obers:1999um} and denoted for
notational coherence here by $f_s(n;\hat g^{-1})$,
\begin{equation}
  \label{F435}
  f_s(n; \hat g^{-1})=\sideset{}{'}\sum_{n_{\alpha}\in\mathbb{Z}^{n}}
  [n_{\alpha}\,\hat g^{\alpha\beta}n_{\beta}]^{-s},\quad\mathfrak{Re}(s)>\frac{n}{2}.
\end{equation}
As shown in \cite{Obers:1999um}, this series satisfies the reflection formula
\begin{equation}
\label{F438}
\pi^{-s}\Gamma(s)f_s(n; \hat g^{-1})=\pi^{s-\frac{n}{2}}
\Gamma\bigg(\frac{n}{2}-s\bigg)f_{\frac{n}{2}-s}(n;\hat g),
\end{equation}
where
\begin{equation}
  \label{eq:206}
  f_s(n; \hat g)=\sideset{}{'}\sum_{m^{\alpha}\in\mathbb{Z}^n}
  [m^{\alpha}\,\hat g_{\alpha\beta}m^{\beta}]^{-s},\quad\mathfrak{Re}(s)>\frac{n}{2}.
\end{equation}
When
applied to \eqref{eq:33a}, we get
\begin{equation}
  \label{F439}
  \zeta_{-\Delta}(s)
  =\frac{\Gamma(\frac{p+q+1}{2}-s)V_{p}}{2^{2s}
    \Gamma(s)\pi^{\frac{p+q+1}{2}}(\text{V}_{q+1})^{\frac{p-2s}{q+1}}}
  f_{\frac{p+q+1}{2}-s}(q+1;\hat g),
\end{equation}
and then, using \eqref{eq:22} together with $p+q=d$,
\begin{equation}
  \label{F441}
  \ln Z_{d,q}(g)=\frac{\Gamma(\frac{d+1}{2})
    V_{d-q}}{2\pi^{\frac{d+1}{2}}(\text{V}_{q+1})^{\frac{d-q}{q+1}}}f_{\frac{d+1}{2}}(q+1;\hat
  g),
\end{equation}
In terms of the completed SL($n,\mathbb{Z}$) Eisenstein series
\begin{equation}
\label{F20}
\xi(z,n;\hat g)=\frac{\Gamma(\frac{z}{2})f_{\frac{z}{2}}(n;\hat g)}{2\pi^{\frac{z}{2}}},
\end{equation}
which satisfies the reflection formula
\begin{equation}
  \label{eq:209}
  \xi(z,n;\hat g^{-1})=\xi(n-z,n;\hat g),
\end{equation}
the partition function becomes
\begin{equation}
  \label{F21.a}
  \boxed{\ln {Z}_{d,q}(g)
    =\frac{V_{d-q}}{(\text{V}_{q+1})^{\frac{d-q}{q+1}}}\xi(d+1,q+1;\hat
    g)}.
\end{equation}

{\bf Remarks:}

(i) The above derivation is only valid for $p>1, q<d$. Indeed, in the case where
there are no large spatial dimensions, $p=0,q=d$, the expression
\begin{equation}
  \label{eq:262a}
  \ln {Z}_{d,d}(g)
  =\xi(d+1,d+1;\hat g),
\end{equation}
diverges. As emphasized in section \ref{sec:zero-mode-contr}, this is due to the
presence of the zero mode, whose contribution cannot be neglected when there are
no large dimensions. A renormalized expression, relying on the canonical
approach of section \ref{sec:canonical-approach-1}, is provided in section
\ref{AppC}.

(ii) The result \eqref{F21.a} can be expressed more simply in terms of the
completed GL$(n,\mathbb Z)$ Eisenstein series $\xi(z,n;g)$, which has the same
definition as the SL$(n,\mathbb Z)$ one in equation \eqref{F20} but with
$\hat g_{\alpha\beta}$ replaced by $g_{\alpha\beta}$, and satisfies the
inversion formula
\begin{equation}
  \label{eq:287}
  \xi(z,n;g^{-1})=\text{V}_n\xi(n-z,n;g), 
\end{equation}
\begin{equation}
  \label{eq:286}
  \ln {Z}_{d,q}(g)= V_{d-q}\text{V}_{q+1}\xi(d+1,q+1;g). 
\end{equation}

(iii) The SL$(n,\mathbb{Z})$ Eisenstein series is invariant under
SL$(n,\mathbb{Z})$ transformations of the lattice vectors defining
$\mathbb{T}^n$: if $S\indices{_\alpha^\beta}\in{\rm SL}(n,\mathbb{Z})$
and
\begin{equation}
  \label{F442}
  {e'}\indices{^a_\alpha}=
  S\indices{_\alpha^\beta}{e}\indices{^a_\beta},\qquad
  g'_{\alpha\beta}=S\indices{_\alpha^{\gamma}}g_{\gamma\delta}S\indices{_{\beta}^{\delta}},
\end{equation}
then 
\begin{multline}
  \label{F443}
  f_{s}(n; \hat g')=\sideset{}{'}\sum_{m^{\alpha}\in\mathbb{Z}^n}[m^{\alpha}\,
  \hat g'_{\alpha\beta}m^{\beta}]^{-s}
  =\sideset{}{'}\sum_{m^{\alpha}\in\mathbb{Z}^n}[m^{\alpha}S\indices{_\alpha^{\gamma}}
  \hat g_{\gamma\delta}S\indices{_{\beta}^{\delta}}m^{\beta}]^{-s}\\=
  \sideset{}{'}\sum_{m'^{\alpha}\in\mathbb{Z}^n}[m'^{\gamma}\hat g_{\gamma\delta}\,
  m'^{\delta}]^{-s}
  =f_{s}(n; \hat g),
\end{multline}
where $m'^{\gamma}=m^{\alpha}S_{\alpha}{}^{\gamma} $. A similar proof holds for
$f_{s}(n;\hat g'^{-1})=f_s(n;\hat g^{-1})$. Since both the volume of $\mathbb T^{q+1}$ and
$\xi(d+1,q+1;\hat g)$, are separately SL$(q+1,\mathbb{Z})$ invariant, so is the
full partition function in \eqref{F21.a}.

(iv) In order to make the connection with the discussion in section
\ref{sec:funct-appr-zeta}, one should introduce, instead of $y^{\alpha}$,
coordinates $y'^{\alpha}=L_{(\alpha)}y^{\alpha}$ with $L_{d+1}=\beta$, so that
the periodicities in \eqref{eq:220} get replaced by
\begin{equation}
  \label{eq:220a}
\phi(y'^{\alpha})=\phi(y'^{\alpha}+L_{\alpha}),\quad \forall \alpha.
\end{equation}
In $y'^{\alpha}$ coordinates, the Euclidean action in \eqref{eq:38}
corresponds to the particular choice of metric given by
${g}'^{ij}=\delta^{ij}$ and ${g}'^{d+1,i}=-\mu^{i}$, where
$i=1,\dots,d$. 

(v) For $p=d$ and $q=0$,
\begin{equation}
  \label{eq:207}
  f_s(1;\hat g)=2\zeta(2s),\quad \xi(z,1;\hat g)=\xi(z),\quad \hat g=1,
\end{equation}
and \eqref{F21.a} reduces to the scalar black body result \eqref{eq:18}.

For $p=d-1$ and $q=1$, \eqref{F21.a} reduces to the partition function on
$\mathbb{R}^{d-1}\times \mathbb{T}^2$ in \eqref{eq:52} covariant under the
SL$(2,\mathbb{Z})/\mathbb{Z}^2$ transformations of the modular parameter in
\eqref{eq:2}. Indeed, when using the parametrization
\begin{equation}
  \begin{split}
  \label{eq:99}
   \hat e\indices{^a_\alpha}&=\frac{1}{\sqrt{L_d\beta}}\begin{pmatrix} L_d & \alpha\\ 0 & \beta 
   \end{pmatrix}=\frac{1}{\sqrt{\tau_2}}\begin{pmatrix} 1 & \tau_1\\ 0 & \tau_2 
   \end{pmatrix},\\
   \hat e\indices{_a^\alpha}&=\frac{1}{\sqrt{L_d\beta}}\begin{pmatrix} \beta & 0\\
     -\alpha & L_d
   \end{pmatrix}=\frac{1}{\sqrt{\tau_2}}\begin{pmatrix} \tau_2 & 0\\ -\tau_1 & 1 
   \end{pmatrix},\\
   \hat g_{\alpha\beta}&=\frac{1}{L_d\beta}\begin{pmatrix}
     L_d^2  & L_d\alpha \\
     L_d\alpha & \alpha^2+\beta^2
   \end{pmatrix}=\frac{1}{\tau_2}\begin{pmatrix} 1  & \tau_1\\
     \tau_1 & \abs{\tau}^2
   \end{pmatrix},
   \\
   \hat g^{\alpha\beta}&=\frac{1}{L_d\beta}\begin{pmatrix}
     \alpha^2+\beta^2 & -L_d\alpha \\
     -L_d\alpha & L_d^2
   \end{pmatrix}=\frac{1}{\tau_2}\begin{pmatrix} \abs{\tau}^2 & -\tau_1\\
     -\tau_1 & 1
   \end{pmatrix},
 \end{split}
\end{equation}
so that
\begin{align}
\label{F437}
  f_s(2;\hat g)
  =\sideset{}{'}\sum_{(m^1,m^2)\in\mathbb{Z}^2}\frac{\tau_2^s}{\abs{m^{2}\tau+m^1}^{2s}}
  =f_s(\tau,\bar{\tau}),
\end{align}
is the real analytic Eisenstein series of \eqref{eq:1}. In addition, since
\begin{equation}
  \label{eq:225}
  \tau=\frac{e\indices{^d_{d+1}}+ie\indices{^{d+1}_{d+1}}}{e\indices{^d_d}+ie\indices{^{d+1}_d}}, 
\end{equation}
the modular transformation \eqref{eq:2} originates from an SL$(2,\mathbb Z)$
transformation of the lattice vectors generating $\mathbb{T}^2$ of the form
\begin{equation}
  \label{eq:226}
  S\indices{_\alpha^\beta}=\begin{pmatrix} d & c \\ b & a    
  \end{pmatrix}.
\end{equation}

(vi) The reflection formula \eqref{F438} reduces to the one satisfied by the
real analytic Eisenstein series $f_s(\tau,\bar{\tau})$ in \eqref{eq:50} for
$n=2$. Indeed, in this case, since $\hat g^{-1}=\epsilon^T \hat g^T\epsilon$,
with $\epsilon\in $ SL$(2,\mathbb Z)$ completely antisymmetric, and
$\epsilon^{12}=1$, it follows in particular that $f_s(2;\hat g)=f_s(2;\hat
g^{-1})$. For $n=1$, formula \eqref{F438} reduces to the reflection formula
\eqref{eq:28} for the Riemann zeta function.

(vii) As shown in Appendix \ref{appD}, the SL$(n,\mathbb Z)$ and GL$(n,\mathbb Z)$
Eisenstein series are eigenfunctions of suitable Laplacians. In order to
understand the interpretation of these relations in terms of the partition
function, we introduce suitable observables, an integrated energy momentum
tensor, 
\begin{equation}
  \label{eq:268}
    {\mathcal{T}}_{\alpha\beta} =\frac{2}{\sqrt{g}} \frac{\partial
      S^E_L[\phi;g]}{\partial g^{\alpha\beta}} =\int
    \text{d}^px\text{d}^{q+1}y\Big[\partial_\alpha\phi\partial_\beta\phi-\frac
    12 g_{\alpha\beta}(\partial_\gamma\phi\partial^\gamma\phi
    +\partial_I\phi\partial^I\phi)\Big],
\end{equation}
as well as integrated bilinears in first order derivatives of the field in the small
direction,
\begin{equation}    
    {\mathcal{D}}_{\alpha\beta}
    =\int\text{d}^px\text{d}^{q+1}y\,\partial_\alpha\phi\partial_\beta\phi.
\end{equation}
When using $\sqrt g=t^{\frac{q+1}{2}}=\text{V}_{q+1}$, these definitions imply that
\begin{equation}
  \label{eq:274}
  \begin{split}
    \frac{\partial S^E_L[\phi;g]}{\partial \hat g_{\alpha\beta}}&=-\frac{\sqrt
      g}{2} t J\indices{_{\gamma\delta}^{\alpha\beta}}\mathcal
    T^{\gamma\delta}
    =-t^{\frac{q-1}{2}}\hat J\indices{^{\alpha\beta\gamma\delta}}
    {\mathcal{D}}_{\gamma\delta},\\
    t \frac{\partial S^E_L[\phi;g]}{\partial t}&=-\frac 12 t^{\frac{q+1}{2}}
    g^{\alpha\beta} \mathcal T_{\alpha\beta},
\end{split}
\end{equation}
where the tensor $\hat J\indices{^{\alpha\beta\gamma\delta}}$ is defined in
Appendix \ref{appD}, with $n=q+1$ here. It then follows from \eqref{eq:264} and
\eqref{F21.a} that
\begin{multline}
  \label{eq:275}
  (q+1)t^{q-1}\hat J\indices{^{\alpha\beta\gamma\delta}}
  \langle \Delta\hat{\mathcal D}_{\alpha\beta}
  \Delta\hat{\mathcal D}_{\gamma\delta} \rangle-q(q+3)t^{\frac{q-1}{2}}
  \hat g^{\alpha\beta}\langle \hat{\mathcal D}_{\alpha\beta}\rangle \\
  =q(d-q)(d+1)\ln Z_{d,q}(g), 
\end{multline}
while 
\begin{equation}
  \label{eq:265}
 g^{\alpha\beta}\langle \hat {\mathcal T}_{\alpha\beta} \rangle
 =(q-d)\frac{1}{\text{V}_{q+1}}\ln Z_{d,q}(g).
\end{equation}

(viii) It is of interest to see whether $\mathrm{SL}(q+1,\mathbb Z)$ invariance
of the partition function in \eqref{F21.a} could have been anticipated from the
path integral definition \eqref{eq:217} before doing the actual computation.
From the viewpoint of Ward identities, when concentrating on transformations of
the coordinates $y^\alpha$ associated with the torus alone, a possible argument
goes as follows. The action $S^E_L[\phi;g]$ is invariant under diffeomorphisms,
$y'^\alpha=y'^\alpha(y^\beta)$, $S^E_L[\phi',g']=S^E_L[\phi,g]$, when $\phi$
transforms as a scalar and $g_{\alpha\beta}$ as a tensor field. A standard
change of integration variables in the path integral then would imply that
$Z[g']=Z[g]$ provided that the path integral measure is invariant. This formal
argument does however not take the boundary conditions into account. Besides the
action $S^E_L[\phi;g]$, the theory is also defined by the boundary conditions
\eqref{eq:220}. When neglecting the zero mode, the path integral measure is
defined in terms of the Fourier coefficients for the periodic paths. It thus
stays invariant if the paths stay periodic with the same periodicities,
$\phi'(y'^\alpha+\delta^\alpha_\beta)=\phi'(y'^\alpha)$, or equivalently,
$\phi(y^\alpha(y'^\beta+\delta^\beta_\gamma))=\phi(y^\alpha(y'^\beta))$. In
turn, this requires
$y^\alpha(y'^\beta+\delta^\beta_\gamma)=y^\alpha(y'^\beta)+m^\alpha$, with
$m^{\alpha}\in\mathbb{Z}$, and then that $y^\alpha=\sum_{n'_\beta\in \mathbb
  Z^{q+1}}Y^\alpha_{n'_\beta} e^{2\pi i n'_\beta y'^\beta}
+S\indices{^\alpha_\gamma}y'^\gamma$ with $S\indices{^\alpha_\gamma}\in
\mathrm{SL}(q+1,\mathbb Z)$. When restricting to
$y^\alpha=Y^\alpha_0+S\indices{^\alpha_\gamma}y'^\gamma$ guarantees that the
associated transformation of the metric remains constant.

\subsection{Canonical approach}
\label{sec:canonical-approach-1}
In order to set up the Hamiltonian analysis, we choose the ADM parametrization
(see \cite{Arnowitt:1962aa} for a review) of the metric adapted to Euclidean
signature,
\begin{equation}
  \label{eq:214}
   g^{\rm ADM}_{\alpha\beta}=
  \begin{pmatrix}
    h_{\iota\kappa} & N_\kappa\\ N_{\iota} & N^2+N_\lambda N^\lambda
  \end{pmatrix},\quad
   g^{\alpha\beta}_{\rm ADM}=
  \begin{pmatrix}
    h^{\iota\kappa} +\frac{N^\iota N^\kappa}{N^2}&
    -\frac{N^\kappa}{N^2}\\ -\frac{N^{\iota}}{N^2}
    & \frac{1}{N^2}
  \end{pmatrix}, 
\end{equation}
where $\iota=d-q+1,\dots,d$ and
$N_\iota= h_{\iota\kappa} N^\kappa$, so that
\begin{multline}
  \label{eq:210}
  S^E_L[\phi;g^{\rm ADM}]=\frac 12 \int_{V_{d+1}} \text{d}^px\text{d}^{q+1}y\,
  \Big[\frac{\sqrt{h}}{N}
  (\partial_{d+1}\phi-N^\iota\partial_{\iota}\phi)^2
   \\ +N\sqrt{h} (h^{\iota\kappa}\partial_\iota\phi\partial_\kappa\phi
  +\partial_I\phi\partial^I\phi)\Big], 
\end{multline}
with associated first order action
\begin{equation}
  \label{eq:211}
  S^E_H[\phi,\pi;g^{\rm ADM}] = \int_{V_{d+1}} \text{d}^px\text{d}^{q+1}y\,
  \Big[-i\pi \partial_{d+1}\phi+N\mathcal H(h)+
    iN^\iota\mathcal H_\iota(h)\Big],
\end{equation}
where 
\begin{equation}
\label{eq:211.5}
\mathcal H(h) =\frac{1}{2\sqrt{h}}
\pi^2 +\frac 12 \sqrt{h}  
(h^{\iota\kappa}\partial_\iota\phi\partial_\kappa\phi+\partial_I\phi\partial^I\phi),
\quad \mathcal H_\iota(h)=  \pi \partial_\iota \phi. 
\end{equation}
  
A moving frame associated to the ADM form of the
metric is
\begin{equation}
  \label{eq:216}
  { e}\indices{_{\rm ADM}^{a}_\alpha}=\begin{pmatrix}
    \theta\indices{^i_\iota} & \theta\indices{^i_\kappa} N^\kappa \\
    0 \hdots 0 & N
  \end{pmatrix},\qquad
  { e}\indices{^{\rm ADM}_{a}^\alpha}=\begin{pmatrix}
    \theta\indices{_i^\iota} & \begin{matrix} 0\\ \vdots \\ 0 
    \end{matrix}
    \\
    -\frac{N^\iota}{N} & \frac{1}{N} 
  \end{pmatrix},
\end{equation}
where
$\theta\indices{^i_\iota}$ is a moving frame for the spatial metric,
\begin{equation}
  \label{eq:215}
  h_{\iota\kappa}=\theta\indices{^i_\iota}
  \delta_{ij}\theta\indices{^j_\kappa}. 
\end{equation}
In this parametrization, if we define $ \mathrm{det}\,\theta\indices{^{i}_\iota}=\sqrt{h}= \text{V}_q$, we have ${\rm det}\, {
  e}\indices{_{\rm ADM}^{a}_\alpha}=N \text{V}_q$ and ${\rm det}\, 
g^{\mathrm{ADM}}_{\alpha\beta}=N^2\text{V}^2_q$.
  
If an explicit parametrization of the spatial metric is needed as well, one may
continue to parametrize both the spatial metric and the associated moving frame
in ADM form with lapse and shift $N_{(q)},N^{\iota_{d-1}}_{(q)}$, with
$\iota_{d-1}=d-q+1,\dots, d-1$ until one reaches $(q)=d-q+1$, in which case only the
lapse $N_{(d-q+1)}$ remains. Note that this yields a parametrization of
$e\indices{_{\rm ADM}^a_\alpha}$ of Iwasawa type,
\begin{equation}
  \label{eq:208}
   e\indices{_{\rm ADM}^a_\alpha}=(\mathcal{K}\mathcal{A}\mathcal{N})\indices{^a_\alpha}, 
\end{equation}
where the SO$(q+1,\mathbb R)$ element $\mathcal{K}$ is the unit
matrix, $\mathcal{K}=\mathbf 1$,
$\mathcal{A}={\rm diag}(N_{(d-q+1)},\dots N_{(d)},N=N_{(d+1)})$, while
$\mathcal{N}$ is upper diagonal with 1's on the diagonal. In this
parametrization, (and also with a more general Iwasawa parametrization
including a non-trivial SO$(q+1,\mathbb R)$ element that does not
contribute to the ADM metric by construction), the volume of the torus
$\mathbb T^{q+1}$ is simply the product of the lapses,
$\text{V}_{q+1}=\prod_{\alpha=d-q+1}^{d+1}N_{(\alpha)}$.

Let us define
\begin{equation}
  \label{eq:243}
  H(h)=\int_{V_d} \text{d}^px\text{d}^d y\, \mathcal{H}(h).
\end{equation}
When using that $\pi$ is a tensor density under spatial coordinate
transformations, 
\begin{equation}
\pi(x^I,y^\iota,y^{d+1})=\sqrt{h}\pi(x^I,x^i,y^{d+1}),\label{eq:255}
\end{equation}
it follows that $H(h)$ is the standard Hamiltonian when expressed
in $x$ coordinates,
\begin{equation}
  \label{eq:232}
  H(h)=\frac 12 \int_{V_{d}} \text{d}^dx \big(\pi^2+\partial_I\phi\partial^I\phi
  +\partial_i\phi\partial^i\phi\big). 
\end{equation}
Furthermore, if we define
\begin{align}
\label{eq:233.a}
  H_{\iota}(h)=\int_{V_d} \text{d}^px\text{d}^dy\,
  \mathcal{H}_{\iota}(h),
\end{align}
using again \eqref{eq:255}, we have in $x$ coordinates
\begin{equation}
  \label{eq:246}
  H_{\iota}(h)=-\theta^{i}{}_{\iota} P_i,\quad
  P_i=-\int_{V_d} \text{d}^dx\,\pi\partial_i\phi.
\end{equation}
It now follows from the usual phase space path integral arguments that, in
operator formalism, the path integral weighted by the first order
Euclidean action \eqref{eq:211}, corresponds to
\begin{equation}
  \label{eq:247}
  {\rm Z}_{d,q}(\beta,\alpha^j,h)
  ={\rm Tr}\, e^{-\beta \hat H(h)+i\alpha^j \hat P_j}
  ={\rm Tr}\, e^{-\beta (\hat H(h)-i\mu^j \hat  P_j)}, 
\end{equation}
provided that
\begin{equation}
  \label{eq:248}
  \beta=N,\quad \alpha^j=\theta\indices{^j_\iota} N^\iota,\quad
  \mu^j=\theta\indices{^j_\iota}\frac{N^\iota}{N}. 
\end{equation}
Note that this result holds for general lattice vectors
$e\indices{^a_\alpha}$, so that the Iwasawa decomposition
\eqref{eq:208}, may also contain a non-trivial rotation matrix.

We now proceed to directly evaluate the trace \eqref{eq:247} in the operator
approach on the spatial manifold $\mathbb{R}^p\times\mathbb{T}^{q}$, with
$d=p+q$. As before, in section \ref{sec:functional-approach-1}, we start by
replacing 
$\mathbb{R}^p$ by $\mathbb S^1_{L_1}\times\dots\times \mathbb S^1_{L_{p}}$. 
An orthonormal basis of functions is  
\begin{equation}
  \label{FF1a}
  e_{n_{I},n_{\iota}}(x)=\frac{1}{\sqrt{V_{d}}}e^{ik_Ix^I}e^{2\pi in_{\iota}\theta\indices{_i^\iota}x^i},
  \qquad k_I=\frac{2\pi n_I}{L_I},\qquad n_I\in\mathbb{Z}^p,\qquad n_{\alpha}\in\mathbb{Z}^{q+1},
\end{equation}
where $V_{d}=V_{d-q}\text{V}_{q}$.
The orthonormality condition is
\begin{multline}
  \label{FF2a}
  (e_{n_{I},n_{\iota}},e_{n'_{I},n'_{\iota}})=\int_{V_{d}}
  \text{d}^{d}x\,e^{*}_{n_{I},n_{\iota}}(x)e_{n'_{I},n'_{\iota}}
(x)  \\=\frac{1}{V_d}\int_{V_{d-q}}
  \text{d}^px\text{d}^qy\sqrt{h}\,e^{i(k'_I-k_I)x^I}e^{2\pi
    i(n'_{\iota}-n_{\iota})y^{\iota}}
  =\prod_{I,\iota}\delta_{n_I,n'_I}\delta_{n_{\iota},n'_{\iota}}.
\end{multline}
The field $\phi$ and its momentum $\pi$ satisfy the following periodicity
conditions along the $x^i$ coordinates
\begin{equation}
\label{FF3}
\phi(x^I,x^i,x^{d+1})=\phi(x^I,x^i+\theta\indices{^i_\iota},x^{d+1}),
\quad \pi(x^I,x^i,x^{d+1})=\pi(x^I,x^i+\theta\indices{^i_\iota},x^{d+1}),
\end{equation}
$\forall \iota=d-q+1,...,d$, and therefore they admit the mode expansions
\begin{equation}
\label{FF4}
  \phi(x^I,x^i)=\sum_{(n_I,n_{\iota})\in\mathbb{Z}^d}
  \phi_{n_I,n_{\iota}}e_{n_{I},n_{\iota}}(x),\quad
  \pi(x^I,x^i)=\sum_{(n_I,n_{\iota})\in\mathbb{Z}^d}
  \pi_{n_I,n_{\iota}}e_{n_{I},n_{\iota}}(x),
\end{equation}
where the Fourier components satisfy the reality conditions
$\phi_{n_I,n_{\iota}}=\phi^*_{-n_I,-n_{\iota}}$ and
$\pi_{n_I,n_{\iota}}=\pi^*_{-n_I,-n_{\iota}}$. Note that, according to the
transformation law \eqref{eq:255}, the Fourier components of
$\pi(x^I,y^{\iota})$ are $\sqrt{h}\pi_{n_I,n_{\iota}}$.
Inserting the mode expansions of $\phi(x^I,y^{\iota})$ and $\pi(x^I,y^{\iota})$
and the ansatz for the metric into the Hamiltonian \eqref{eq:243}, we get
\begin{equation}
\label{FF6}
  H(h)=\frac{1}{2}\sum_{(n_I,n_{\iota})\in\mathbb{Z}^d}
  (\pi_{n_I,n_{\iota}}\pi^*_{n_I,n_{\iota}}+\omega^2_{k_i}\phi_{n_I,n_{\iota}}
  \phi^*_{n_I,n_{\iota}}),
\end{equation}
with 
\begin{equation}
  \label{eq:221}
  \omega_{k_i}=\sqrt{k_Ik^I+(2\pi)^2n_{\iota}h^{\iota\kappa}n_{\kappa}}.
\end{equation}
Similarly, using \eqref{eq:246}, the mode expansion of the momenta is 
\begin{equation}
\label{FF7}
P_{i}=-i\theta_i{}^{\iota}\sum_{(n_I,n_{\iota})
  \in\mathbb{Z}^d}(2\pi n_{\iota})\pi^*_{n_I,n_{\iota}}\phi_{n_I,n_{\iota}}.
\end{equation}
Introducing oscillator variables
\begin{equation}
\label{FF8}
a_{n_I,n_{\iota}}=
\sqrt{\frac{\omega_{k_i}}{2}}\big[\phi_{n_I,n_{\iota}}
+\frac{i}{\omega_{k_i}}\pi_{n_I,n_{\iota}}\big],
\end{equation}
the quantized Hamiltonian without the $(n_I,n_{\iota})=(0,\dots,0)$ mode and the
linear momenta are given by
\begin{equation}
\label{FF9}
\begin{split}
  \hat{H'}(h) &= \sideset{}{'}\sum_{(n_I,n_{\iota})\in\mathbb{Z}^d}
  \omega_{k _i}\hat{a}^{\dagger}_{n_I,n_{\iota}}\hat{a}_{n_I,n_{\iota}}
  +E^{d,q}_0(h),\\
  \hat{P}_i& =\theta_i{}^{\iota}
  \sideset{}{'}\sum_{(n_i,n_{\iota})\in\mathbb{Z}^d}(2\pi i n_{\iota})
  \hat{a}^{\dagger}_{n_I,n_{\iota}}\hat{a}_{n_I,n_{\iota}},
\end{split}
\end{equation}
where $E^{d,q}_0(h)$ is the Casimir energy on
$\mathbb{R}^p\times\mathbb{T}^q$. Introducing
$\hat h_{\iota\kappa}=(\text{V}_q)^{-\frac{2}{q}}h_{\iota\kappa}$
and
$\hat \theta^{i}{}_{\iota}=(\text{V}_q)^{-\frac{1}{q}}\theta^{i}{}_{\iota}$
with unit determinants, it is explicitly given by
\begin{equation}
\label{FF13}
E^{d,q}_0(h)=-\frac 12 \frac{V_{d-q}}{(\text{V}_q)^{\frac{d-q+1}{q}}}
\frac{\Gamma(\frac{d+1}{2})}{\pi^{\frac{d+1}{2}}}f_{\frac{d+1}{2}}(q;\hat
h)
=-\frac{V_{d-q}}{(\text{V}_q)^{\frac{d-q+1}{q}}}\xi(d+1,q;\hat
h),
\end{equation}
and reduces to \eqref{eq:64} in the case where $q=2$ and where
$h$ is diagonal. 

We can now evaluate the partition function
\eqref{eq:247},
\begin{multline}
  \label{FF10} {\rm Z}_{d,q}(\beta,\alpha^j,h)
  =\mathrm{Tr}\,e^{-\beta\hat{H'}(h)-i\alpha^j
    \hat{P}_j}\\=e^{-\beta
    E^{d,q}_0(h)}\sideset{}{'}\prod_{(n_I,n_{\iota})\in\mathbb{Z}^d}
  \sum_{N_{k_i}}e^{(-\beta\omega_{k_i}+2\pi
    i\alpha^j\theta_j{}^{\iota}
    n_{\iota})N_{k_i}}\\=\sideset{}{'}\prod_{(n_I,n_{\iota})\in\mathbb{Z}^d}\frac{e^{-\beta
      E^{d,q}_0(h)}}{1-e^{-\beta\omega_{k_i}+2\pi
      i\alpha^j\theta_j{}^{\iota} n_{\iota}}}.
\end{multline}
Taking the logarithm and turning the sums over $n_{I}$ into integrals gives
\begin{multline}
\label{FF11}
\ln {\rm Z}_{d,q}(\beta,\alpha^j,h)=-\beta E^{d,q}_0(h)\\-\frac{V_{d-q}}{(2\pi)^p}\frac{2\pi^{\frac{p}{2}}}{\Gamma(\frac{p}{2})}
\sum_{n_{\iota}\in\mathbb{Z}^q}\int_0^{\infty}\text{d}k\,
k^{p-1}\ln[1-e^{-\beta\sqrt{k^2+(2\pi)^2n_{\iota}h^{\iota\kappa}n_{\kappa}}+2\pi
  i\alpha^j\theta_j{}^{\iota} n_{\iota}}].
\end{multline}
The rest of the computation proceeds as in section \ref{sec:canon-appr-extend}
with the substitutions $d\rightarrow p+1$, $k^2_d\rightarrow (2\pi)^2n_{\iota}h^{\iota\kappa}n_{\kappa}$ and $\alpha k_d\rightarrow
2\pi\alpha^j\theta_j{}^{\iota}n_{\iota}$. After using
the reduplication formula \eqref{eq:57}, the result is 
\begin{multline}
\label{FF12}
\ln {\rm Z}_{d,q}(\beta,\alpha^j,h)=-\beta E^{d,q}_0(h)+\xi(d-q+1)\frac{V_{d-q}}{\beta^{d-q}}\\
+2\frac{V_{d-q}}{\beta^{\frac{d-q-1}{2}}}
\sideset{}{'}\sum_{n_{\iota}\in\mathbb{Z}^q}\sum_{l\in\mathbb{N^*}}
\bigg(\frac{\sqrt{n_{\iota}h^{\iota\kappa}n_{\kappa}}}{l}\bigg)^{\frac{d-q+1}{2}}
K_{\frac{d+1-q}{2}}(2\pi
l\beta\sqrt{n_{\iota}h^{\iota\kappa}n_{\kappa}})e^{2\pi i l n_{\iota}
  \alpha^j\theta_j{}^{\iota}}.
\end{multline}
When taking into account \eqref{FF13}, equation \eqref{FF12} becomes, in terms
of the dimensionless parameter $b=(\text{V}_q)^{-\frac{1}{q}}\beta$,
\begin{multline}
\label{FF14}
\boxed{\ln{\rm Z}_{d,q}(\beta,\alpha^j,h)=
  \frac{V_{d-q}}{(\text{V}_q)^{\frac{d-q}{q}}}
  \bigg[b\xi(d+1,q;\hat h)+\xi(d+1-q)b^{-(d-q)}}\\
\boxed{+2b^{-\frac{d-q-1}{2}}\sideset{}{'}\sum_{n_{\iota}\in\mathbb{Z}^q}
  \sum_{l\in\mathbb{N}^*}\bigg(\frac{\sqrt{n_{\iota}\hat h^{\iota\kappa}
      n_{\kappa}}}{l}\bigg)^{\frac{d+1-q}{2}}
  K_{\frac{d+1-q}{2}}(2\pi lb\sqrt{n_{\iota} \hat h^{\iota\kappa}n_{\kappa}})
  e^{2\pi i l n_{\iota} N^{\iota}}\bigg]}.
\end{multline}

{\bf Remarks:}

(i) For $q=1$, this expression correctly reduces to \eqref{eq:81} with the
identifications $V_1=L_d$, $b=\tau_2$ and $N^1=\tau_1$.

(ii) Using the ADM parametrization, $\text{V}_{q+1}=\text{V}_q\beta$ and
hence the result for the partition function in \eqref{F21.a} reads
\begin{equation}
  \label{eq:230}
  \ln {\rm Z_{d,q}}(\beta,\alpha^j,h)=
  \frac{V_{d-q}}{(\text{V}_q\beta)^{\frac{d-q}{q+1}}}
  \xi(d+1,q+1;\hat g^{\mathrm{ADM}}),
\end{equation}
where
\begin{equation}
  \label{eq:234}
  \hat g^{\mathrm{ADM}}_{\alpha\beta}
  =b^{-\frac{2}{q+1}}\left(\begin{matrix}\hat h_{\iota \kappa} & \hat h_{\kappa\lambda}N^{\lambda}\\\hat h_{\iota \lambda}N^{\lambda}
      & b^2+\hat h_{\lambda \xi}N^{\lambda}N^{\xi}
    \end{matrix}\right),
\end{equation}
\begin{equation}
  \label{eq:259}
  \hat g^{\alpha\beta}_{\mathrm{ADM}}=b^{-\frac{2q}{q+1}}\left(\begin{matrix}b^{2}\hat{h}^{\iota\kappa}+N^{\iota}N^{\kappa} & -
      N^{\kappa}\\-
      N^{\iota} &1
    \end{matrix}\right). 
\end{equation}
Comparing \eqref{eq:230} and \eqref{FF14}, suggests that the Fourier transform
of the completed SL($n,\mathbb{Z}$) Eisenstein series for an $n$-dimensional
metric $\hat g^{\rm ADM}_{\alpha\beta}$ with unit determinant parametrized in
ADM form is
\begin{multline}
  \label{FF15}
  \xi(z,n;\hat g^{\rm ADM})=b^{\frac{z}{n}}\xi(z,n-1;\hat h)
  +b^{\frac{(n-z)(n-1)}{n}}\xi(z-n+1)\\
  +2b^{\frac{2(z-n)-(z-n-1)n}{2n}}
  \sideset{}{'}\sum_{n_{\iota}\in\mathbb{Z}^{n-1}}\sum_{l\in\mathbb{N}^*}
  \bigg(\frac{\sqrt{n_{\iota}\hat{h}^{\iota\kappa}n_{\kappa}}}{l}\bigg)^{\frac{z-n+1}{2}}
  \\K_{\frac{z-n+1}{2}}(2\pi
  lb\sqrt{n_{\iota}\hat{h}^{\iota\kappa}n_{\kappa}})e^{2\pi i l n_{\iota}
    N^{\iota}}.
\end{multline}
in the sense that the equality of \eqref{eq:230} and \eqref{FF14} corresponds to
the case $z=d+1$ and $n=q+1$. 

(iii) At low temperature/small box $b\gg 1$, the leading term in the expansion
to the partition function is directly related to the Casimir energy. The leading
correction is the contribution of the modes with spatial frequencies
$n_\iota=0$. It coincides with the black body result \eqref{eq:18} of a massless
scalar field in $d-q$ spatial dimensions. It is independent of $\text{V}_{q}$
and does not contribute to the Casimir pressure,
\begin{equation}
  \label{eq:279}
  {\rm p}_{d,q}(\beta,\alpha^j,
  h)=\frac{1}{V_{d-q}}\frac{\partial(\beta^{-1}\ln {\rm Z}_{d,q}(\beta,\alpha^j,
   h))}{\partial \text{V}_q}.
\end{equation}
The asymptotic behavior of the modified Bessel functions in \eqref{eq:78}
implies that all other terms are exponentially suppressed. The
low-temperature/small box expansion of the Casimir pressure is thus,
\begin{equation}
  \label{eq:281}
  {\rm p}^{\rm low}_{d,q}(\beta,\alpha^j,
  h)=-\frac{d-q+1}{q}\frac{1}{\text{V}_q^{\frac{d+1}{q}}}\xi(d+1,q;\hat{h})
  + \dots, 
\end{equation}
where the dots denote exponentially suppressed
terms. For the low-temperature/small box expansion of the entropy
\begin{equation}
  \label{eq:280}
  {\rm S}_{d,q}(\beta,\alpha^j,h)
  =(1-\beta\partial_\beta)\ln {\rm Z}_{d,q}(\beta,\alpha^j,h),
\end{equation}
on the other hand, the first term in \eqref{FF14} proportional to the Casimir
energy drops out since it is linear in $\beta$, while the leading term now comes
from the lower dimensional scalar field, i.e., the modes with $n_\iota=0$,
\begin{equation}
  \label{eq:65bis}
  {\rm S}^{\rm low}_{d,q}(\beta,\alpha^j,h)
    =(d+1-q)\xi(d+1-q)\frac{V_{d-q}}{\beta^{d-q}}+\dots. 
\end{equation}
Since the leading terms in the low temperature expansions \eqref{eq:281},
\eqref{eq:65bis} do not depend on $\alpha^j$, they are the same when starting
from $\ln Z_{d,q}(\beta,\mu^j,h)$ with $\alpha^j=\beta\mu^j$.

(iv) In order to discuss the high temperature/large box limit, we start by
using the reflection formula \eqref{eq:209} on the partition function in
\eqref{F21.a} with a metric in ADM parametrization gives
\begin{align}
  \label{FF16}
  \ln Z_{d,q}({g}^{\mathrm{ADM}})=\frac{V_{d-q}}{(\beta \text{V}_q)^{\frac{d-q}{q+1}}}
  \xi(q-d,q+1;\hat g^{-1}_{\mathrm{ADM}}).
\end{align}
For the Fourier transform in \eqref{FF15}, we need to parametrize $\hat
g^{-1}_{\mathrm{ADM}}$ in ADM form of a metric with unit determinant rather than
the ADM form of an inverse metric with unit determinant. This implies that one
now parametrizes the frame and co-frame as
\begin{equation}
  \label{eq:235}
  { e}\indices{^{\rm ADM}_{a}^\alpha}=\begin{pmatrix}
   \theta\indices{^D_i^\iota} &\theta\indices{^D_i^\kappa} N^D_\kappa \\
    0 \hdots 0 & N_D
  \end{pmatrix},\qquad
  { e}\indices{_{\rm ADM}^{a}_\alpha}=\begin{pmatrix}
   \theta\indices{_D^i_\iota} & \begin{matrix} 0\\ \vdots \\ 0 
    \end{matrix}
    \\
    -\frac{N^D_\iota}{N_D} & \frac{1}{N_D} 
  \end{pmatrix},
\end{equation}
and one parametrizes the associated metric and its inverse as
\begin{equation}
  \label{eq:239}
   g_{\rm ADM}^{\alpha\beta}=
  \begin{pmatrix}
   h\indices{_D^{\iota\kappa}} & N^\kappa_D\\ N^{\iota}_D & N^2_D+N^D_\lambda N^\lambda_D
  \end{pmatrix},\quad
   g^{\rm ADM}_{\alpha\beta}=
  \begin{pmatrix}
   h\indices{^D_{\iota\kappa}} +\frac{N_\iota^D N^D_\kappa}{N^2_D}&
    -\frac{N^D_\kappa}{N^2_D}\\ -\frac{N^D_{\iota}}{N^2_D}
    & \frac{1}{N^2_D}
  \end{pmatrix}, 
\end{equation}
with $N^\iota_D= h\indices{_D^{\iota\kappa}} N_\kappa^D$ and
$h\indices{_D^{\iota\kappa}}=\theta \indices{^D_i^\iota}\delta^{ij}\theta\indices{^D_j^\kappa}$.
Comparing to the usual ADM parametrization in \eqref{eq:214}, it follows that
the dual and standard ADM parameters are related as 
\begin{equation}
  \label{eq:244}
  \begin{split}
  h\indices{_D^{\iota\kappa}}&=h\indices{^{\iota\kappa}}
  +\frac{N^\iota N^\kappa}{N^2},\quad
  \frac{1}{N_D^2}=N^2+N_\lambda N^\lambda,\quad  N^\iota_D=-\frac{N^\iota}{N^2},\\
 h\indices{^D_{\iota\kappa}}&=h\indices{_{\iota\kappa}}
  -\frac{N_\iota N_\kappa}{N^2+N_\lambda N^\lambda},
\end{split}
\end{equation}
and then also as,
\begin{equation}
  \label{eq:245}
  \frac{1}{N^2}=N^2_D+N^D_\lambda N^\lambda_D,
  \quad N_\iota=-\frac{N_{\iota}^D}{N_D^2},\quad
  h\indices{_{\iota\kappa}}=h\indices{^D_{\iota\kappa}}
  +\frac{N_\iota^D N^D_\kappa}{N^2_D}. 
\end{equation}
From \eqref{eq:239}, it follows that the determinant of
${g}^{\rm ADM}_{\alpha\beta}$ which is $N^2 \text{V}_q^2$ in the
standard parametrization, is given by
$\mathrm{det}\, {g}^{\rm ADM}=N^{-2}_D\mathrm{det}\,
h^D \equiv N^{-2}_D (\text{V}_q^D)^2$ in the dual
parametrization. As a consequence, $\text{V}_q^D=\text{V}_q N N_D$, and using
\eqref{eq:244},
\begin{align}
\label{eq:245.1}
\text{V}_q^D=\text{V}_q\frac{N}{\sqrt{N^2+N_{\lambda}N^{\lambda}}}.
\end{align}
Introducing the adimensional parameter $b_D= N_D (\text{V}^D_q)^{\frac{1}{q}}$,
we have for the normalized metric,
\begin{align}
\label{eq:245.2}
  &\hat g^{\mathrm{ADM}}_{\alpha\beta}=(b_D)^{\frac{-2q}{q+1}}\left(\begin{matrix}b_D^2\hat{h}^D_{\iota \kappa}
      +N^D_{\iota}N^D_{\kappa}& -N_{\kappa}^D
      \\-N^D_{\iota} & 1
    \end{matrix}\right),\\ 
    \label{eq:245.3}
    &\hat g^{\alpha\beta}_{\mathrm{ADM}}
  =(b_D)^{-\frac{2}{q+1}}\left(\begin{matrix}\hat h^{\iota\kappa}_D & N^{\kappa}_D
      \\N^{\iota}_D & b_D^2+\hat h^D_{\iota \kappa}N^{\iota}_DN^{\kappa}_D
    \end{matrix}\right).
\end{align}
In terms of the dual chemical potentials $\mu^j_D=\frac{\theta\indices{_D^j_\iota} N^\iota_D}{N
_D}=\frac{\alpha^j_D}{N_D}$, it follows that equations \eqref{eq:244} are equivalent to
\begin{equation}
  \label{FF25}
  \begin{split}
    b_D&=\frac{1}{b}\frac{1}{(1+\mu_i\mu^i)^{\frac{q+1}{2q}}},\quad
    \hat \theta^j_D{}_{\iota}b_D\mu^D_j
    =-\frac{1}{b}\frac{1}{(1+\mu_i\mu^i)}\hat \theta^j{}_{\iota}\mu_j,\\
  \hat h_{\iota\kappa}^D&=(1+\mu_i\mu^i)^{\frac{1}{q}}
  (\hat h_{\iota\kappa}-\hat \theta^j{}_{\iota}\hat{\theta}^k{}_{\kappa}
  \frac{\mu_j\mu_k}{1+\mu_i\mu^i}), 
\end{split}
\end{equation}
and that $\text{V}_q^D=\text{V}_q(1+\mu_i\mu^i)^{-\frac 12}$. 
Note in particular that in the case $q=1$, the first two equations reduce to the
$S$ transformation of \eqref{eq:83} while the last equation reduces to a trivial
identity.

We are now in a position to apply the Fourier transform \eqref{FF15} at $z=q-d$,
$n=q+1$, to \eqref{FF16}, 
\begin{multline}
\label{FF17}
\ln Z_{d,q}({g^{\mathrm{ADM}}})
=\frac{V_{d-q}}{(\beta \text{V}_q)^{\frac{d-q}{q+1}}}
\bigg[b_{D}^{\frac{q-d}{q+1}}\xi(q-d,q;\hat h^{-1}_{D})
+b_D^{\frac{q(d+1)}{q+1}}\xi(-d)
\\+2b_D^{\frac{dq+2q-d}{2(q+1)}}
\sideset{}{'}\sum_{m^{\iota}\in\mathbb{Z}^q}
\sum_{l\in\mathbb{N}^*}\bigg(\frac{\sqrt{m^{\iota}
    \hat h^D_{\iota\kappa}m^{\kappa}}}{l}\bigg)^{-\frac{d}{2}}
K_{-\frac{d}{2}}(2\pi l
b_{D}\sqrt{m^{\iota}\hat h_{\iota\kappa}^Dm^{\kappa}})
e^{2\pi i l m^{\iota} N_{\iota}^D}\bigg],
\end{multline}
or, after using the reflection formulas on the first two terms,
$K_\nu(z)=K_{-\nu}(z)$, and simplifying,
\begin{multline}
\label{FF17a}
\boxed{\ln Z_{d,q}({g^{\mathrm{ADM}}}) =\frac{V_{d-q}}{(\text{V}^D_q)^{\frac{d-q}{q}}}
\bigg[b_D^d\xi(d+1)+\xi(d,q;\hat
h_{D})} \\\boxed{+2b_D^{\frac{d}{2}}
\sideset{}{'}\sum_{m^{\iota}\in\mathbb{Z}^q}
\sum_{l\in\mathbb{N}^*}\bigg(\frac{\sqrt{m^{\iota}
    \hat{h}^D_{\iota\kappa}m^{\kappa}}}{l}\bigg)^{-\frac{d}{2}} 
K_{\frac{d}{2}}(2\pi l b_{D}\sqrt{m^{\iota}\hat{h}_{\iota\kappa}^Dm^{\kappa}})
e^{2\pi i l m^{\iota}N_{\iota}^D}\bigg]},
\end{multline}
where $N^D_\iota= \hat{\theta}^{i}_D{}_{\iota}b_D\mu_i^D$. The
high-temperature/large volume limit is now defined by $b\ll1$ so that $b_D\sim
b^{-1}\gg 1$. This agrees with the discussion after \eqref{eq:84} in the case
$q=1$. Furthermore, in that case, the appropriate powers of $(1+\mu^2)$ that
relate the high and low temperature expansions may be obtained from the
definitions of $b_D$ and $\text{V}_q^D$ above. The asymptotic behaviour of the
Bessel function in \eqref{eq:78} implies that all the terms appearing in the
sums in \eqref{FF17a} are exponentially suppressed. We thus get in the
high-temperature/large volume limit,
\begin{multline}
\label{FF26}
\ln Z_{d,q}^{\mathrm{high}}(\beta,\mu^i, h)=
\frac{V_{d-q}\text{V}_q}{\beta^d}\xi(d+1)(1+\mu_i\mu^i)^{-\frac{d+1}{2}}
\\+\frac{V_{d-q}}{(\text{V}_q)^{\frac{d-q}{q}}}
(1+\mu_i\mu^i)^{\frac{d-q}{2q}}
\xi(d,q;\hat h_D)+\dots,
\end{multline}
The first term of this expression is the black body term, suitably corrected by
the coupling of the linear momenta, while the second reduces, in the $q=1$ case,
to the second of \eqref{eq:188}. Note that, while in the $q=1$ case, the
high-temperature limit of the partition function could be simply obtained by
using modular covariance \eqref{eq:55} of $\ln\mathcal{Z}_d(\tau,\bar{\tau})$
under the S-transformation $\tau'=-\frac{1}{\tau}$ as shown in \eqref{eq:84} and
\eqref{eq:188}, in the case $q>1$, we had to use a construction involving the
dual parametrization in this derivation here. The associated expansion of the
Casimir pressure and the entropy are
\begin{multline}
  \label{eq:283}
  p^{\rm high}_{d,q}(\beta,\mu^j,
  h)=\frac{\xi(d+1)}{\beta^{d+1}}(1+\mu_i\mu^i)^{-\frac{d+1}{2}}
  \\-\frac{(d-q)\xi(d,q;\hat h_D)}{q\beta(\text{V}_q)^{\frac{d}{q}}}
  (1+\mu_i\mu^i)^{\frac{d-q}{2q}}
  +\dots,
\end{multline}
and
\begin{multline}
  \label{eq:284}
  S^{\rm high}_{d,q}(\beta,\mu^j,h)=
  \frac{V_{d-q}\text{V}_q}{\beta^d}(d+1)\xi(d+1)(1+\mu_i\mu^i)^{-\frac{d+1}{2}}
  \\+\frac{V_{d-q}}{(\text{V}_q)^{\frac{d-q}{q}}}
(1+\mu_i\mu^i)^{\frac{d-q}{2q}}
\xi(d,q;\hat h_D)+\dots, 
\end{multline}
respectively. The associated asymptotic micro-canonical density of
states is now
\begin{equation}
  \label{eq:284.1}
  \ln \rho_d(E,\mu^j,h)
  \approx (d+1)(\frac{E}{d})^{\frac{d}{d+1}}[\xi(d+1)V_{d-q}\text{V}_q]^{\frac{1}{d+1}}
 \frac{1}{\sqrt{1+\mu_i\mu^i}},
\end{equation}
that can be also written as 
\begin{equation}
\label{eq:284.2}
  \ln \rho_d(E,\mu^j,h)
  \approx (d+1)[\frac{E(\text{V}_q)^{\frac{1}{q}}}{d}]^{\frac{d}{d+1}}
  (\frac{\pi c_{d,q}}{6})^{\frac{1}{d+1}}
 \frac{1}{\sqrt{1+\mu_i\mu^i}},
\end{equation}
where
\begin{align}
\label{eq:284.3}
c_{d,q}=\frac{6\xi(d+1)}{\pi}\frac{V_{d-q}}{(\text{V}_q)^{\frac{d-q}{q}}},
\end{align}
reduces to \eqref{eq:185} for $q=1$. Note however that, unlike what
happens in the $q=1$ case, here $c_{d,q}$ cannot be related to the
Casimir energy $E_0^{d,q}(h)$.

(v) The partition function contains additional integrated
observables through the dependence of the Hamiltonian on the parameters encoded
in $h^{\iota\kappa}$. Defining, 
\begin{multline}
  \label{eq:241}
    {}^{(q)}{\mathcal{T}}_{\iota\kappa} =\frac{2}{\sqrt{h}} \frac{\partial S^E_H[\phi,\pi;g^{\mathrm{ADM}}]}{\partial h^{\iota\kappa}} \\  =
    \int_{V_{d-q}} \text{d}^px \text{d}^{q+1}y N\Big[ \partial_\iota\phi\partial_\kappa\phi-\frac
    12 h_{\iota\kappa}\big(-\frac{1}{h}\pi^2+h^{\lambda\xi} \partial_\lambda\phi\partial_\xi\phi+\partial_I\phi\partial^I\phi\big)\Big],
  \end{multline}
which correspond to 
\begin{multline}
    \label{eq:233}
    {}^{(q)}{\mathcal{T}}_{ik}=\sqrt{h}\theta\indices{_i^\iota}\theta\indices{_k^\kappa}
    {}^{(q)}{\mathcal{T}}_{\iota\kappa}\\ =\int_{V_{d+1}}\text{d}^{d+1}x\Big[
    \partial_i\phi\partial_k\phi-\frac 12 \delta_{ik}\big(-\pi^2
    +\partial_i\phi\partial^i\phi+\partial_I\phi\partial^I\phi\big)\Big],
\end{multline}
in $x$ coordinates, it follows that 
\begin{equation}
  \label{eq:242}
  \langle {}^{(q)}\widehat{{\mathcal{T}}}_{\iota\kappa} \rangle = -\frac{2}{\sqrt{h}}
  \frac{\partial }{\partial h^{\iota\kappa}}\ln Z_{d,q}(g). 
\end{equation}
As in the covariant case, one may
consider instead the variable $v$ defined by 
$h_{\iota\kappa}=v \hat h_{\iota\kappa}$ with
$\text{V}_q=\sqrt{h}=(v)^{\frac{q}{2}}$, $N=b(\text{V}_q)^{\frac
  1q}=b(v)^{\frac{1}{2}}$, 
and the integrated observables
\begin{equation}    
  {}^{(q)}{\mathcal{D}}_{\iota\kappa}
  = \int\text{d}^px\text{d}^{q+1}y\,\partial_\iota\phi\partial_\kappa\phi,
\end{equation}
such that  
\begin{equation}
  \label{eq:274a}
  \begin{split}
    \frac{\partial S^E_H[\phi,\pi;g^{\mathrm{ADM}}]}{\partial \hat h_{\iota\kappa}}&
    =-\frac{\sqrt{h}}{2}
    v{}^{(q)}J\indices{_{\xi\lambda}^{\iota\kappa}}\mathcal
    {}^{(q)}\mathcal{T}^{\xi\lambda}
    =-\frac 12 (v)^{\frac{q-1}{2}} b {}^{(q)}\hat J\indices{^{\iota\kappa\xi\lambda}}
    {}^{(q)}{\mathcal{D}}_{\xi\lambda},\\
    v\frac{\partial S^E_H[\phi,\pi;g^{\mathrm{ADM}}]}{\partial v}&
    =-\frac 12 (v)^{\frac{q}{2}}h^{\iota\kappa} {}^{(q)}\mathcal T_{\iota\kappa}.
  \end{split}
\end{equation}
Notice that the last observable is directly related to the Casimir pressure. It
then follows from \eqref{eq:264} that
\begin{equation}
  \label{eq:314}
 h^{\iota\kappa} \langle  {}^{(q)}\hat {\mathcal T}_{\iota\kappa}  \rangle=(q-d)
  \frac{1}{\text{V}_q}\ln Z_{d,q}(g^{\mathrm{ADM}}).
\end{equation}
More generally, the variables of the Hamiltonian formalism may be chosen as
$v$, $\hat h_{\iota\kappa}$,$N^\iota$, $b$, while those of the
Lagrangian formalism as $t$, $\hat g_{\alpha\beta}$. When using the ADM
parametrization \eqref{eq:234} together with $t=v(b)^{\frac{2}{q+1}}$, the
equivalence of the Lagrangian and Hamiltonian path integrals implies not only
\begin{equation}
  \label{eq:315}
  g^{\alpha\beta}\langle  \hat {\mathcal T}_{\alpha\beta} \rangle
  =\frac{1}{(v)^{\frac 12} b}h^{\iota\kappa} \langle {}^{(q)}
  \hat {\mathcal T}_{\iota\kappa}  \rangle, 
\end{equation}
in agreement with \eqref{eq:314} and \eqref{eq:265}, but yields more generally
the relations between Lagrangian and Hamiltonian observables inside correlation
functions. In particular, when expressed in terms of Hamiltonian observables,
the relation \eqref{eq:275} will be equivalent to applying the ADM form of the
SL$(q+1)$ Laplacian in \eqref{eq:311} to \eqref{eq:230}. Indeed, to use this
Laplacian, it is the spacetime volume $\text{V}_{q+1}$ that is kept fixed, so
that this Laplacian acts on \eqref{eq:230} with fixed pre-factor
${V_{d-q}}(\text{V}_{q+1})^{-\frac{d-q}{q+1}}\xi(d+1,q+1;g^{\mathrm{ADM}})$ and
the powers of $b$ are those in \eqref{FF15} with $z=d+1$, $n=q+1$ rather than
those in \eqref{FF14}.

(vi) In the case $q=1$, the issue of modular covariance of $\ln \mathcal
Z_d(\tau,\bar\tau)$ in \eqref{eq:52.2} versus SL$(2,\mathbb Z)$ invariance of
$\ln Z_d(g)$ in \eqref{F21.a} can be clarified when working with
homogeneous coordinates. A general vielbein characterizing $\mathbb T^2$ and its
Iwasawa decomposition may be written in terms of two complex variables
$z_1=x_1+iy_1$, $z_2=x_2+iy_2$ as
\begin{multline}
  \label{eq:237}
  { e}\indices{^a_\alpha}=\begin{pmatrix} x_1 & x_2\\ y_1 & y_2 
  \end{pmatrix}\\=\begin{pmatrix}
    \frac{x_1}{|z_1|} & -\frac{y_1}{|z_1|}\\ \frac{y_1}{|z_1|} & \frac{x_1}{|z_1|}
  \end{pmatrix}\sqrt{{\mathfrak{Im}}(\bar z_1 z_2)}
  \begin{pmatrix}
    \frac{|z_1|}{\sqrt{\mathfrak{Im}(\bar z_1 z_2)}} & 0\\ 0 &
    \frac{\sqrt{\mathfrak{Im}(\bar z_1 z_2)}}{|z_1|}
  \end{pmatrix}
  \begin{pmatrix}
    1 & \frac{{\mathfrak{Re}}(\bar z_1 z_2)}{|z_1|^2}\\ 0 & 1
  \end{pmatrix},
\end{multline}
where we assume ${\mathfrak{Im}}(\bar z_1 z_2)>0$.
This gives
\begin{equation}
   g_{\alpha\beta}=
\begin{pmatrix}
  \abs{z_1}^2 & \mathfrak{Re}(\bar z_1z_2)\\ \mathfrak{Re}(\bar z_1z_2) & \abs{z_2}^2
\end{pmatrix},\quad  g^{\alpha\beta}=\frac{1}{(\mathfrak{Im}(\bar z_1z_2))^2}
\begin{pmatrix}
  \abs{z_2}^2 & -\mathfrak{Re}(\bar z_1z_2)\\ -\mathfrak{Re}(\bar
  z_1z_2) & \abs{z_1}^2  
\end{pmatrix}
\label{eq:229},
\end{equation}
where the relation
\begin{equation}
  \label{eq:256}
  \abs{z_2}^2=\frac{(\mathfrak{Re}(\bar z_1z_2))^2
    +(\mathfrak{Im}(\bar z_1z_2))^2}{\abs{z_1}^2},
\end{equation}
holds. Note that we have added the matrix element
$y_1= e\indices{^{d+1}_d}$, which was absent in \eqref{eq:99},
so that the earlier discussion is recovered when $x_1=L_d$, $y_1=0$,
$x_2=\alpha$, $y_2=\beta$.

Under an SL$(2,\mathbb Z)$ transformation as in \eqref{eq:226}, $(
e')\indices{^a_\alpha}= e\indices{^a_\beta}S\indices{_\alpha^\beta}$, we have
\begin{equation}
  \label{eq:238}
  \begin{pmatrix}
    x'_1 & x'_2\\ y'_1 & y'_2
  \end{pmatrix}=
  \begin{pmatrix}
    d x_1 +c x_2  & b x_1  +a x_2 \\ d y_1 +c y_2 & b y_1+ a y_2
  \end{pmatrix},
\end{equation}
while the modular parameter, which according to \eqref{eq:225} is now given by 
\begin{equation}
  \tau = \frac{z_2}{z_1}, 
\label{eq:240}
\end{equation}
continues to transform as in \eqref{eq:2}. The ADM form of the metric
\eqref{eq:229} is obtained from the change of variables
\begin{equation}
  \label{eq:236}
  N=\frac{\mathfrak{Im}(\bar z_1z_2)}{\abs{z_1}},\quad N^1=
  \frac{\mathfrak{Re}(\bar z_1z_2)}{\abs{z_1}^2},\quad h_{11}=N^2_{(1)}
  =\abs{z_1}^2. 
\end{equation}
When using \eqref{eq:248}, it follows that temperature and the chemical
potential are given in terms of the parameters $z_1,z_2$ characterizing the
torus by the non-linear relations
\begin{equation}
  \label{eq:213}
  \beta=N=\frac{\mathfrak{Im}(\bar z_1z_2)}{\abs{z_1}},\quad
  \alpha =\sqrt{h_{11}} N^1=
  \frac{\mathfrak{Re}(\bar z_1z_2)}{\abs{z_1}}.
\end{equation}
For notational convenience, we also introduce
\begin{equation}
  \label{eq:258}
  \gamma=\sqrt{h_{11}}=\abs{z_1}. 
\end{equation}
In these terms,
\begin{equation}
  \label{eq:249}
  H(\gamma)=\frac 12 \int_{V_{d-1}}\text{d}^px \text{d}^{2}y
  \Big[\frac{1}{\gamma}(\pi^2
  +\partial_1\phi\partial_1\phi)
  +\gamma\partial_I\phi\partial^I\phi\Big],
\end{equation}
and
\begin{equation}
  \label{eq:250}
  {}^{(1)}{\mathcal{T}}_{11}=-\frac 12 \int_{V_{d-1}}
  \text{d}^px\text{d}^{2}y\,\beta\Big[\pi^2 + \partial_1\phi\partial_1\phi
  - \gamma^2\partial_I\phi\partial^I\phi\Big]. 
\end{equation}
Writing 
\begin{equation}
  \label{eq:251}
  {\rm Z_d}(\beta,\alpha,\gamma)={\rm Tr}\, e^{-\beta\hat H(\gamma)+i\alpha P_d},
\end{equation}
it follows from \eqref{eq:230} that
\begin{equation}
  \label{eq:253}
\boxed{\ln {\rm Z_{d,1}}(\beta,\alpha,\gamma)
    =\frac{V_{d-1}}{(\gamma\beta)^{\frac{d-1}{2}}}\xi(d+1,2;\hat g^{\mathrm{ADM}})}, 
\end{equation}
where the ADM metric with unit determinant is expressed in terms of the chemical
potentials $\beta,\alpha,\gamma$ as 
\begin{equation}
  \label{eq:254}
  \hat g^{\mathrm{ADM}}_{\alpha\beta}=
  \begin{pmatrix} \displaystyle{\frac{\gamma}{\beta}} &\displaystyle{\frac{\alpha}{\beta}} \\
    \displaystyle{\frac{\alpha}{\beta}} & \displaystyle{\frac{\beta}{\gamma}
      +\frac{\alpha^2}{\beta\gamma}}
  \end{pmatrix}.    
\end{equation}
The transformation laws of the chemical potentials under SL$(2,\mathbb Z)$
transformation can be obtained from those of $z_1,z_2$ in \eqref{eq:238} and are
explicitly given by
\begin{equation}
  \label{eq:257}
  \begin{split}
    \gamma'&=\sqrt{d^2\gamma^2+2cd\gamma\alpha+c^2(\alpha^2+\beta^2)},\\
    \beta'&=\frac{\beta\gamma}{\sqrt{d^2\gamma^2+2cd\gamma\alpha+c^2(\alpha^2+\beta^2)}},\\
    \alpha'&=\frac{db\gamma^2+ac(\alpha^2+\beta^2)
        +(ad+bc)\alpha\gamma}{\sqrt{d^2\gamma^2+2cd\gamma\alpha+c^2(\alpha^2+\beta^2)}}.
  \end{split}
\end{equation}
It follows that \eqref{eq:253} is fully SL$(2,\mathbb Z)$ invariant
under these transformations, whereas it is covariant under modular
transformations of the modular parameter $\tau$ alone.

Finally, we note that in terms of the modular parameter \eqref{eq:240}, the ADM metric with
unit determinant keeps the same form as in \eqref{eq:99},  
\begin{equation}
  \label{eq:260}
  \hat g^{\mathrm{ADM}}_{\alpha\beta}=\frac{1}{\tau_2}
  \begin{pmatrix} 1 & \tau_1 \\ \tau_1 & \abs{\tau}^2
  \end{pmatrix}.  
\end{equation}
In the particular case where $x_1=L_d=\gamma$, $y_1=0$, $x_2=\alpha$,
$y_2=\beta$, the partition function in \eqref{eq:253}, when expressed in terms
of $\tau$ and $\bar{\tau}$, reduces to \eqref{eq:52.2} as it should.

\subsection{Worldline approach}
\label{sec:worldline-approach}
In the world-line approach, one considers a particle on the skewed
Euclidean torus ${\mathbb T}^{q+1}$. The Hamiltonian is
\begin{equation}
\label{W2}
\hat{\mathrm H}_{\mathbb T^{q+1}}=\delta^{a
  b}\hat{p}_a\hat{p}_{b}={g}^{\alpha\beta}\hat{p}_{\alpha}\hat{p}_{\beta},   
\end{equation}
$\hat{p}_{\alpha}=-i\partial_{\alpha}$.  The basis of eigenfunctions
of the Laplacian on $\mathbb{T}^{q+1}$ are the
$e_{n_{\alpha}}(y)=\langle y|p \rangle$ given in \eqref{FF1}, where
$|y\rangle\equiv |y^{\alpha}\rangle$ and
$|p\rangle\equiv |p_{\alpha}\rangle$ with $p_\alpha=2\pi n_{\alpha}$.
The heat kernel is
\begin{multline}
\label{W5}
K_{\mathbb{T}^{q+1}}(y'^\alpha,y^\alpha;t)=\langle
y'|e^{-t\hat{\mathrm H}_{\mathbb{T}^{q+1}}}|y\rangle=\sum_{n_{\alpha}\in\mathbb{Z}^{d+1}}
\sum_{n'_{\alpha}\in\mathbb{Z}^{d+1}}\langle
y'|p'\rangle\langle p'|e^{-t\hat{\mathrm H}_{\mathbb{T}^{q+1}}}|p\rangle\langle
p|y\rangle\\=\frac{1}{\text{V}_{q+1}}\sum_{n_{\alpha}\in\mathbb{Z}^{d+1}}e^{2\pi
  i
  n_{\alpha}(y'^\alpha-y^\alpha)}e^{-t(2\pi)^2{g}^{\alpha\beta}n_{\alpha}n_{\beta}}
=\frac{1}{\text{V}_{q+1}}\vartheta_{d+1}(y'^{\alpha}-y^{\alpha}|it4\pi{g}^{-1}),
\end{multline}
where a resolution of the identity in terms of momentum eigenstates has been
used and the result has been expressed in terms of the Riemann theta function in
\eqref{F5}. Including also the contribution of the large $d-q$ dimensions, the
heat kernel on $\mathbb{R}^p\times\mathbb{T}^{q+1}$ is
\begin{multline}
  \label{eq:269}
  \boxed{K_{\mathbb{R}^p\times\mathbb{T}^{q+1}}(x'^I,y'^\alpha,x^I,y^\alpha;t)}\\
  \boxed{=\frac{1}{(4\pi t)^{\frac{p}{2}}\text{V}_{q+1}}
    e^{-\frac{1}{4t}(x'_I-x_I)(x'^I-x^I)}
  \vartheta_{q+1}(y'^{\alpha}-y^{\alpha}|it4\pi{g}^{-1})}.
\end{multline}
Its trace reduces to
\begin{equation}
\label{W6}
\mathrm{Tr}\,e^{-t\hat{\mathrm H}_{\mathbb{R}^p\times\mathbb{T}^{q+1}}}
=\frac{V_{p}}{(4\pi t)^{\frac{d-q}{2}}}
\sum_{n_{\alpha}\in\mathbb{Z}^{q+1}}e^{-t(2\pi)^2{g}^{\alpha\beta}n_{\alpha}n_{\beta}}
=\frac{V_{p}}{(4\pi t)^{\frac{d-q}{2}}}\theta_{q+1}(t4\pi{g}^{-1}),
\end{equation}
while the partition function becomes 
\begin{multline}
\label{W7}
\boxed{\ln
Z_{d,q}({g})}=\frac{V_{d-q}}{2}\sideset{}{'}\sum_{n_{\alpha}\in\mathbb{Z}^{q+1}}
\int_0^{\infty}\frac{\text{d}t}{t}\frac{1}{(4\pi
  t)^{\frac{d-q}{2}}}e^{-t(2\pi)^2{g}^{\alpha\beta}n_{\alpha}n_{\beta}}\\
\boxed{=\frac{1}{2}\frac{V_{d-q}}{(\text{V}_{q+1})^{\frac{d-q}{q+1}}}
\int_0^{\infty}\text{d}t\,t^{\frac{q-d}{2}-1}(\theta_{q+1}(t\hat g^{-1})-1)}.
\end{multline}

Alternatively, a path integral computation for the heat kernel on
$\mathbb{T}^{q+1}$ that includes paths winding around the cycles of the torus
yields
\begin{equation}
  \label{eq:270}
  K_{\mathbb{T}^{q+1}}(y'^{\alpha},y^{\alpha};t)
  =\frac{1}{(4\pi t)^{\frac{q+1}{2}}}\sum_{m^{\alpha}\in\mathbb{Z}^{q+1}}
  e^{-\frac{1}{4t}{g}_{\alpha\beta}(y'^{\alpha}-y^{\alpha}+m^{\alpha})(y'^{\beta}-y^{\beta}+m^{\beta})},
\end{equation}
which can be shown to be equivalent to \eqref{W5} by using the Poisson summation
formula \eqref{W11}. For $\mathbb{R}^p\times\mathbb{T}^{q+1}$, one finds now
\begin{multline}
  \label{eq:269a}
  \boxed{K_{\mathbb{R}^p\times\mathbb{T}^{q+1}}(x'^I,y'^\alpha,x^I,y^\alpha;t)=}
  \\\boxed{\sum_{m^{\alpha}\in\mathbb{Z}^{q+1}}
    \frac{1}{(4\pi t)^{\frac{d+1}{2}}}e^{-\frac{1}{4t}[(x'_I-x_I)(x'^I-x^I)+
      (y'^{\alpha}-y^{\alpha}+m^{\alpha}){g}_{\alpha\beta}
      (y'^{\beta}-y^{\beta}+m^{\beta})]},}
\end{multline}
with trace 
\begin{align}
\label{W5.1}
  \mathrm{Tr}\,e^{-t\hat{{\rm H}}_{\mathbb{R}^p\times\mathbb{T}^{q+1}}}
  =\frac{V_{d-q}\text{V}_{q+1}}{(4\pi
  t)^{\frac{d+1}{2}}}
  \sum_{m^{\alpha}\in\mathbb{Z}^{q+1}}e^{-\frac{1}{4t}m^{\alpha}{g}_{\alpha\beta}m^{\beta}},
\end{align}
while the partition function, after changing the integration variable
$t'=\frac{1}{4\pi t}( \text{V}_{q+1})^{\frac{2}{q+1}}$,
is given by 
\begin{equation}
  \label{F24}
  \boxed{\ln {Z}_{d,q}(g)=\frac{V_{d-q}}{2(\text{V}_{q+1})^{\frac{d-q}{q+1}}}
    \int_0^{\infty}\text{d}t\,t^{\frac{d+1}{2}-1}\big(\theta_{q+1}(t\hat
    g)-1\big)}.
\end{equation}
When comparing this expression with equation \eqref{F21.a}, it is consistent
with the suggestion that the completed SL$(n,\mathbb{Z})$ Eisenstein series is
related to the Mellin transform of the Riemann theta function
$\theta_{n}(t\hat g)$ minus 1,
\begin{equation}
\label{F25}
\xi(z,n;\hat g)=\frac 12 \int_0^{\infty}\text{d}t\,t^{\frac{z}{2}-1}
\big(\theta_{n}(t\hat g)-1\big),\quad \mathfrak{Re}(z)>1.
\end{equation}
in the sense that the equality of \eqref{F21.a} and \eqref{F24} corresponds to
the case $z=d+1$, $n=q+1$.

From the world-line approach, the Fourier transform and the high
temperature/large box expansion may also be obtained directly without
necessarily passing through canonical quantization by the following
generalization of the approach of \cite{Obers:1999um}, section 2.4. The argument
of the exponential, $-\pi t n_\alpha \hat g^{\alpha\beta} n_\beta$ in
\eqref{W7}, when using the dual ADM parametrization in \eqref{eq:245.3} together
with $N^\iota_D=\hat \theta^D_i{}^\iota b_D\mu_D^i$, becomes
\begin{multline}
  \label{eq:293}
  -\pi tb_D^{-\frac{2}{q+1}}\big[n_\iota \hat h^{\iota\kappa}_Dn_\kappa+2n_\iota 
  N^\iota_D n_{d+1}+(b^2_D+N^\iota_DN_\iota^D)n_{d+1}^2\big]\\
  =-\pi tb_D^{-\frac{2}{q+1}}\big[(n_\iota+N_\iota^Dn_{d+1})\hat
  h^{\iota\kappa}_D(n_\kappa+N^D_\kappa n_{d+1})+b^2_Dn_{d+1}^2\big].
\end{multline}
One then splits the sum over $n_\alpha\in \mathbb Z^{q+1}/(0,\dots,0)$. The
first part is given by $n_{d+1}=0$. After the change of variables
$t'=(b_D)^{-\frac{2}{q+1}}t$ and using
$\text{V}_{q+1}=(\text{V}_q)^{\frac{q+1}{q}}b$, the result is
\begin{equation}
  \label{eq:294}
  \frac{V_{d-q}}{(\text{V}_q)^{\frac{d-q}{q}}}\frac{1}{(bb_D)^{\frac{d-q}{q+1}}}
  \xi(q-d,q;\hat g_D^{-1}). 
\end{equation}
After using the reflection formula \eqref{eq:209} and also the first
of \eqref{FF25}, this reduces to the second term of \eqref{FF17a}.

For the second part, since it involves $n_{d+1}\neq 0$, one may use a Poisson
resummation for the free sum over $n_\iota\in \mathbb Z^q$ to get
\begin{multline}
  \label{eq:298}
  \frac{V_{d-q}}{2(\text{V}_q)^{\frac{d-q}{q}}}
  \frac{(b_D)^{\frac{q}{q+1}}}{(b)^{\frac{d-q}{q+1}}}\int_0^\infty
  \text{d}t\,t^{-\frac{d}{2}-1}\sideset{}{'}\sum_{n_{d+1}\in\mathbb Z}\sum_{m^\iota\in \mathbb
    Z^{q}}\\
  e^{-\pi t^{-1}b_D^{\frac{2}{q+1}}[(m^\iota-itb_D^{-\frac{2}{q+1}}N^\iota_D
    n_{d+1})\hat h_{\iota\kappa}^D
    (m^\kappa-itb_D^{-\frac{2}{q+1}}N^\kappa_D n_{d+1})]-\pi
    tb_D^{-\frac{2}{q+1}}
    (b^2_D+N_\iota^DN^\iota_D)n_{d+1}^2}. 
\end{multline}
If all $m^\iota$ vanish, this reduces to
\begin{equation}
  \label{eq:299}
  \frac{V_{d-q}}{2(\text{V}_q)^{\frac{d-q}{q}}}
  \frac{(b_D)^{\frac{q}{q+1}}}{(b)^{\frac{d-q}{q+1}}}\int_0^\infty
  \text{d}t\,t^{-\frac{d}{2}-1}\sideset{}{'}\sum_{n_{d+1}\in\mathbb Z}
  e^{-\pi tb_D^{\frac{2q}{q+1}}n_{d+1}^2}
  =\frac{V_{d-q}}{(\text{V}_q)^{\frac{d-q}{q}}}(1+\mu_i\mu^i)^{\frac{d-q}{2q}}b^d_D\xi(-d). 
\end{equation}
where the change of integration variables $t'=t (b_D)^{\frac{2q}{q+1}}$ has been
used. After using the reflection formula \eqref{eq:28}, this agrees with the
``corrected'' black body result, that is to say first term of \eqref{FF17a} or
of \eqref{FF26}.

The remaining sum is
\begin{multline}
  \label{eq:298a}
  \frac{V_{d-q}}{2(\text{V}_q)^{\frac{d-q}{q}}}(1+\mu_1\mu^i)^{\frac{d-q}{2q}}
  b_D^{\frac{d}{q+1}}\sideset{}{'}\sum_{n_{d+1}\in\mathbb Z}\sideset{}{'}\sum_{m^\iota\in \mathbb
    Z^{q}}\\\int_0^\infty
  dt\,t^{-\frac{d}{2}-1}e^{-\pi t^{-1}b_D^{\frac{2}{q+1}}m^\iota\hat{h}_{\iota\kappa}^Dm^\kappa
    -\pi t(b_D)^{\frac{2q}{q+1}}n_{d+1}^2+2\pi i
    m^\iota N_\iota^D n_{d+1}}. 
\end{multline}
When using (8.432.6) of \cite{GRADSHTEYN1980904}, 
\begin{equation}
  \label{eq:289}
  \int^\infty_0\text{d}t\,
  t^{-\nu-1}e^{-t-\frac{z^2}{4t}}=2(\frac{z}{2})^{-\nu}K_\nu(z), 
\end{equation}
it follows that 
\begin{multline}
  \label{eq:288}
  \int_0^\infty
  \text{d}t\,t^{-\frac{d}{2}-1}e^{-\pi t^{-1}b_D^{\frac{2}{q+1}}m^\iota{}^{(q)}\hat
    g_{\iota\kappa}^Dm^\kappa
    -\pi tb_D^{\frac{2q}{q+1}}n_{d+1}^2}\\
  =2(b_D)^{\frac{d(q-1)}{2(q+1)}}\Big(\frac{\sqrt{m^\iota\hat{h}_{\iota\kappa}^Dm^\kappa}}{n_{d+1}}\Big)^{-\frac{d}{2}}
K_{\frac{d}{2}}(2\pi n_{d+1} b_D\sqrt{m^\iota\hat
    h_{\iota\kappa}^Dm^\kappa}), 
\end{multline}
the remaining sum agrees with the last line of \eqref{FF17a}.

\subsection{Partition function on $\mathbb{T}^{d+1}$ and harmonic anomaly}
\label{AppC}

The partition function on $\mathbb{T}^{d+1}$, computed as in \eqref{F21.a}, is
given by the divergent expression 
\begin{equation}
\ln Z^{\rm div}_{d,d}(g)=\xi(d+1,d+1;\hat
g).\label{eq:302}
\end{equation}
It may be regularized by studying the behavior of $\xi(\epsilon+d+1,d+1;\hat g)$
in the limit $\epsilon\to 0$,
\begin{equation}
  \ln Z^{\rm reg}_{d+\epsilon,d}(g)= \xi(d+1+\epsilon,d+1;\hat g).\label{eq:271}
\end{equation}
According to \eqref{FF15}, its Fourier expansion is given by
\begin{multline}
  \ln Z^{\rm reg}_{d+\epsilon,d}(g^{\rm ADM})=b^{1+\frac{\epsilon}{d+1}}
  \xi(d+1+\epsilon,d;\hat{h})
+b^{-\frac{\epsilon d}{d+1}}\xi(1+\epsilon)
+2b^{\frac{2\epsilon-(\epsilon-1)(d+1)}{2(d+1)}}\\
\sideset{}{'}\sum_{n_{\iota}\in\mathbb{Z}^{d}}\sum_{l\in\mathbb{N}^*}
\bigg(\frac{\sqrt{n_{\iota}\hat{h}^{\iota\kappa}n_{\kappa}}}{l}\bigg)^{\frac{\epsilon+1}{2}}
K_{\frac{\epsilon+1}{2}}(2\pi
lb\sqrt{n_{\iota}\hat h^{\iota\kappa}n_{\kappa}})e^{2\pi i l n_{\iota}
  N^{\iota}}.\label{eq:301a}
\end{multline}
For $\epsilon\to 0$, the first and last terms are finite 
and given respectively by $b\xi(d+1,d;\hat{h})$ and, when using that
$K_{\frac{1}{2}}(x)=\sqrt{\frac{\pi}{2x}}e^{-x}$,
\begin{equation}
\label{C3}
\sideset{}{'}\sum_{n_{\iota}\in\mathbb{Z}^d}\sum_{l\in\mathbb{N}^*}\frac{e^{2\pi
    i l(n_{\iota}N^{\iota}+i
    b\sqrt{n_{\iota}\hat{h}^{\iota\kappa}n_{\kappa}})}}{l}
\\=-\sideset{}{'}\sum_{n_{\iota}\in\mathbb{Z}^d}
\ln(1-e^{2\pi i (n_{\iota}N^{\iota}
  +i b\sqrt{n_{\iota}\hat{h}^{\iota\kappa}n_{\kappa}})}).
\end{equation}
The second term is divergent for $\epsilon\to 0$. When using
\eqref{eq:205}, it is explicitly given by
\begin{multline}
  \label{eq:301}
  b^{-\frac{\epsilon d}{d+1}}\xi(\epsilon+1)=
  \frac{\Gamma(\frac{\epsilon+1}{2})}{\pi^{\frac{\epsilon+1}{2}}}b^{-\frac{\epsilon d}{d+1}}\zeta(\epsilon+1)
  \\=\frac{1}{\epsilon}+\gamma+\frac 12(
  \frac{\Gamma'(\frac{1}{2})}{\Gamma(\frac{1}{2})}-\sqrt\pi \ln \pi)
  -\frac{d}{d+1}\ln b+O(\epsilon).
\end{multline}
Putting the terms back together, and using
$\frac{\Gamma'(\frac{1}{2})}{\Gamma(\frac{1}{2})}=-2\ln 2-\gamma$, it follows that
\begin{multline}
  \label{eq:303}
  \ln Z^{\rm reg}_{d+\epsilon,d}(g^{\rm ADM})=b\xi(d+1,d;\hat{h})
  -\sideset{}{'}\sum_{n_{\iota}\in\mathbb{Z}^d} \ln(1-e^{2\pi i
    (n_{\iota}N^{\iota}
    +i b\sqrt{n_{\iota}\hat{h}^{\iota\kappa}n_{\kappa}})})\\
  +\frac{1}{\epsilon}+\frac{\gamma-\sqrt{\pi}\ln\pi}{2}-\ln 2-\frac{d}{d+1}\ln
  b +O(\epsilon).
\end{multline}
The renormalized partition function is then defined by subtracting the pole,
together with irrelevant purely numerical factors, and then taking the limit,
\begin{multline}
  \label{eq:304}
  \boxed{\ln Z^{\rm ren}_{d,d}(g^{\rm ADM})}=\lim_{\epsilon\to 0}\Big[\ln Z^{\rm
    reg}_{d+\epsilon,d}(g^{\rm ADM})-
  (\frac{1}{\epsilon}+\frac{\gamma-\sqrt{\pi}\ln\pi}{2}-\ln
  2)]\\
  \boxed{=b\xi(d+1,d;\hat h)
    -\sideset{}{'}\sum_{n_{\iota}\in\mathbb{Z}^d} \ln(1-e^{2\pi i
      (n_{\iota}N^{\iota} +i b\sqrt{n_{\iota}\hat{h}^{\iota\kappa}n_{\kappa}})})-\frac{d}{d+1}\ln b}.
\end{multline}

{\bf Remarks:}

(i) The first term contains the Casimir energy for $\mathbb{T}^{d}$. Together
with the second, their exponential is what replaces the expression in terms of
the Dedekind $\eta$ function and its complex conjugate in \eqref{eq:95}. The
last term is a non-trivial contributions due to the pole.

(ii) As in \cite{Obers:1999um}, Proposition 2, one notes that the renormalized
result continues to be ${\rm SL}(d+1,\mathbb Z)$ invariant since the residue of
the pole and the numerical factors do not depend on the parameters of the torus.

(iii) When applying \eqref{eq:264} to the formal divergent expression $\ln Z^{\rm
  div}_{d,d}(g)$ in \eqref{eq:302}, it would follow naively that
$\Delta_{\mathrm{SL}(d+1)}\ln Z^{\rm div}_{d,d}=0$. However, for the regularized
partition function one finds,
\begin{equation}
  \label{eq:272}
  \Delta_{\mathrm{SL}(d+1)}\ln Z^{\rm reg}_{d+\epsilon,d}(g^{\rm ADM})
  =\frac{\epsilon(d+1+\epsilon) d}{4(d+1)}\ln Z^{\rm reg}_{d+\epsilon,d}(g^{\rm ADM}).
\end{equation}
Using now that $\Delta_{\mathrm{SL}(d+1)}$ does not depend on $\epsilon$, while the
subtracted terms do not involve parameters of $g^{\rm ADM}_{\alpha\beta}$, the
definition of the renormalized partition function in terms of the regularized
one in the first line of \eqref{eq:304}, then yields the harmonic anomaly
\cite{Obers:1999um},
\begin{equation}
  \label{eq:290}
  \boxed{\Delta_{\mathrm{SL}(d+1)}\ln Z^{\rm ren}_{d,d}(g^{\rm ADM})=\frac{d}{4}}. 
\end{equation}
As a consequence, one now finds 
\begin{equation}
  \label{eq:275a}
  (d+1)t^{d-1}\hat J\indices{^{\alpha\beta\gamma\delta}}
  \langle \Delta\hat{\mathcal D}_{\alpha\beta}
  \Delta\hat{\mathcal D}_{\gamma\delta} \rangle-d(d+3)t^{\frac{d-1}{2}}
  \hat g^{\alpha\beta}\langle \hat{\mathcal D}_{\alpha\beta}\rangle 
  =d(d+1), 
\end{equation}
rather than the naive result of $0$ on the right hand side that one would get by
extrapolating \eqref{eq:275} to $q=d$.

When directly applying the ADM form of the Laplacian in equation \eqref{eq:310}
to the renormalized result, it follows that the harmonic anomaly comes purely
from the $-\frac{d}{d+1}\ln b$ term, and as a consequence,
\begin{equation}
  \label{eq:312}
  \Delta_{\mathrm{SL}(d+1),\mathrm{ADM}}\Big[b\xi(d+1,d;\hat h)
  -\sideset{}{'}\sum_{n_{\iota}\in\mathbb{Z}^d} \ln(1-e^{2\pi i
    (n_{\iota}N^{\iota} +i b\sqrt{n_{\iota}\hat{h}^{\iota\kappa}n_{\kappa}})})\Big]=0. 
\end{equation}

\section{Outlook}
\label{sec:discussion}

For a free massless boson on flat backgrounds, we have obtained exact analytic
results for the partition function in terms of well-studied modular forms. High
respectively low temperature expansions can be re-summed to all orders and
various conjectures about modular invariance in higher dimensions can be
explicitly checked. The detailed understanding that has been achieved should
prove useful in more non-trivial situations where approximation methods adapted
to either the high or low temperature regime have to be applied.

From the viewpoint of boundary conditions, the most symmetric approach consists
in taking the field periodic in all spacetimes dimensions, or more generally, to
compute the partition function on $\mathbb T^{d+1}$. The detailed analysis shows
however that the presence of large spatial dimensions acts as a regulator so
that certain divergences and zero mode issues do not show up in that case.

It would be interesting to generalize the computation of free field
partition functions here to that of thermal correlators in interacting
theories (see e.g.~\cite{Petkou2018,Petkou:2021zhg} for recent work in
this context) and to understand what survives of the present
considerations in the case of non-flat backgrounds.

From a mathematical perspective, relations \eqref{eq:158}, \eqref{eq:72},
\eqref{F10}, \eqref{F25} \eqref{FF15} between zeta functions or Eisenstein
series and their Fourier series or (inverse) Mellin transforms that follow from
the equivalence of various quantization methods hold for integer $z=d+1$ or half
integer variables $s=\frac{d+1}{2}$. In order to access more general complex
values of $z$ or $s$, one may think about the computation of these partition
functions in the framework of dimensional regularization, as discussed in the
last section.

More obvious generalizations include the study of the massive case as well as
that of fermions. Finally, it would also be interesting to understand from first
principles if there is a reason why the same Eisenstein series occur in string
theory and in these scalar field partition functions and why the partition
function for a massless scalar field on $R^{d-q}\times T^{q+1}$ should be an
eigenfunction of the $SL(q+1)$ Laplacian.

\section*{Acknowledgements}
\label{sec:acknowledgements}

\addcontentsline{toc}{section}{Acknowledgments}

The authors are grateful to A.~Kleinschmidt for helpful suggestions
and to T.~Petkou and A.~Aggarwal for useful discussions. This work is
supported by the F.R.S.-FNRS Belgium through conventions FRFC PDR
T.1025.14 and IISN 4.4503.15, as well as a research fellowship for
M.~Bonte. The work of F.~Alessio has been partially supported by a fellowship from the
``Fondazione Angelo Della Riccia''. G.~B. is grateful to the Erwin
Schr\"odinger International Institute for Mathematics and Physics for
hospitality during the final stages of this work.

\appendix

\section{$\mathrm{GL}(n,\mathbb R)/\mathrm{SO}(n,\mathbb R)$ and
  $\mathrm{SL}(n,\mathbb R)/\mathrm{SO}(n,\mathbb R)$ Laplacians}
\label{appD}

In this appendix, we provide some background on the Laplacians that are used in
section \ref{sec:canonical-approach-1}.

One starts from the (generalized)
DeWitt supermetric \cite{Dewitt:1967yk} (see also
e.g.~\cite{Giulini:1993ct,Hamber:1997ut})
\begin{equation}
  \label{eq:273}
  G^{\alpha\beta\gamma\delta}(\eta,\lambda;n)=\frac {g^{\frac{\eta}{2}}}{2} (g^{\alpha\gamma}g^{\beta\delta}
  +g^{\alpha\delta}g^{\beta\gamma}-2\lambda g^{\alpha\beta}g^{\gamma\delta}),
\end{equation}
with $\lambda=1,\eta=1,n=3$ in \cite{Dewitt:1967yk} and $\lambda=0=\eta$ in
\cite{Obers:1999um}. The inverse supermetric exists when $\lambda\neq
\frac{1}{n}$ and is given by
\begin{equation}
  G_{\alpha\beta\gamma\delta}(\eta,\lambda;n)=\frac {g^{\frac{-\eta}{2}}}{2} (g_{\alpha\gamma}g_{\beta\delta}
  +g_{\alpha\delta}g_{\beta\gamma}-\frac{2\lambda}{\lambda n-1} g_{\alpha\beta}g_{\gamma\delta}),
  \label{eq:275b}
\end{equation}
\begin{equation}
  \label{eq:274b}
  G_{\alpha\beta\mu\nu}(\eta,\lambda;n)G^{\mu\nu\gamma\delta}(\eta,\lambda;n)
  =\frac{1}{2}(\delta^\gamma_\alpha\delta^\delta_\beta+\delta^\delta_\alpha\delta_\beta^\gamma).
\end{equation}
For the determinant
$G(\eta,\lambda;n)={\rm det}\, G^{\alpha\beta\gamma\delta}(\eta,\lambda;n)$, one uses that
\begin{align}
  \label{eq:268a}
  -\delta \ln
  G(\eta,\lambda;n)=G^{\alpha\beta\gamma\delta}(\eta,\lambda;n)\delta
  G_{\alpha\beta\gamma\delta}
  (\eta,\lambda;n)=
  (n+1)(1-\frac{n\eta
    }{4})\delta \ln g,
\end{align}
to infer that 
\begin{equation}
  G(\eta,\lambda;n)=ag^{-(n+1)(1-\frac{n\eta}{4})},\label{eq:276}
\end{equation}
for some constant $a$. One then finds $G(1,1;3)=ag^{-1}$ as in
\cite{Dewitt:1967yk} and $G(0,0;n)=ag^{-(n+1)}$ for the case relevant to
\cite{Obers:1999um} and to us here. For simplicity, we now denote
$G^{\alpha\beta\gamma\delta}(0,0;n)=G^{\alpha\beta\gamma\delta}$ in the following.

According to \cite{Obers:1999um}, the metric on $\mathrm{GL}(n,\mathbb
R)/\mathrm{SO}(n,\mathbb R)$ is given by the supermetric
\begin{equation}
  \label{eq:273a}
  ds^2_{\mathrm{GL}(n)} = G^{\alpha\beta\gamma\delta}\text{d} g_{\alpha\beta}\text{d}g_{\gamma\delta},\quad
  G^{\alpha\beta\gamma\delta}=\frac{1}{2}(g^{\alpha\gamma}g^{\beta\delta}
  +g^{\alpha\delta}g^{\beta\gamma}).
\end{equation}
The inverse is 
$G_{\alpha\beta\gamma\delta}$, with indices on the supermetric lowered and
raised with $g_{\alpha\beta}$ and its inverse. In particular,
\begin{equation}
  \label{eq:252}
  G_{\alpha\beta}^{\gamma\delta}=\delta^{\gamma\delta}_{\alpha\beta}
  =\frac 12 (\delta^\gamma_\alpha\delta^\delta_\beta+\delta^\gamma_\beta\delta^\delta_\alpha).
\end{equation}
Since the determinant of the supermetric is 
$G=ag^{-(n+1)}$,  the associated Laplacian is
\begin{equation}
  \label{eq:277}
  \Delta_{\mathrm{GL}(n)} =G^{-\frac 12}\frac{\partial}{\partial g_{\alpha\beta}}(G^{\frac 12}
  G_{\alpha\beta\gamma\delta})
  \frac{\partial}{\partial g_{\gamma\delta}}, \quad
  \frac{\partial g_{\alpha\beta}}{\partial g_{\gamma\delta}}
  =\delta^{\gamma\delta}_{\alpha\beta},
\end{equation}
or explicitly,
\begin{equation}
  \label{eq:228}
  \Delta_{\mathrm{GL}(n)}=G_{\alpha\beta\gamma\delta}
  \frac{\partial^2}{\partial g_{\alpha\beta}\partial g_{\gamma\delta}}
  +\frac{n+1}{2}g_{\alpha\beta}\frac{\partial}{\partial g_{\alpha\beta}}. 
\end{equation}
When applied to the Eisenstein series, one finds 
\begin{equation}
  \label{eq:262}
  \begin{split}
  \Delta_{\mathrm{GL}(n)} f_s(n;g)&=\frac{s(2s-n+1)}{2} f_s(n;g),\\
  \Delta_{\mathrm{GL}(n)}\xi(z,n;g)&=\frac{z(z-n+1)}{4} \xi(z,n;g).
\end{split}
\end{equation}

Consider now the change of variables from $g_{\alpha\beta}$ to $t,\hat
g_{\alpha\beta}$, 
\begin{equation}
  \label{eq:292}
  g_{\alpha\beta}=t\hat g_{\alpha\beta}, 
\end{equation}
with $t=g^{\frac{1}{n}}$, $\hat g=1$. It follows that
\begin{equation}
  \label{eq:295}
  \begin{split}
    &\frac{\partial t}{\partial
      g_{\alpha\beta}}=\frac{1}{n} t g^{\alpha\beta},\quad \frac{\partial
      g_{\alpha\beta}}{\partial t}=\hat g_{\alpha\beta},\quad 
    g_{\alpha\beta}\frac{\partial \hat g_{\gamma\delta}}{\partial
      g_{\alpha\beta}}=0= g^{\alpha\beta}
    \frac{\partial  g_{\alpha\beta}}{\partial \hat g_{\gamma\delta}}, \\
    &\frac{\partial \hat g_{\alpha\beta}}{\partial g_{\gamma\delta}}=\frac{1}{t}
    J\indices{_{\alpha\beta}^{\gamma\delta}},\
    \frac{\partial g_{\alpha\beta}}{\partial \hat g_{\gamma\delta}}=t
    J\indices{_{\alpha\beta}^{\gamma\delta}},\ J\indices{_{\alpha\beta}^{\gamma\delta}}=
    \frac{1}{2}(\delta^\gamma_\alpha\delta^\delta_\beta
    +\delta^\delta_\alpha\delta_\beta^\gamma-\frac{2}{n} g^{\gamma\delta} g_{\alpha\beta}). 
    \end{split}
\end{equation}
with
\begin{equation}
  \label{eq:297}
  \begin{split}
    J\indices{_{\alpha\beta}^{\gamma\delta}}J\indices{_{\gamma\delta}^{\rho\sigma}}
    =J\indices{_{\alpha\beta}^{\rho\sigma}},\quad g^{\alpha\beta}J\indices{_{\alpha\beta}^{\gamma\delta}}=0=
    J\indices{_{\alpha\beta}^{\gamma\delta}}g_{\gamma\delta},\\
    J\indices{_{\alpha\beta}^{\alpha\beta}}=\frac{n(n+1)-2}{2},\quad J\indices{_{\alpha}^{\beta\gamma\alpha}}
    =\frac{n(n+1)-2}{2n}g^{\beta\gamma}.
  \end{split}
\end{equation}
Let $\hat G^{\alpha\beta\gamma\delta}, \hat
J\indices{_{\alpha\beta}^{\gamma\delta}}$ be the same expressions as $\hat
G^{\alpha\beta\gamma\delta}, \hat J\indices{_{\alpha\beta}^{\gamma\delta}}$ with
$g_{\alpha\beta}$ and its inverse replaced by $\hat g_{\alpha\beta}$ and its
inverse, including for raising and lowering indices. In particular $\hat
J\indices{_{\alpha\beta}^{\gamma\delta}}=
J\indices{_{\alpha\beta}^{\gamma\delta}}$. In the new parametrization, the
supermetric becomes
\begin{equation}
  \label{eq:296}
  ds^2_{\mathrm{GL}(n)}=\hat g^{\alpha\gamma}\hat g^{\beta\delta}\text{d}\hat g_{\alpha\beta}\text{d}\hat g_{\gamma\delta}
  +n \text{d}(\ln t)^2=\hat J^{\alpha\beta\gamma\delta}\text{d}\hat g_{\alpha\beta}\text{d}\hat g_{\gamma\delta}
  +n (\text{d}\ln t)^2,
\end{equation}
so that the metric on $\mathrm{SL}(n,\mathbb R)/\mathrm{SO}(n,\mathbb R)$ is
given by
\begin{equation}
  \label{eq:300}
  ds^2_{\mathrm{SL}(n)}=\hat J^{\alpha\beta\gamma\delta}\text{d}\hat g_{\alpha\beta}\text{d}\hat g_{\gamma\delta}
  =J^{\alpha\beta\gamma\delta}\text{d} g_{\alpha\beta}\text{d} g_{\gamma\delta}.
\end{equation}
It follows from \eqref{eq:296} that the associated Laplacian satisfies
\begin{equation}
  \label{eq:227}
  \Delta_{\mathrm{GL}(n)}=\Delta_{\mathrm{SL}(n)}+\frac{1}{n}(t\frac{\partial}{\partial t})^2, 
  \end{equation}
and is thus explicitly given by
\begin{equation}
  \begin{split}
  \Delta_{\mathrm{SL}(n)}=J_{\alpha\beta\gamma\delta}
  \frac{\partial^2}{\partial g_{\alpha\beta}\partial g_{\gamma\delta}}
  +\frac{n(n+1)-2}{2n}g_{\alpha\beta}\frac{\partial}{\partial g_{\alpha\beta}}\\
  =\hat J_{\alpha\beta\gamma\delta}
  \frac{\partial^2}{\partial \hat g_{\alpha\beta}\partial \hat g_{\gamma\delta}}.
\end{split}
\label{eq:263} 
\end{equation}
It follows that
\begin{equation}
  \label{eq:264}
  \begin{split}
  \Delta_{\mathrm{SL}(n)}f_s(n;\hat g)&=\frac{s(n-1)(2s-n)}{2n} f_s(n;\hat g),\\
  \Delta_{\mathrm{SL}(n)}\xi(z,n;\hat g)&=\frac{z(n-1)(z-n)}{4n} \xi(z,n;\hat g).
\end{split}
\end{equation}

In particular, setting $n=2$, yields $\Delta_{\mathrm{SL}(2)}f_s(2;\hat
g)=\frac{s(s-1)}{2} f_s(2;\hat g)$. Since
\begin{multline}
  \label{eq:276a}
  \text{d}s^2_{\mathrm{SL}(2)}=(\hat g^{11})^2(\text{d}\hat g_{11})^2+4\hat g^{11}\hat g^{12}\text{d}\hat g_{11} \text{d}\hat g_{12}
  +2(\hat g^{12})^2\text{d}\hat g_{11} \text{d}\hat g_{22}\\+2[\hat g^{11}\hat g^{22}+(\hat g^{12})^2](\text{d}\hat g_{12})^2
  +4\hat g^{12}\hat g^{22}\text{d}\hat g_{12} \text{d}\hat g_{22}+(\hat g^{22})^2(\text{d}\hat g_{22})^2,
\end{multline}
reduces to
\begin{equation}
  \label{eq:306}
  \text{d}s^2_{\mathrm{SL}(2)}=\frac{2}{\tau_2^2}(\text{d}\tau^2_1+\text{d}\tau^2_2),
\end{equation}
when using the parametrization in the last two lines of \eqref{eq:99},
it follows that
\begin{equation}
  \label{eq:307}
  \Delta_{\mathrm{SL}(2)}=\frac{\tau^2_2}{2}(\partial^2_{\tau_1}+\partial^2_{\tau_2})
    =\frac 12 \Delta_{\mathbb H}. 
\end{equation}

In order to work out the supermetric and the Laplacian in ADM parametrization,
we start with the decomposition
\begin{multline}
  \label{eq:313}
  g^{\alpha\gamma}g^{\beta\delta}\text{d}g_{\alpha\beta}\text{d}g_{\gamma\delta}
  =g^{\iota\xi}g^{\kappa\lambda}\text{d}g_{\iota\kappa}\text{d}g_{\xi\lambda}
  +4g^{\iota\xi}g^{\kappa,q+1}\text{d}g_{\iota\kappa}dg_{\xi,q+1}
  \\+2g^{\iota,q+1}g^{\kappa,q+1}\text{d}g_{\iota\kappa}\text{d}g_{q+1,q+1}
  +2(g^{\iota\kappa}g^{q+1,q+1}+g^{\iota,q+1}g^{\kappa,q+1})\text{d}g_{\iota,q+1}\text{d}g_{\kappa,q+1}
  \\+4g^{\iota,q+1}g^{q+1,q+1}\text{d}g_{\iota,d+1}\text{d}g_{q+1,q+1}
  +(g^{q+1,q+1})^2(\text{d}g_{q+1,q+1})^2.
\end{multline}

In terms of the ADM parametrization \eqref{eq:214}, the supermetric becomes
\begin{multline}
  \label{eq:267}
  \text{d}s^2_{\mathrm{GL}(q+1),\mathrm{ADM}}=G^{\alpha\beta\gamma\delta}_{\mathrm{ADM}}\text{d}g^{\mathrm{ADM}}_{\alpha\beta}
  \text{d}g^{\mathrm{ADM}}_{\gamma\delta}\\
= {}^{(q)}G^{\iota\kappa\lambda\xi}\text{d}h_{\iota\kappa}\text{d}h_{\lambda\xi}+\frac{2}{N^2}
h_{\iota\kappa} \text{d}N^\iota dN^\kappa+4(\text{d}\ln N)^2. 
\end{multline}
Since $g_{\alpha\beta}^{\mathrm{ADM}}=(N\sqrt{h})^2
\hat g^{\mathrm{ADM}}_{\alpha\beta}$, it follows from \eqref{eq:296} that
\begin{multline}
  \label{eq:308}
  \text{d}s^2_{\mathrm{SL}(q+1),\mathrm{ADM}}=\hat
  G^{\alpha\beta\gamma\delta}_{\mathrm{ADM}}\text{d}\hat g^{\mathrm{ADM}}_{\alpha\beta}\text{d}\hat
  g^{\mathrm{ADM}}_{\gamma\delta}
  =
  {}^{(q)}G^{\iota\kappa\lambda\xi}\text{d}h_{\iota\kappa}\text{d}h_{\lambda\xi}\\
  +\frac{2}{N^2}
 h_{\iota\kappa} \text{d}N^\iota
  \text{d}N^\kappa+4(\text{d}\ln N)^2
  -\frac{4}{q+1}(\text{d}\ln N+ \frac 12 \text{d}\ln h)^2. 
\end{multline}
Since $h_{\iota\kappa}=(h)^{\frac{1}{q}}\hat
  h_{\iota\kappa}$ it follows by using again \eqref{eq:296} that
\begin{multline}
    \label{eq:309}
    \text{d}s^2_{\mathrm{SL}(q+1),\mathrm{ADM}}=
   {}^{(q)} \hat G^{\iota\kappa\lambda\xi}\text{d}\hat{h}_{\iota\kappa}\text{d}\hat h_{\lambda\xi}\\
    +\frac{2}{N^2}
    h_{\iota\kappa} \text{d}N^\iota
    \text{d}N^\kappa+4(\text{d}\ln N)^2
    -\frac{4}{q+1}(\text{d}\ln N+ \frac 12 \text{d}\ln h)^2
    +\frac{1}{q}(\text{d}\ln h)^2.
\end{multline}
Changing variables to $b,\text{V}_q$ with
$N=b(\text{V}_q)^{\frac{1}{q}}$, $h=(\text{V}_q)^2$, gives
\begin{equation}
  \label{eq:310}
  ds^2_{\mathrm{SL}(q+1),\mathrm{ADM}}=
  {}^{(q)}\hat G^{\iota\kappa\lambda\xi}\text{d}\hat{h}_{\iota\kappa}\text{d}\hat h_{\lambda\xi}+\frac{2}{b^2}
  \hat h_{\iota\kappa} \text{d}N^\iota d N^\kappa
  +\frac{4q}{q+1}( \text{d}\ln b)^2.
\end{equation}
The determinant is given by
\begin{equation}
  \label{eq:317}
  \hat G_{\rm ADM}={}^{(q)}\hat G\frac{2^{q+2}}{b^{2(q+1)}}\frac{4q}{q+1},
\end{equation}
where ${}^{(q)}\hat G$ does not depend on either $b$ or $N^\iota$. The
associated Laplacian is then given by
\begin{equation}
  \label{eq:311}
  \Delta_{\mathrm{SL}(q+1),\mathrm{ADM}}=\Delta_{\mathrm{SL}(q)}
  +\frac{b^2\hat h^{\iota\kappa}}{2}\frac{\partial}{\partial N^\iota}
  \frac{\partial}{\partial N^\kappa}+\frac{q+1}{4q}\Big[b^2
  \frac{\partial^2}{\partial b^2}-(q-1)b\frac{\partial}{\partial b}\Big]. 
\end{equation}
When taking into account that $\Delta_{\mathrm{SL}(1),\mathrm{ADM}}$ vanishes
and identifying $b=\tau_2$, $N^1=\tau_1$, this formula correctly reduces to the
SL$(2,\mathbb Z)$ Laplacian \eqref{eq:307}. It may be used recursively to
construct an explicit expression of the SL$(q+1,\mathbb Z)$ Laplacian by using a
similar parametrization of $\hat h_{\iota\kappa}$ in terms of variables
$b_{(q)},N^{\iota_{d-1}}_{(q)}$.

\section{Trigonometric expansions}
\label{AppB}

In this appendix, we derive expansions for the Epstein zeta function and the
Eisenstein series in terms of trigonometric functions.

The Epstein zeta function is defined as
\begin{equation}
\label{E27.1}
  \zeta(s;a_1,...,a_p)=\sideset{}{'}\sum_{n_1,...,n_p\in\mathbb{Z}^p}
  \frac{1}{(a_1n_1^2+...+a_pn_p^2)^s},
\end{equation}
and satisfies the functional relation
\begin{equation}
\label{E27.2}
  \Gamma(s)\zeta(s;a_1,...,a_p)=\frac{\pi^{2s-\frac{p}{2}}}{\sqrt{a_1...a_p}}
  \Gamma\bigg(\frac{p}{2}-s\bigg)
  \zeta\bigg(\frac{p}{2}-s;\frac{1}{a_1},...,\frac{1}{a_p}\bigg).
\end{equation}
The real analytic Eisenstein series $f_{s}(\tau,\bar{\tau})$ and the Epstein
zeta function are related by
\begin{equation}
\label{E27.3}
  f_{s}(i\tau_2,-i\tau_2)
  =\sideset{}{'}\sum_{(n,m)\in\mathbb{Z}^2}\frac{\tau_2^s}{(n^2 \tau_2^2+m^2)^s}
  =\tau_2^s\zeta(s;\tau_2^2,1).
\end{equation}

The Fourier transform of the real analytic Eisenstein series in \eqref{eq:72},
in the particular case of purely imaginary $\tau$, may be directly obtained from
a Sommerfeld-Watson transform, which we briefly review first. Consider the
series
\begin{equation}
\label{E27.4}
S=\sum_{m\in\mathbb{Z}}f(m),
\end{equation}
and the complex function 
\begin{equation}
\label{E27.5}
F(z)=\pi f(z)\cot(\pi z),
\end{equation}
having simple poles in $z=m$, $m\in \mathbb{Z}$, with residues
\begin{equation}
\label{E27.6}
\mathrm{Res}\big[F(z)\big]|_{z=m}=f(m).
\end{equation}
Assuming that
\begin{equation}
\label{E27.7}
f(z)\stackrel{\abs{z}\rightarrow\infty}{\xrightarrow{\hspace*{1.3cm}}}
\frac{1}{\abs{z}^{1+\epsilon}},\hspace{1cm}\epsilon > 0,
\end{equation}
the integral along a circle $C_{R}$ of radius $R$ centred at the origin vanishes
as $R\rightarrow \infty$
\begin{equation}
\label{E27.8}
0=\oint_{C_{\infty}} F(z)dz=2\pi i\bigg[\sum_{m\in\mathbb{Z}}
\mathrm{Res}\big[F(z)\big]|_{z=m}
+\sum_{i}\mathrm{Res}\big[F(z)\big]|_{z=z_i}\bigg],
\end{equation}
where we used the residue theorem and where $z_i$ are the other poles of $F(z)$, at
points different from $z=m$. It follows that
\begin{equation}
\label{E27.9}
\sum_{m\in\mathbb{Z}}f(m)=-\sum_{i}\mathrm{Res}\big[F(z)\big]|_{z=z_i}.
\end{equation}
In the particular case of the series,
\begin{align}
\label{E27.10}
S(a)=\sum_{m\in\mathbb{Z}}\frac{1}{a^2+m^2}.
\end{align}
one considers the function $F(z)=\frac{\pi}{z^2+a^2}\cot(\pi z)$ which has two
other simple poles in $z=\pm i a$ with residues
\begin{equation}
\label{E27.11}
\mathrm{Res}\left.\left[\frac{\pi}{z^2+a^2}
    \cot(\pi z)\right]\right|_{z=\pm  i a}
=-\frac{\pi}{2a}\coth(\pi a).
\end{equation}
Using \eqref{E27.9}, we thus have
\begin{equation}
\label{E27.12}
\sum_{m\in\mathbb{Z}}\frac{1}{a^2+m^2}=\frac{\pi}{a}\coth(\pi a), 
\end{equation}
which implies that 
\begin{multline}
\label{E27.13}
\sum_{m\in\mathbb{Z}}\frac{1}{(a^2+m^2)^s}=\frac{(-)^{s-1}}{2^{s-1}(s-1)!}
\left(\frac{1}{a}\frac{\D}{\D
    a}\right)^{s-1}\sum_{m\in\mathbb{Z}}\frac{1}{a^2+m^2}\\
=\frac{(-)^{s-1}}{2^{s-1}(s-1)!}
  \left(\frac{\D}{\D x}\right)^{s-1}\bigg(\coth(\pi \sqrt{2x})\frac{\pi}{\sqrt{2x}}\bigg)
  \\
  =  \sum_{m=0}^{s-1}
  \frac{\pi^{2s-2m-\frac{1}{2}}(-)^{m}\Gamma(s-\frac 12-m)}
    {2^m\Gamma(m+1)\Gamma(s-m)(\pi a)^{2s-2m-1}}\frac{d^m}{dx^m}
\coth (\pi\sqrt{2x}),                                           
\end{multline}
where $x=\frac 12 a^2$, and we used the Leibniz rule together with
\begin{equation}
  \label{eq:100}
  \frac{\text{d}^n}{\text{d}x^n}x^{-\frac 12}=(-)^{n}\frac{\Gamma(\frac 12 +n) x^{-\frac 12 -n}}{\sqrt \pi},
\end{equation}
for $n=s-1-m$ have been used. In order to evaluate $\frac{\text{d}^n}{\text{d}x^n} \coth
(\pi\sqrt{2x})$ we use formula (3n) of \cite{10.2307/2310518}, which gives for
$m\geq 1$, 
\begin{equation}
  \label{eq:104}
  \frac{\text{d}^m}{\text{d}x^m}(f\circ g)(x)=\sum_{r=1}^m (f^{(r)}\circ g)
  (\sum^r_{u=1}\frac{(-)^{r-u}g^{r-u}}{u!(r-u)!}
    \frac{\text{d}^m}{\text{d}x^m}g^u),
\end{equation}
with $f(x)=\coth x$ and $g=\pi\sqrt{2x}$ and where the exponent $(r)$
denotes the $r$-th derivative of the function with
respect to its argument, and also 
\begin{equation}
  \label{eq:102}
  \frac{\text{d}^m}{dx^m}x^{\frac u2}=\frac{\Gamma(1+\frac u2) x^{\frac u2 -m}}{\Gamma(1+\frac u2 -m)}.
\end{equation}
When used together with the reduplication formula \eqref{eq:57} at $z=u+1$, we
get 
\begin{multline}
  \label{eq:105}
  \sum_{m\in\mathbb{Z}}\frac{1}{(a^2+m^2)^s}=\frac{\pi^{2s}}{(\pi
    a)^{2s-1}}\Big[\frac{\Gamma(s-\frac 12)}{\pi^{\frac 12}\Gamma(s)}\coth (\pi
  a)+\sum_{m=1}^{s-1}
  \frac{(-)^m\Gamma(s-m-\frac 12)}{\Gamma(m+1)\Gamma(s-m)}
  \\\sum_{r=1}^m
  (\pi a)^{r}\coth ^{(r)}(\pi
  a)(\sum_{u=1}^r\frac{(-)^{r-u}}{2^u\Gamma(\frac{{u+1}}2)\Gamma(r-u+1)
    \Gamma(1+\frac u2-m)})\Big].
\end{multline}
Note that the higher derivatives of $\coth(x)$ may be further expressed in terms of
polynomials of $\coth(x)$. Even though we will not explicitly need it here,
other useful relations are obtained by starting from \eqref{E27.12} at
$a=il\tau$ to show that
\begin{equation}
  \label{eq:109}
  \sum_{m\in\mathbb{Z}}\frac{1}{(m+l\tau)}=
  \pi  \cot(\pi l \tau)
\end{equation}
and then also 
\begin{equation}
  \label{eq:106}
  \sum_{m\in\mathbb{Z}}\frac{1}{(m+l\tau)^k}=
  \frac{(-)^{k-1}}{(k-1)!}\frac{\pi}{l^{k-1}}\frac{\text{d}^{k-1}}{\text{d}\tau^{k-1}}
  \cot(\pi l \tau).
\end{equation}

The Epstein zeta function $\zeta(s;\tau_2^2,1)$ can be decomposed as 
\begin{equation}
\label{E27.14}
\zeta(s;\tau_2^2,1)=\sideset{}{'}\sum_{m\in\mathbb{Z}}
\frac{1}{m^{2s}}+\sideset{}{'}\sum_{n\in\mathbb{Z}}
\sum_{m\in\mathbb{Z}}\frac{1}{(n^2\tau_2^2+m^2)^s}.
\end{equation}
Applying the first of \eqref{E27.13} to the last term in \eqref{E27.14} with $a=n\tau_2$
gives,
\begin{equation}
\label{E27.15}
\zeta(s;\tau_2^2,1)=2\zeta(2s)+\frac{(-)^{s-1}}{2^{s-1}(s-1)!}
\sideset{}{'}\sum_{n\in\mathbb{Z}}\frac{1}{n^{2s-1}}
\bigg[\left(\frac{1}{\tau_2}\frac{\D}{\D \tau_2}\right)^{s-1}\frac{\pi}{\tau_2}
\coth(\pi n \tau_2)\bigg],
\end{equation}
Similarly, the analogous expansion
for the real analytic Eisenstein series, which we give here without proof, is
\begin{equation}
  \label{E27.16}
  \tau_2^{-s}f_s(\tau,\bar\tau)=2\zeta(2s)
  +\left[\frac{i(-)^{s-1}}{2^s(s-1)!}\sideset{}{'}
    \sum_{n\in\mathbb{Z}}\frac{1}{n^{2s-1}}
    \left(\frac{1}{\tau_2}\frac{\D}{\D \tau_2}\right)^{s-1}
    \frac{\pi}{\tau_2}\cot(\pi n \tau)+\mathrm{c.c.}\right].
\end{equation}
The right hand side of \eqref{E27.16} correctly reduces to \eqref{E27.15} when
$\tau=i\tau_2$.
Alternatively, one may use \eqref{eq:105}, to write
\begin{multline}
  \label{eq:107}
  \zeta(s;\tau_2^2,1)=2\zeta(2s)+\sideset{}{'}\sum_{n\in\mathbb{Z}}
  \frac{\pi^{2s}}{(\pi
    n\tau_2)^{2s-1}}\Big[\frac{\Gamma(s-\frac 12)}{\pi^{\frac 12}\Gamma(s)}\coth (\pi
  n\tau_2)+\sum_{m=1}^{s-1}
  \frac{(-)^m\Gamma(s-m-\frac 12)}{\Gamma(m+1)\Gamma(s-m)}
  \\\sum_{r=1}^m
  (\pi n\tau_2)^{r}\coth ^{(r)}(\pi
  n\tau_2)(\sum_{u=1}^r\frac{(-)^{r-u}}{2^u\Gamma(\frac{{u+1}}2)\Gamma(r-u+1)
    \Gamma(1+\frac u2-m)})\Big],
\end{multline}
and also
\begin{multline}
  \label{eq:108}
  \tau_2^{-s}f_s(\tau,\bar\tau)=2\zeta(2s)+\sideset{}{'}\sum_{n\in\mathbb{Z}}
  \frac{\pi^{2s}}{2(\pi
    n\tau_2)^{2s-1}}\Big[\frac{\Gamma(s-\frac 12)}{\pi^{\frac 12}\Gamma(s)}i\cot (\pi
  n\tau)+\sum_{m=1}^{s-1}
  \frac{(-)^m\Gamma(s-m-\frac 12)}{\Gamma(m+1)\Gamma(s-m)}
  \\\sum_{r=1}^m
  (i\pi n\tau_2)^{r}i\cot^{(r)}(\pi
  n\tau)(\sum_{u=1}^r\frac{(-)^{r-u}}{2^u\Gamma(\frac{{u+1}}2)\Gamma(r-u+1)
    \Gamma(1+\frac u2-m)})+{\rm c.c.}\Big].
\end{multline}

Let us briefly consider in more detail the simplest case of $d=3$, $s=2$, which
is directly relevant for standard Casimir physics. In this case, \eqref{eq:108}
gives
\begin{equation}
  \label{eq:98}
  f_2(\tau,\bar\tau)=\frac{\pi^4\tau_2^2}{2}\Big[\frac{2}{45}+
  \sum_{l\in \mathbb N^*}\big(
  \frac{i\cot{\pi l\tau }}{(\pi l\tau_2)^3}
  -\frac{1}{(\pi l\tau_2)^2\sin^2{\pi l\tau}}
  +{\rm c.c.}\big)\Big], 
\end{equation}
while the partition function is
\begin{equation}
  \label{eq:97}
  \boxed{\ln\mathcal Z_3(\tau,\bar\tau)= \frac{\pi^2L_1L_2\tau_2}{4L_3^2}
    \Big[\frac{2}{45}+\sum_{l\in \mathbb N^*}\big(
    \frac{i\cot{\pi l\tau }}{(\pi l\tau_2)^3}-\frac{1}{(\pi l\tau_2)^2\sin^2{\pi l\tau}}
    +{\rm c.c.}\big)\Big]}.
\end{equation}
This could also have been obtained starting from the canonical approach. Indeed,
the integral in \eqref{eq:70} can be done by direct integrations by parts so
that one finds instead of \eqref{eq:71},
\begin{multline}
  \label{eq:89}
  \ln\mathcal Z_3(\tau,\bar\tau)= \frac{\pi^2L_1L_2\tau_2}{90L_3^2}
  +\frac{L_1L_2}{2\pi (L_3\tau_2)^2}\zeta(3)\\
  +\frac{L_1L_2}{2\pi\beta^2}\sum_{n_3\in\mathbb{N}^*}\sum_{l\in\mathbb{N}^*}
  \Big[\frac{(1+2\pi ln_3\tau_2)}{l^3}
  ({e^{2\pi i n_3 l\tau}}+{e^{-2\pi i n_3 l\bar\tau}})\Big]. 
\end{multline}
The above result now follows when using that $\zeta(3)=\sum_{l\in \mathbb
  N^*}(\frac{1}{2} l^{-3}+\frac{1}{2} l^{-3})$, and that
\begin{equation}
  \label{eq:96}
  \begin{split}
    \sum_{n_3\in\mathbb N^*} e^{2\pi i n_3 l\tau}+\frac{1}{2}
    =\frac{1}{1-e^{2\pi i l\tau}}-\frac{1}{2}=
  \frac{i}{2}\cot{\pi l \tau},\\
  \sum_{n_3\in\mathbb N^*} 2\pi n_3 e^{2\pi i n_3 l\tau}
  =\frac{1}{il}\frac{\text{d}}{\text{d}\tau}\frac{1}{1-e^{2\pi il\tau}}
  =-\frac{\pi}{2\sin^2{\pi l\tau}}.
\end{split}
\end{equation}
If the chemical potential $\alpha$ vanishes,
$\tau=i\tau_2=i\frac{\beta}{L_3}$,
we now have
\begin{equation}
  \label{eq:103}
  \zeta(2;\tau_2^2,1)=
  \sideset{}{'}\sum_{l,m}\frac{1}{(l^2+m^2\tau_2^2)^2}=\pi^4\Big[\frac{1}{45}
  +\sum_{l\in \mathbb N^*}\big(
  \frac{\coth{\pi l\tau_2 }}{(\pi l \tau_2)^3}+\frac{1}{(\pi l \tau_2)^2
    \sinh^2{\pi l \tau_2}}\big)\Big], 
\end{equation}
and
\begin{equation}
  \label{eq:101}
  \ln{\mathcal Z}_3(\tau_2)= \frac{L_1L_2\tau_2}{2\pi^2L_3^2}\zeta(2;\tau_2^2,1).
\end{equation}
which is the form under which the result appears in the Casimir literature when
subtraction is done only for the term containing the vacuum energy.

\vfill
\pagebreak

\providecommand{\href}[2]{#2}\begingroup\raggedright\endgroup


\begin{thebibliography}{10}

\bibitem{Obers:1999um}
N.~Obers and B.~Pioline, \emph{{Eisenstein series and string thresholds}},
  \href{https://doi.org/10.1007/s002200050022}{\emph{Commun. Math. Phys.}
  {\bfseries 209} (2000) 275}
  [\href{https://arxiv.org/abs/hep-th/9903113}{{\ttfamily hep-th/9903113}}].

\bibitem{Petropoulos2012}
P.M.~Petropoulos and P.~Vanhove, \emph{Gravity, strings, modular and
  quasimodular forms},  \href{https://arxiv.org/abs/1206.0571v1}{{\ttfamily
  1206.0571v1}}.

\bibitem{Williams:2014udz}
F.L.~Williams and K.~Kirsten, eds., \emph{{A Window into Zeta and Modular
  Physics}}, Cambridge University Press (2014).

\bibitem{fleig_gustafsson_kleinschmidt_persson_2018}
P.~Fleig, H.P.A.~Gustafsson, A.~Kleinschmidt and D.~Persson, \emph{Eisenstein
  Series and Automorphic Representations: With Applications in String Theory},
  Cambridge Studies in Advanced Mathematics, Cambridge University Press (2018),
  \href{https://doi.org/10.1017/9781316995860}{10.1017/9781316995860}.

\bibitem{Bluemlein2019}
J.~Bl{\"u}mlein, C.~Schneider and P.~Paule, eds., \emph{Elliptic Integrals,
  Elliptic Functions and Modular Forms in Quantum Field Theory}, Texts \&
  Monographs in Symbolic Computation, Springer International Publishing (2019).

\bibitem{Ambjorn:1981xw}
J.~Ambjorn and S.~Wolfram, \emph{{Properties of the Vacuum. 1. Mechanical and
  Thermodynamic}},
  \href{https://doi.org/10.1016/0003-4916(83)90065-9}{\emph{Annals Phys.}
  {\bfseries 147} (1983) 1}.

\bibitem{Plunien:1986ca}
G.~Plunien, B.~Muller and W.~Greiner, \emph{{The Casimir Effect}},
  \href{https://doi.org/10.1016/0370-1573(86)90020-7}{\emph{Phys. Rept.}
  {\bfseries 134} (1986) 87}.

\bibitem{Gies:2003cv}
H.~Gies, K.~Langfeld and L.~Moyaerts, \emph{{Casimir effect on the worldline}},
  \href{https://doi.org/10.1088/1126-6708/2003/06/018}{\emph{JHEP} {\bfseries
  06} (2003) 018} [\href{https://arxiv.org/abs/hep-th/0303264}{{\ttfamily
  hep-th/0303264}}].

\bibitem{Plunien:1987fr}
G.~Plunien, B.~Muller and W.~Greiner, \emph{{Casimir Energy at Finite
  Temperature}}, {\emph{Physica} {\bfseries 145A} (1987) 202}.

\bibitem{doi:10.1142/2065}
E.~Elizalde, S.D.~Odintsov, A.~Romeo, A.A.~Bytsenko and S.~Zerbini, \emph{Zeta
  Regularization Techniques with Applications}, World Scientific (1994),
  \href{https://doi.org/10.1142/2065}{10.1142/2065},
  [\href{https://arxiv.org/abs/https://www.worldscientific.com/doi/pdf/10.1142/2065}{{\ttfamily
  https://www.worldscientific.com/doi/pdf/10.1142/2065}}].

\bibitem{Bordag:2009zzd}
M.~Bordag, G.L.~Klimchitskaya, U.~Mohideen and V.M.~Mostepanenko,
  \emph{{Advances in the Casimir effect}}, vol.~145 of
  \emph{Int.Ser.Monogr.Phys.}, Oxford University Press (2009).

\bibitem{Cappelli:1988vw}
A.~Cappelli and A.~Coste, \emph{{On the Stress Tensor of Conformal Field
  Theories in Higher Dimensions}},
  \href{https://doi.org/10.1016/0550-3213(89)90414-8}{\emph{Nucl. Phys. B}
  {\bfseries 314} (1989) 707}.

\bibitem{Belavin:1984vu}
A.A.~Belavin, A.M.~Polyakov and A.B.~Zamolodchikov, \emph{{Infinite conformal
  symmetry in two-dimensional quantum field theory}},
  \href{https://doi.org/10.1016/0550-3213(84)90052-X}{\emph{Nucl. Phys.}
  {\bfseries B241} (1984) 333}.

\bibitem{Cardy:1986ie}
J.L.~Cardy, \emph{{Operator Content of Two-Dimensional Conformally Invariant
  Theories}}, \href{https://doi.org/10.1016/0550-3213(86)90552-3}{\emph{Nucl.
  Phys.} {\bfseries B270} (1986) 186}.

\bibitem{CAPPELLI1987445}
A.~Cappelli, C.~Itzykson and J.-B.~Zuber, \emph{Modular invariant partition
  functions in two dimensions},
  \href{https://doi.org/https://doi.org/10.1016/0550-3213(87)90155-6}{\emph{Nuclear
  Physics B} {\bfseries 280} (1987) 445 }.

\bibitem{Cardy:1991kr}
J.L.~Cardy, \emph{{Operator content and modular properties of higher
  dimensional conformal field theories}},
  \href{https://doi.org/10.1016/0550-3213(91)90024-R}{\emph{Nucl. Phys.}
  {\bfseries B366} (1991) 403}.

\bibitem{Dolan1998}
L.~Dolan and C.R.~Nappi, \emph{A modular invariant partition function for the
  five-brane}, \href{https://doi.org/10.1016/S0550-3213(98)00537-9}{\emph{Nucl.
  Phys. B} {\bfseries 530} (1998) 683}
  [\href{https://arxiv.org/abs/hep-th/9806016}{{\ttfamily hep-th/9806016}}].

\bibitem{Shaghoulian:2015kta}
E.~Shaghoulian, \emph{{Modular forms and a generalized Cardy formula in higher
  dimensions}}, \href{https://doi.org/10.1103/PhysRevD.93.126005}{\emph{Phys.
  Rev. D} {\bfseries 93} (2016) 126005}
  [\href{https://arxiv.org/abs/1508.02728}{{\ttfamily 1508.02728}}].

\bibitem{Shaghoulian:2016gol}
E.~Shaghoulian, \emph{{Modular Invariance of Conformal Field Theory on
  $S^1×S^3$ and Circle Fibrations}},
  \href{https://doi.org/10.1103/PhysRevLett.119.131601}{\emph{Phys. Rev. Lett.}
  {\bfseries 119} (2017) 131601}
  [\href{https://arxiv.org/abs/1612.05257}{{\ttfamily 1612.05257}}].

\bibitem{Horowitz:2017ifu}
G.T.~Horowitz and E.~Shaghoulian, \emph{{Detachable circles and
  temperature-inversion dualities for CFT$_{d}$}},
  \href{https://doi.org/10.1007/JHEP01(2018)135}{\emph{JHEP} {\bfseries 01}
  (2018) 135} [\href{https://arxiv.org/abs/1709.06084}{{\ttfamily
  1709.06084}}].

\bibitem{Barnich:2019xhd}
G.~Barnich, \emph{{Black hole entropy from nonproper gauge degrees of freedom:
  The charged vacuum capacitor}},
  \href{https://doi.org/10.1103/PhysRevD.99.026007}{\emph{Phys. Rev.}
  {\bfseries D99} (2019) 026007}.

\bibitem{GlennBarnich}
G.~Barnich and M.~Bonte, \emph{{Soft degrees of freedom, Gibbons-Hawking
  contribution and entropy from Casimir effect}},  in \emph{Proceedings, 11th
  International Symposium on Quantum Theory and Symmetries (QTS2019): Montreal,
  Canada, July, 1-5, 2019}, M.B.~Paranjape, R.~MacKenzie, Z.~Thomova,
  P.~Winternitz and W.~Witczak-Krempa, eds., CRM Series in Mathematical
  Physics, 2021, \href{https://doi.org/10.1007/978-3-030-55777-5}{DOI}
  [\href{https://arxiv.org/abs/1912.12698}{{\ttfamily 1912.12698}}].

\bibitem{Deutsch1979}
D.~Deutsch and P.~Candelas, \emph{Boundary effects in quantum field theory},
  \href{https://doi.org/10.1103/PhysRevD.20.3063}{\emph{Phys. Rev. D}
  {\bfseries 20} (1979) 3063}.

\bibitem{Alessio:2020okv}
F.~Alessio and G.~Barnich, \emph{{Modular invariance in finite temperature
  Casimir effect}}, \href{https://doi.org/10.1007/JHEP10(2020)134}{\emph{JHEP}
  {\bfseries 10} (2020) 134}
  [\href{https://arxiv.org/abs/2007.13334}{{\ttfamily 2007.13334}}].

\bibitem{Alessio2021}
F.~Alessio, G.~Barnich and M.~Bonte, \emph{{Gravitons in a Casimir box}},
  \href{https://doi.org/10.1007/JHEP02(2021)216}{\emph{JHEP} {\bfseries 02}
  (2021) 216} [\href{https://arxiv.org/abs/2011.14432}{{\ttfamily
  2011.14432}}].

\bibitem{Lutken:1988ge}
C.~Lutken and F.~Ravndal, \emph{{A Symmetry in the Finite Temperature Casimir
  Effect}}, \href{https://doi.org/10.1088/0305-4470/21/16/002}{\emph{J. Phys.
  A} {\bfseries 21} (1988) L793}.

\bibitem{PhysRevD.40.4191}
F.~Ravndal and D.~Tollefsen, \emph{{Temperature inversion symmetry in the
  Casimir effect}}, \href{https://doi.org/10.1103/PhysRevD.40.4191}{\emph{Phys.
  Rev. D} {\bfseries 40} (1989) 4191}.

\bibitem{Wotzasek:1989gq}
C.~Wotzasek, \emph{{On the Casimir Effect and the Temperature Inversion
  Symmetry}}, \href{https://doi.org/10.1088/0305-4470/23/9/023}{\emph{J. Phys.
  A} {\bfseries 23} (1990) 1627}.

\bibitem{Ravndal:1990mu}
F.~Ravndal and C.~Wotzasek, \emph{{Temperature inversion symmetry in the
  Gross-Neveu model}},
  \href{https://doi.org/10.1016/0370-2693(90)91253-8}{\emph{Phys. Lett. B}
  {\bfseries 249} (1990) 266}.

\bibitem{Brown:1969na}
L.S.~Brown and G.J.~Maclay, \emph{{Vacuum stress between conducting plates: An
  Image solution}}, \href{https://doi.org/10.1103/PhysRev.184.1272}{\emph{Phys.
  Rev.} {\bfseries 184} (1969) 1272}.

\bibitem{ROBERGE1986734}
A.~Roberge and N.~Weiss, \emph{{Gauge theories with imaginary chemical
  potential and the phases of QCD}},
  \href{https://doi.org/https://doi.org/10.1016/0550-3213(86)90582-1}{\emph{Nuclear
  Physics B} {\bfseries 275} (1986) 734 }.

\bibitem{Alford:1998sd}
M.G.~Alford, A.~Kapustin and F.~Wilczek, \emph{{Imaginary chemical potential
  and finite fermion density on the lattice}},
  \href{https://doi.org/10.1103/PhysRevD.59.054502}{\emph{Phys. Rev. D}
  {\bfseries 59} (1999) 054502}
  [\href{https://arxiv.org/abs/hep-lat/9807039}{{\ttfamily hep-lat/9807039}}].

\bibitem{PhysRevD.75.025003}
F.~Karbstein and M.~Thies, \emph{How to get from imaginary to real chemical
  potential}, \href{https://doi.org/10.1103/PhysRevD.75.025003}{\emph{Phys.
  Rev. D} {\bfseries 75} (2007) 025003}.

\bibitem{Hawking:1976ja}
S.W.~Hawking, \emph{{Zeta Function Regularization of Path Integrals in Curved
  Space-Time}}, \href{https://doi.org/10.1007/BF01626516}{\emph{Commun. Math.
  Phys.} {\bfseries 55} (1977) 133}.

\bibitem{DeWitt:1975ys}
B.S.~DeWitt, \emph{Quantum field theory in curved space-time}, {\emph{Phys.
  Rept.} {\bfseries 19} (1975) 295}.

\bibitem{Dowker:1975tf}
J.~Dowker and R.~Critchley, \emph{{Effective Lagrangian and Energy Momentum
  Tensor in de Sitter Space}},
  \href{https://doi.org/10.1103/PhysRevD.13.3224}{\emph{Phys. Rev. D}
  {\bfseries 13} (1976) 3224}.

\bibitem{Kapusta:1981aa}
J.I.~Kapusta, \emph{{Bose-Einstein Condensation, Spontaneous Symmetry Breaking,
  and Gauge Theories}},
  \href{https://doi.org/10.1103/PhysRevD.24.426}{\emph{Phys.Rev.} {\bfseries
  D24} (1981) 426}.

\bibitem{kirsten2001spectral}
K.~Kirsten, \emph{Spectral Functions in Mathematics and Physics}, CRC Press
  (2001).

\bibitem{enwiki:1029909638}
{Wikipedia contributors}, ``Theta function --- {Wikipedia}{,} the free
  encyclopedia.''
  \url{https://en.wikipedia.org/wiki/Theta_function#Integral_representations},
  2021.

\bibitem{BALIAN1978165}
R.~Balian and B.~Duplantier, \emph{{Electromagnetic waves near perfect
  conductors. II. Casimir effect}},
  \href{https://doi.org/https://doi.org/10.1016/0003-4916(78)90083-0}{\emph{Annals
  of Physics} {\bfseries 112} (1978) 165 }.

\bibitem{Ford:1988gt}
L.~Ford, \emph{{Spectrum of the Casimir Effect}},
  \href{https://doi.org/10.1103/PhysRevD.38.528}{\emph{Phys. Rev. D} {\bfseries
  38} (1988) 528}.

\bibitem{Polchinski:1985zf}
J.~Polchinski, \emph{{Evaluation of the One Loop String Path Integral}},
  \href{https://doi.org/10.1007/BF01210791}{\emph{Commun. Math. Phys.}
  {\bfseries 104} (1986) 37}.

\bibitem{ItzyksonZuber1986}
C.~Itzykson and J.-B.~Zuber, \emph{Two-dimensional conformal invariant theories
  on a torus}, {\emph{Nuclear Physics B} {\bfseries 275} (1986) 580}.

\bibitem{Itzykson:1989sy}
C.~Itzykson and J.~Drouffe, \emph{{Statistical Field Theory. Volume 2: Strong
  coupling, Monte Carlo methods, conformal field theory, and random systems}},
  Cambridge University Press (1989).

\bibitem{DiFrancesco:1997nk}
P.~Di~Francesco, P.~Mathieu and D.~Senechal, \emph{Conformal field theory},
  Springer Verlag (1997).

\bibitem{henkel1999conformal}
M.~Henkel, \emph{Conformal Invariance and Critical Phenomena}, Texts and
  monographs in physics, Springer (1999).

\bibitem{Callan:1970ze}
C.G.~Callan, Jr., S.R.~Coleman and R.~Jackiw, \emph{{A New improved energy -
  momentum tensor}},
  \href{https://doi.org/10.1016/0003-4916(70)90394-5}{\emph{Annals Phys.}
  {\bfseries 59} (1970) 42}.

\bibitem{Arnowitt:1962aa}
R.~Arnowitt, S.~Deser and C.~Misner, \emph{The dynamics of general relativity},
   in \emph{Gravitation, an Introduction to Current Research}, pp.~227--265,
  Wiley, New York (1962).

\bibitem{GRADSHTEYN1980904}
I.~Gradshteyn and I.~Ryzhik, \emph{8-9 - special functions},  in \emph{Table of
  Integrals, Series, and Products}, I.~GRADSHTEYN and I.~RYZHIK, eds., pp.~904
  -- 1080, Academic Press (1980),
  \href{https://doi.org/https://doi.org/10.1016/B978-0-12-294760-5.50020-9}{DOI}.

\bibitem{Petkou2018}
A.C.~Petkou and A.~Stergiou, \emph{Dynamics of finite-temperature conformal
  field theories from operator product expansion inversion formulas},
  \href{https://doi.org/10.1103/PhysRevLett.121.071602}{\emph{Phys. Rev. Lett.}
  {\bfseries 121} (2018) 071602}
  [\href{https://arxiv.org/abs/1806.02340}{{\ttfamily 1806.02340}}].

\bibitem{Petkou:2021zhg}
A.C.~Petkou, \emph{{Thermal one-point functions and single-valued
  polylogarithms}},
  \href{https://doi.org/10.1016/j.physletb.2021.136467}{\emph{Phys. Lett. B}
  {\bfseries 820} (2021) 136467}
  [\href{https://arxiv.org/abs/2105.03530}{{\ttfamily 2105.03530}}].

\bibitem{Dewitt:1967yk}
B.S.~De{W}itt, \emph{{Quantum Theory of Gravity. 1. {T}he Canonical Theory}},
  {\emph{Phys. Rev.} {\bfseries 160} (1967) 1113}.

\bibitem{Giulini:1993ct}
D.~Giulini, \emph{{What is the geometry of superspace?}},
  \href{https://doi.org/10.1103/PhysRevD.51.5630}{\emph{Phys. Rev. D}
  {\bfseries 51} (1995) 5630}
  [\href{https://arxiv.org/abs/gr-qc/9311017}{{\ttfamily gr-qc/9311017}}].

\bibitem{Hamber:1997ut}
H.W.~Hamber and R.M.~Williams, \emph{{On the measure in simplicial gravity}},
  \href{https://doi.org/10.1103/PhysRevD.59.064014}{\emph{Phys. Rev. D}
  {\bfseries 59} (1999) 064014}
  [\href{https://arxiv.org/abs/hep-th/9708019}{{\ttfamily hep-th/9708019}}].

\bibitem{10.2307/2310518}
M.~McKiernan, \emph{On the nth derivative of composite functions}, {\emph{The
  American Mathematical Monthly} {\bfseries 63} (1956) 331}.

\end{thebibliography}

\end{document}